%% file: main.tex
\documentclass[onecolumn]{aastex61}
\usepackage{amsmath,amstext}
\usepackage[T1]{fontenc}
\usepackage{apjfonts} 
\usepackage[figure,figure*]{hypcap}
\usepackage{xifthen}

\shorttitle{accepted to ApJ 11/15/2019}
\shortauthors{Lee et al.}

\newcommand{\ORION}{\texttt{ORION2}}

\newcommand{\alfvenmach}{\mathcal{M}_{\rm A}}
\newcommand{\mach}{\mathcal{M}}
\newcommand{\cs}{c_{\rm s}}
\newcommand{\css}{c^2_{\rm s}}
\newcommand{\va}{v_{\rm A}}
\newcommand{\virial}{\alpha_{\rm vir}}
\newcommand{\vrms}{v_{\rm rms}}
\newcommand{\beq}[1][]{%
\ifthenelse{\isempty{#1}{}}{\begin{equation}}{\begin{equation}\label{#1}}}
\newcommand{\eeq}{\end{equation}}
\newcommand{\bcenter}{\begin{center}}
\newcommand{\ecenter}{\end{center}}
\newcommand{\vv}[1]{\mathbf{#1}}
\newcommand{\change}[1]{ {\bf \color{red}  #1 } }

\renewcommand{\change}[1]{ { #1 } }

\renewcommand{\nat}{Natur}

\begin{document}

\title{The Formation and Evolution of Wide-Orbit Stellar Multiples In Magnetized Clouds}

\author{Aaron T. Lee}
\affiliation{Department of Physics and Astronomy, St. Mary's College of California, Moraga, CA 94575, USA; atl8@stmarys-ca.edu}
\author{Stella S. R. Offner}
\affiliation{Department of Astronomy, University of Texas Austin, Austin, TX, 78712, USA}
\author{Kaitlin M. Kratter}
\affiliation{Department of Astronomy, University of Arizona, Tucson, AZ 85721, USA}
\author{Rachel A. Smullen}
\affiliation{Department of Astronomy, University of Arizona, Tucson, AZ 85721, USA}
\author{Pak Shing Li}
\affiliation{Department of Astronomy, University of California Berkeley, Berkeley, CA 94720, USA}

\begin{abstract}
Stars rarely form in isolation. Nearly half of the stars in the Milky Way have a companion, and this fraction increases in star-forming regions. However, why some dense cores and filaments form bound pairs while others form single stars remains unclear. We present a set of three-dimensional, gravo-magnetohydrodynamic simulations of turbulent star-forming clouds, aimed at understanding the formation and evolution of multiple-star systems formed through large scale ($\gtrsim 10^3$ AU) turbulent fragmentation. We investigate three global magnetic field strengths, with global mass-to-flux ratios of $\mu_\phi=2$, 8, and 32. The initial separations of protostars in multiples depends on the global magnetic field strength, with stronger magnetic fields (e.g., $\mu_\phi=2$) suppressing fragmentation on smaller scales. The overall multiplicity fraction (MF) is between $0.4-0.6$ for our strong and intermediate magnetic field strengths, which is in agreement with observations. The weak field case has a lower fraction. The MF is relatively constant throughout the simulations, even though stellar densities increase as collapse continues. While the MF rarely exceeds 60\% in all three simulations, over 80\% of all protostars are part of a binary system at some point. We additionally find that the distribution of binary spin mis-alignment angles is consistent with a randomized distribution. In all three simulations, several binaries originate with wide separations and dynamically evolve to $\lesssim 10^2$ AU separations. We show that a simple model of mass accretion and dynamical friction with the gas can explain this orbital evolution.

\end{abstract}



\input{intro.tex}

\input{methods.tex}
\input{results.tex}
\input{discussion.tex}

\acknowledgments
\change{The authors thank the anonymous referee for substantive comments that improved the quality of this paper.} We additionally thank Hope Chen, Brandt Gaches, David Guszejnov for helpful discussions and suggestions. ATL, SSRO, and KMK acknowledge support from NASA grant NNX15AT05G. KMK additionally acknowledges support from NASA grant 80NSSC18K0726 and RCSA award ID 26077. RS acknowledges support from an NSF Graduate Fellowship. PSL acknowledges support from NASA ATP grant NNX17AK39G. Computations were done on the Massachusetts Green High Performance Computing Center (GHPCC) and the Texas Advance Computing Center (TACC). The plotting and visualization of simulation data was done using the package {\it yt} of \citet{turkyt}.

\facilities{Massachusetts Green High Performance Computing Center, Texas Advance Computing Center -- Lonestar5, Stampede2}  
\software{{\it yt} \citep{turkyt}, \ORION\ \citep{li12} }

\bibliographystyle{yahapj}

\input{main.bbl}

\appendix
\input{appendix1.tex}

\end{document}

%% file: intro.tex
%
{
\section{Introduction}
 
  With nearly half of Solar-type stars in the field having a companion \citep{Raghavan2010,duchene13,moedistefano17} and an even higher incidence of multiplicity in star forming regions \citep{Ghez1993,Leinert1993,chen13,tobin16}, multiplicity is a ubiquitous feature of stellar populations. Multiplicity affects our ability to predict supernova occurrence rates \citep{moedistefano17},  assess stellar populations' role in reionization \citep{ConroyKratter2012,Maetal2016,Rosdahl2018}, estimate the number of potentially stable planetary systems \citep{Takeda_2008,Xie2011paper1,marzarigallina2016}, and calculate an accurate stellar census in distant galaxies \citep{eldridge2009}. The formation and evolution of binary stars and larger multiple-star systems is a central problem in astrophysics.  
 
  The statistics of multiple stars in both the main sequence and pre-main sequence phases are becoming better characterized \citep{Duquennoy1991,duchene13}, but there is not yet a consensus on the physical processes that determine multiplicity. Trends relating stellar mass, age, environment, and separation can be important tools for identifying formation and evolution mechanisms. Observations of star-forming clouds suggest multiplicity rates decrease starting from the young Class I phase down to the older field population \citep{chen13,tobin16}. Multiplicity may be not only common but certain for the youngest Class 0 sources, though the observations are limited because of the resolution needed to probe these deeply embedded protostars \citep{Larson1972binary,chen13}. In both clouds and the field, more massive stars have higher multiplicity frequencies  \citep{kraushillenbrand2007,Raghavan2010,SanaMassive2011}.  However, high stellar density regions (e.g., the ONC) have lower binary fractions than low-mass star forming regions \citep[e.g., Taurus,][]{king2012paper2,king2012paper1,MoeStefano2013},  though among low-mass regions there is a large dispersion in multiplicity statistics \citep{Correiaetal2006,krausetal2011}. Across star-forming regions ranging 5 Myrs to 100 Myrs in age, there is evidence for only slight evolution in the multiplicity fraction. It is clear that multiplicity is primarily determined during the star formation process \citep{Kounkel2019}. 
  
  One of the primary reasons for the lack of a comprehensive binary formation theory is the large range of important, highly non-linear physics involved in star formation. Stars form in large ($\sim$10 parsecs) clouds that undergo gravitational collapse \citep{mckeeostriker07}. Through the interplay of supersonic turbulence, self-gravity, stellar feedback, and magnetic fields, all whose total energies are comparable, filamentary structures form \citep{Arzoumanian2011}. Rotation encourages fragmentation at large and small scales \citep[e.g.,][]{Boss1986v}. Within these, dense cores ($\sim$0.1 parsec) condense into stars. Further fragmentation of the core may lead to binary systems \citep{Boss1988, Burkert:2000,Fisher04,Offner2010,Offner16}. Increased turbulence can increase fragmentation, even at the core scales, while magnetic fields will lower the overall star formation rate \citep{PriceBate2008MNRAS,PriceBate2009MNRAS,padoan12}. How the large-scale properties of natal clouds connect to the frequency and characteristics of formed binary systems remains an open question we begin to investigate in this work. 
  
  Several theories of binary formation have been proposed \citep{Tohline2002,Kratter2011ASPC}. In the turbulent fragmentation models, binary formation occurs when non-linear perturbations cause sub-regions of cores to collapse relative to the background gas, forming individual condensations \citep{Pringle1989,Goodwin2004,Padoan:2002,Offner2010}. The resulting bound objects can have initial separations as large as $10^3-10^4$ AU. In disk fragmentation models, stars form massive protostellar disks that become susceptible to gravitational instabilities \citep{adamsetal1989,Bonnell1994,Whitworth1995,Stamatellos2009,Kratteretal2010}. If a disk becomes unstable, and the gas cools efficiently, it can fragment to produce one or more co-planar companions that accrete from the parent disk, and, depending on the formation epoch, the natal core. Finally, dynamical interactions between initially unbound objects can result in binary formation  \citep{batebonnellbromm,goodwinkroupa2005,MoeckelBally2007}. In this scenario, eventual capture occurs through, for example, dynamical friction with the gas \citep{Indulekha2013} and interactions between slow-moving neighboring gas cores \citep{Tokovinin2017}, or from dynamical interactions involving three or more bodies \citep{Reipurth2012}. \change{This mechanism may play an important role in cases of highly clustered environments where close interactions are common. Finally, star-disk interactions may capture passing protostars to form a binary system, though without including the self-gravity of the cloud, this mechanism is ineffective \citep{ClarkePringle1991}. On the other hand, when the cloud is undergoing global collapse, rapid orbital evolution amongst protostars embedded in generally-converging cores increase the capture rate \citep[][also this work]{bate2018}. However, in this case, the term ``capture'' is ambiguous, given that the local environment of the protostars tends to be gravitationally bound, at least marginally. 
  }
  
  The relative occurrence rate for these mechanisms is debated. For example, observations of Perseus identified a bi-modal distribution of binary separations for Class 0 and Class I objects, with peaks around $10^2$ AU and $3\times10^3$ AU \citep{tobin16}. These peaks are consistent with disk instabilities and turbulent fragmentation scenarios, respectively, which has led to an interpretation of two primary channels for short-period and long-period binaries \citep[e.g.,][]{pineda15,Tobin2016Nature}. The distribution of separations in the field, however, resembles a single-peaked Gaussian, peaking between 10 and 100 AU \citep{Raghavan2010}. The origin of a bi-model distribution, as well as the transition to the field distribution, may be explained by dynamical evolution. Dynamical evolution can evolve wide-orbit binaries to smaller separations in much less than a gravitational free-fall time of the cloud \citep{Offner2010,Offner16}. Other star-forming regions show a distribution of separations that looks closer to the field distribution, which can also be explained by the two above modes of fragmentation plus dynamical evolution \citep{MoeKratter2018}. 
  
  Drawing robust conclusions about dynamical evolution from single-time snapshots of different star-forming  regions is challenging. Instead, large magneto-hydrodynamical simulations of star cluster formation are necessary to better understand the physical processes associated with binary formation as well as how global cloud properties affect multiplicity. \change{Such simulations are now common, but multiplicity studies have typically focused on the distribution at the end of a simulation \citep[e.g.,][]{krumholz12,LiIRDC2018} or have tracked only global averages thoughout the simulation \citep[e.g.,][]{cunningham18}.  } The goal of this paper is to determine how turbulence and magnetic fields effect the formation and evolution of multiple-star system formed through turbulent fragmentation. In these simulations, we identify when and where multiples arise as well as how they evolve during the Class 0/I phase. We initialize three turbulent boxes with varying magnetic field strengths. Our simulations achieve a resolution of either $\Delta x_{\rm f} = 50$ AU, or 25 AU, which is sufficient to resolve the formation of wide-orbit binaries ($\gtrsim 500$ AU) but excludes disk formation mechanisms, since higher resolutions ($\sim$AU) would be required. 
  
 In Section~\ref{sec:methods} we discuss our numerical approach and setup. The results are displayed in Section~\ref{sec:results} and are summarized and placed in a larger context in Section~\ref{sec:summary}.

} 

%% file: methods.tex
%
{
%
\newcommand{\p}{\partial}
%
%
\section{Methods}\label{sec:methods}

Each simulation proceeds through two distinct phases: (1) a driving phase, which generates turbulent initial conditions, and (2) a collapse phase, which follows the gravitational collapse and subsequent star formation. In this section, we discuss the numerical methods used during each phase and our choice of initial conditions. Fundamental parameters of the runs are summarized in Table~\ref{tab:parameters}.

\begin{table}
\bcenter
\caption{Simulation Parameters and Derived Quantities} 
\begin{tabular}{|c|r|r|r|r||r|r|r|r|r|r|} \hline
\phantom{3} & Run Name$^a$ & $\Delta x_{\rm f}$$^b$ & $\alpha_{\rm vir}$ & $\mu_\phi$  & $B_0$$^c$  & $\beta$ & $\alfvenmach$  & $B_{\rm rms}$$^d$  & $\beta_{\rm rms}$ & $\mathcal{M}_{\rm A,rms}$ \\ \hline
{Box Size\ \ $L$}   & MU2  & 50  & 1 & 2   & 25.5 &  0.07 & 1.23 & 31.7 &  0.045 &  0.99 \\ 
 2 parsecs  & MU8  & 50  & 1 & 8   & 6.35 &  1.12 & 4.94 & 20.7 & 0.106 & 1.52 \\ \cline{0-0} 
{Initial Density\ \ $\bar\rho$ } & MU32  & 50  & 1 & 32    & 1.59 &  17.85 & 19.74 & 10.0 & 0.452 & 3.15 \\  
$5.08\times10^{-21}$ g/cm$^3$  & MU2HR  & 25  & 1 & 2   & 25.5 &  0.07 & 1.23 & 31.7 &  0.045 &  0.99 \\  \hline
\end{tabular} \label{tab:parameters}
\footnotetext{\ Written as `MUXY', where X = the value of $\mu_\phi$, and Y = blank for our normal resolution runs with 4 AMR levels or `HR' for our 5-level run. }
\footnotetext{\ Finest cell size in AU. Assumes a base grid of $512^3$. The AMR refinement improves resolution by a factor of two for each AMR level.}
\footnotetext{\ All field strengths are measured in micro-Gauss ($\mu$G). }
\footnotetext{\ Post turbulence driving. The values of $\beta_{\rm rms}$ and $\mathcal{M}_{\rm A,rms}$ are calculated using this value.}
\ecenter\end{table}

\subsection{Numerical Methodology}

We carry out three simulations using the ideal magnetohydrodynamics (MHD) adaptive mesh refinement (AMR) code \ORION\ \citep{li12}. These simulations include self-gravity, ideal magnetohydrodynamics \citep{li12}, and gravitating and accreting sink particles \citep{Krumholz04,myers13,lee14}. \ORION\ also allows for protostellar feedback through outflows \citep{cunningham11} and reprocessing of protostellar radiation through the flux-limited diffusion approximation \citep{krumholz07}, but we reserve exploring how such processes impact multiplicity for a future paper. \ORION\ uses the Chombo library for AMR and load balancing, and an extended version of the Constrained Transport scheme from PLUTO \citep{mignone12} to evolve the system. One benefit of the constrained transport scheme is that $\nabla\cdot\mathbf{B}=0$ is guaranteed to machine accuracy without having to include this constraint in advancing the solution. Details regarding the implementation of these modules in \ORION\ can be found in the references. Below we discuss some of these algorithms in brief, highlighting any parameter choices and assumptions we made. 

\ORION\ solves the following set of equations
\begin{eqnarray}
\frac{\p\rho}{\p t} &=& - \nabla\cdot(\rho\mathbf{v}) - \sum_p \dot{M}_{\rm p}\, W(\mathbf{x}-\mathbf{x}_{\rm p}) \label{eqn:continuity}\\
\frac{\p \rho\mathbf{v}}{\p t} &=& -\nabla\cdot(\rho\mathbf{v}\mathbf{v}) - \nabla \left(P + \frac{B^2}{8\pi} \right) + \frac{1}{4\pi} \mathbf{B}\cdot\nabla\mathbf{B} - \rho \nabla(\phi+\phi_{\rm p}) - \sum_p  \dot{\mathbf{p}}_{\rm p} W(\mathbf{x}-\mathbf{x}_{\rm p})  \label{eqn:euler} \\
\frac{\p\rho e}{\p t} &=& -\nabla\left[ \left(\rho e+P+\frac{B^2}{8\pi}\right)\mathbf{v} - \frac{1}{4\pi} \mathbf{B}(\mathbf{v}\cdot\mathbf{B}) \right]-\rho\mathbf{v}\cdot\nabla(\phi+\phi_{\rm p}) - \sum_p  \dot{\mathcal{E}}_{\rm p} W(\mathbf{x}-\mathbf{x}_{\rm p}) \label{eqn:energy} \\
\frac{\p \mathbf{B}}{\p t} &=& \nabla \times (\mathbf{v}\times\mathbf{B}) \label{eqn:induction}
\end{eqnarray}
These equations are the conservation equations for the mass density ($\rho$), momentum density ($\rho \mathbf{v}$), and energy density ($e$). In addition there is the induction equation for the magnetic field ($\mathbf{B}$). The summations include the contribution from the particles, which can accrete mass, momentum, and energy from the gas. The vectors $\mathbf{x}_{\rm p}$ are the particle locations, and the accretion zone is four finest cells in radius. How mass, momentum, and energy is removed from the gas within the accretion cells ($\dot{M}_{\rm p}$, $\mathbf{\dot{p}}_{\rm p}$, $\dot{\mathcal{E}}_{\rm p}$) depends on a weighting kernel $W \propto \exp(-|\mathbf{r}^2_{\rm p}|/\Delta x^2_{\rm f})$, where $\mathbf{r}_{\rm p}$ is the radial distance from the sink particle to the cell center. Physical constants have their usual symbol assignments. These are closed by the equation of state
\begin{equation}
P = \frac{\rho k_{\rm B} T_{\rm g}}{\mu m_{\rm H}} = (\gamma-1)\rho e\ , \label{eqn:idealeos}
\end{equation}
where $\mu = 2.33$, appropriate for contemporary molecular clouds. Here $m_{\rm H}$ is the hydrogen atom mass. The first equality of Equation (\ref{eqn:idealeos}) is the ideal gas law, which relates the pressure $P$ to the gas temperature $T_{\rm g}$. The second equality relates $P$ to the internal energy through the standard thermodynamic relation. The latter equality requires knowledge of the adiabatic index $\gamma$, the ratio of constant volume and pressure heat capacities. For these runs, we make the isothermal approximation and set $\gamma = 1.0001$.  

\ORION\ evaluates the gravitational potential of the gas $\phi$ and of the sink particles $\phi_{\rm p}$ by solving Poisson's equation and direct sum, respectively:
\begin{eqnarray}
\nabla^2 \phi &=& 4\pi G \rho \label{eqn:poisson}\ ,\\
\nabla \phi_{\rm p} &=& \sum_{{\rm particles\ } {\rm p}} \frac{G m_{\rm p}}{|\vv{r}_{\rm p}|^2 + \Delta x^2_{\rm f} }\, \hat{\vv{r}}_{\rm p} \label{eqn:sinkpotential}
\end{eqnarray}
where the distance vectors are $ \vv{r}_{\rm p} = \vv{x}_{\rm p}-\vv{x}$ and $\hat{\vv{r}}_{\rm p} = \vv{r}_{\rm p}/|\vv{r}_{\rm p}|$. The magnitude of the acceleration from sink particles on the gas and other sink particles is softened by one finest grid cell length. 

Independent of the grid, the sink particles are updated via the $N$-body equations
\begin{eqnarray}
\frac{d m_{\rm p}}{dt} &=& \sum_{{\rm cells\ } i}\dot{M}_i  W(\vv{x}_i-\vv{x}_{\rm p})\ , \\
\frac{d \vv{x}_{\rm p}}{d t} &=& \vv{v}_{\rm p}\ , \\
\frac{d \vv{v}_{\rm p}}{d t} &=& -\nabla\phi - \nabla\phi_{\rm p} + \sum_{{\rm cells\ } i}\frac{\dot{\vv{p}}_i}{m_{\rm p}} \ ,
\end{eqnarray}
where the `cell' sum is done over the accretion zone of particle ${\rm p}$. Here these quantities are evaluated at the particle's location $\vv{x}_{\rm p}$ and $\vv{x}_i$ is evaluated at cell centers. The acceleration contribution from the particles is the same sum as in Equation (\ref{eqn:sinkpotential}), skipping the term corresponding to the sink particle being updated. Sink particles can also merge if several criteria are met: (1) The sink particles are less than $ 4 \Delta x_{\rm f}$ from each other and (2) the least massive sink is less than $0.04 M_\odot$ in mass. This mass limit roughly corresponds to the mass where hydrostatic cores (potentially AU in size) undergo contraction toward condensed protostars (stellar radii in size), after which mergers would become less likely even if the protostars passed within $4\Delta x_{\rm f}\approx200$ AU of each other \citep{Masunaga:2000}. Periodic boundary conditions are used on all gas variables and on the gravitational potentials.

\subsection{Initial Conditions}

Simulations of low-mass star formation must use initial conditions that are consistent with the observed physical properties of star-forming clouds. As first shown by \citet{larson81} and confirmed by many studies, the non-thermal motions within the molecular cloud obey a linewidth-size relation, which relates the non-thermal velocity dispersion of the gas ($\sigma_{\rm nt}$) and the size of the cloud:
\beq[eqn:larson] \sigma_{\rm nt} = 0.72\, \sigma^* \left(\frac{R}{\text{1 pc}}\right)^{1/2}\ \ \text{km/s} \ . \eeq
Here $R$ is the radius of the cloud and $\sigma^*$ is a normalization factor \citep[$\approx1$ for typical star forming clouds,][]{heyerlarson08}. These non-thermal motions are supersonic on large scales, and they dominate the total one-dimensional dispersion. For turbulent non-gravitating flows, $\sigma_{\rm nt}$ relates to the local rms velocity and the sonic Mach number by $\mach \equiv v_{\rm rms}/\cs = \sqrt{3}\sigma_{1D}/\cs \approx \sqrt{3}\sigma_{\rm nt}/\cs$.\footnote{The full relationship between $\sigma_{1D}$ and $\sigma_{\rm nt}$ is $\sigma^2_{1D} = (\css+\sigma^2_{\rm nt}) = (1+3/\mach^2)\,\sigma^2_{\rm nt}$. For supersonic flows, $\sigma_{1D}\approx \sigma_{\rm nt}$.} 
Here $\cs$ is the isothermal sound speed, and $\cs=0.188\, (T/10 \text{ K})^{1/2}$ km/s for contemporary molecular hydrogen gas (which includes helium).

When considering the relative importance of thermal and bulk motions compared to gravity in molecular clouds, galactic and extragalactic clouds are  observed to be virialized with virial parameters
\beq[eqn:virial] \virial = \frac{5 \sigma^2_{1D} R}{ G M_{\rm gas}} \eeq
approximating unity \citep{solomon87,bolatto08,wongetal11}, though the value of $\alpha_{\rm vir}$ can exhibit a wide range \citep[e.g.,][finds a median value of 1.9]{heyerlarson08}.\footnote{For a non-magnetized cloud, $\virial=1$ corresponds to virial equilibrium, and $\virial=2$ is marginally gravitationally bound.} 

Our turbulent initial conditions follow the scaling relations derived in the Appendix of \citet{mckeeliklein10}. These equations derive properties of a turbulent box in terms of four dimensionless numbers: the large-scale value of $\mach$, $\alpha_{\rm vir}$, $\sigma^*$, and $T$. We will initialize all of our turbulent boxes with $\mach = 6.6$ and $T=10$ K, motivated by observations \citep{mckeeostriker07}. By setting $\sigma^*=1$, we restrain the problem of characterizing the turbulent properties of the simulations to one free parameter, $\alpha_{\rm vir}$, which we set to unity.

Additionally, the gas will be magnetized. One way to quantify the strength of the magnetic field is $\mu_\phi$, the ratio of the cloud's mass to the critical mass that can be supported by magnetic fields against gravitational collapse:
\beq[eqn:muphi] \mu_\phi = \frac{M_{\rm gas}}{M_\phi} = \frac{2\pi \sqrt{G} M_{\rm gas}}{\Phi}\ , \eeq
where $\Phi = B_\perp$ times the area perpendicular to the magnetic field. Clouds with $\mu_\phi<1$ are supported against collapse. For simulation boxes, $\Phi=B L^2$ for box size $L$. This can be related to the plasma $\beta$, the ratio of the thermal pressure to the magnetic pressure, or the Alfv\'en Mach number $\alfvenmach = \vrms/\va$, the ratio of the local rms gas velocity to the Alfv\'en velocity, $\va=B/\sqrt{4\pi\rho}$:
\beq[beta] 
\beta = \frac{\rho\css}{B^2/8\pi} = 2\left(\frac{\alfvenmach}{\mach}\right)^2 = \frac{2}{\pi}\left( \frac{1/\sqrt{\rho G}}{L/\cs}\right)^2\, \mu^2_\phi\ .
\eeq
We perform three simulations, using either $\mu_\phi=2$, 8, or 32. 

Following the relations described in \citet{mckeeliklein10}, the simulation box size is $L=2$ pc with an initial mass density of $\bar\rho=5.08\times10^{-21}$ g/cm$^3$ (total gas mass $M_{\rm gas,0}=601 M_\odot$, or $\bar n_{\rm H}=2.17\times10^3$ cm$^{-3}$ using a mean molecular weight of 1.4 times the mass of a hydrogen atom). The initial magnetic field is $B_0=50.9\times10^{-6}/\mu_\phi$ Gauss. These simulations are run for one free-fall time of the gas
\beq[eqn:freefalltime]
t_{\rm ff} = \sqrt{\frac{3\pi}{32G\bar{\rho}}} = 9.34\times 10^5\ \text{years}\ .
\eeq
 
At the end of the turbulent driving phase, derived quantities such as the rms magnetic field strength are also calculated. These values are displayed in Table \ref{tab:parameters}. Because all our initial conditions are super-Alfvenic, there is amplification of the initial magnetic field during the driving phase \citep{federrathetal11a,federrathetal11b}. 

To commence the turbulence driving phase, we initialize a $512^3$ unigrid domain with uniform gas properties and stir the gas according to the prescription of \citet{Dubinski:1995}. This method drives the gas in Fourier space by using a flat power spectrum in the range $1\leq kL/2\pi \leq 2,$ where $k$ is the wavenumber. The ratio of imposed compressive perturbations to divergence-free (solenoidal) perturbations is 1:2. Each driving run uses the same random number generator seed. During the driving phase, self-gravity, sink particles, and AMR are not employed. The gas is advanced for two gas crossing times \citep{maclow99}, which sets up a turbulent power spectrum consistent with supersonic turbulence. At the end of the driving phase, driving is turned off while the remaining physics is turned on and AMR is allowed.

\subsection{Grid Refinement Criteria and Sink Creation}

The power of AMR codes comes from their ability to refine only the areas of interest. This is particularly useful in simulations of star formation, which have large dynamic ranges but which only a small volume requires high resolution to accurately track the collapse of gas into stars. During the simulation, grids can be added and removed based on criteria set by the user.

Our simulations use a base grid (level $l=0$) of $512^3$ and allows up to $l_{\rm max} = 4$ levels of refinement, giving cell sizes $\Delta x_l = L/512/2^l$ and an effective  resolution of $\Delta x_{\rm f} \equiv \Delta x_4 = 50$ AU. One of these simulations is also rerun with five levels of refinement instead. We refine cells on the base and first levels where there are sharp density gradients, namely, we refine a cell when the density gradient $\nabla \rho / \rho$ exceeds 1.0. For all levels, we refine cells of high density to ensure the Jeans length $\sim \pi c_{\rm s}/\sqrt{G\rho}$ is resolved by at least 8 cells. That is, if the density of a cell on level $l < l_{\rm max}$ exceeds $\rho_{\rm TJ}(J=8,l)$, where 
\begin{equation}\label{eqn:jeansrho}
\rho_{\rm TJ}(J,l) = \frac{\pi c^2_{\rm s}}{G J^2 \Delta x^2_l}\ ,
\end{equation}
then the cell is refined to level $l+1$ \citep{truelove98}. This Jeans criterion is employed primarily on the higher levels in regions undergoing gravitational collapse. The refined grid is redrawn every two coarse time steps by recursively calling the above criterion for each level. Finally, sink particles are created on the finest level if the density exceeds $\rho_{\rm TJ}(4,l_{\rm max})$ \citep{Krumholz04}. The sink particle's initial mass is equal to the excess mass in that cell $m_{\rm p,0} = [\rho - \rho_{\rm TJ}(4,l_{\rm max})]\Delta x^3_{\rm f}$. In a region undergoing collapse, many sink particles can form at once. However, sink particles within $4\Delta x_{\rm f}$ of each other merge assuming our above criteria are met.

} 

%% file: results.tex
%
{
\section{Results}\label{sec:results}

In this section, we describe the results of our three $\Delta x_{\rm f}=50$ AU runs, which we label as MU2, MU8, and MU32, depending on the value of $\mu_\phi$ (see Table \ref{tab:parameters}). Section \ref{ssec:globalevo}  describes the evolution of the large-scale morphology and overall star formation during the gravitational collapse phase. This section also comments on our high resolution run MU2HR. Section \ref{ssec:multipleresults} begins by describing our criterion to define multiplicity and then discusses multiplicity statistics and the time evolution of multiple-star systems. 


\subsection{Global Evolution}\label{ssec:globalevo}

The global morphology of the gas depends on the relative strength of the magnetic field to both the strength of the turbulent driving and gravity. During the driving phase, gas collides and shocks, dissipating energy to smaller scales. Magnetic fields resist this compression and density contrasts are overall reduced for strong magnetic fields compared to weaker fields. Once gravity is turned on, gas collapses preferentially along the magnetic field direction, particularly for strong fields, forming denser filamentary structures perpendicular to the field. While all runs were driven with the same large-scale sonic Mach number {$\cal M$}, the Alfv\'en Mach number {${\cal M}_{\rm A}=v_{\rm rms}/v_{\rm A}$} increases with decreasing field strength. This allows for stronger shocks perpendicular to the field direction, further enhancing the density contrast in MU32 compared to MU2. The end-of-simulation gas structure for all three runs is shown in Figure \ref{fig:finaloutput}, which plots six column density snapshots after one free-fall time of the collapse phase. The column density is defined as 
\begin{equation}\label{eqn:columndensity}
\Sigma_{\rm g} = \int^{L/2}_{-L/2} \rho\, dx\ .
\end{equation}
The top panels integrate along the $x$-axis, while the bottom panels integrate along the $z$-axis. The arrow in the upper-left corner of the rows shows the initial direction of the magnetic field. All panels share the same horizontal axis (in this case, the $y$-axis from the simulation outputs). The dots show protostars that have formed in the simulation. Note that the denser filamentary structures are nearly perpendicular to the original direction of the field in the strong field case, whereas the orientation of the filaments is more random in the other two runs.

\begin{figure}\label{fig:finaloutput}
\bcenter
\includegraphics[scale=0.5]{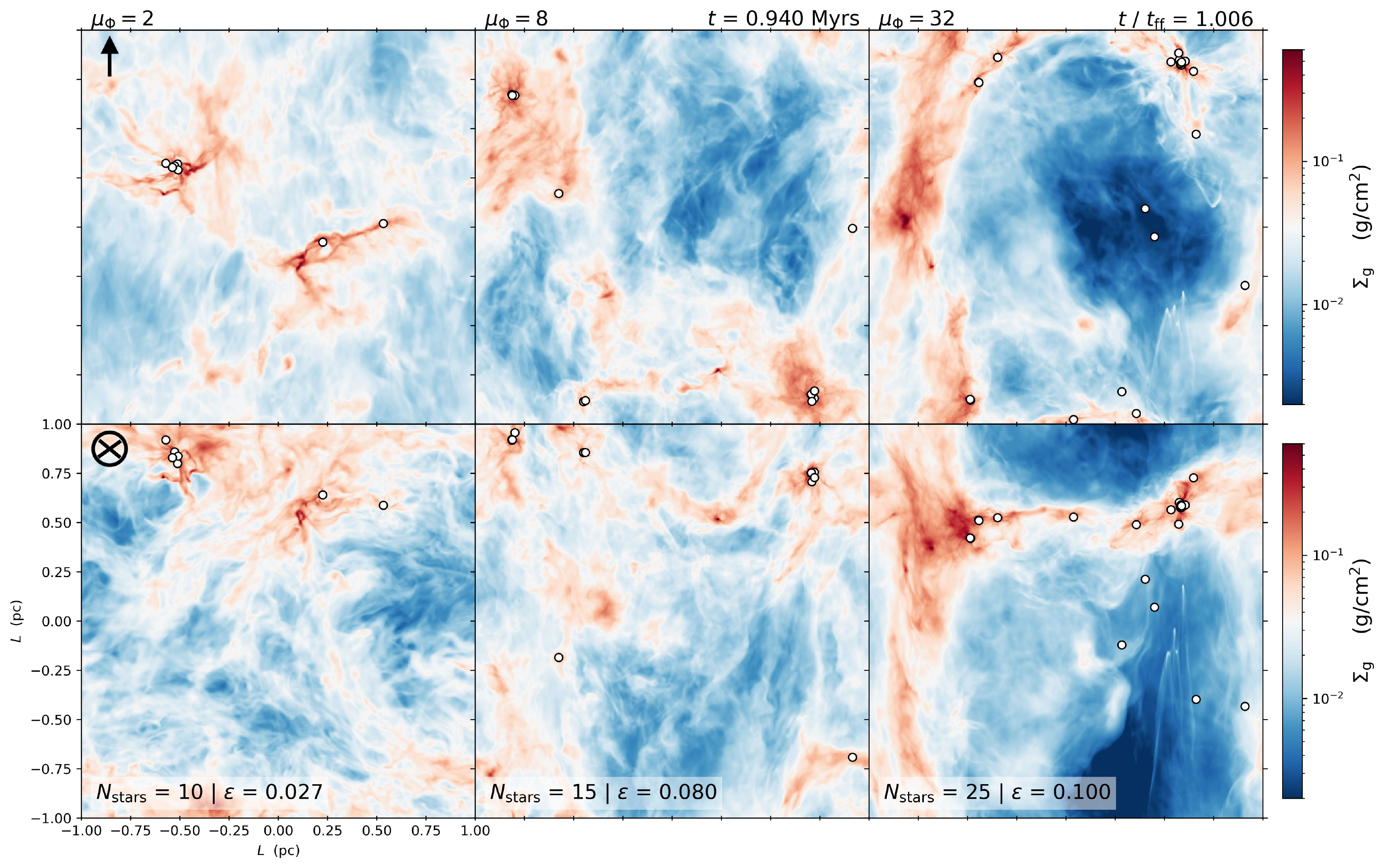}
\ecenter
\caption{Column density $\Sigma_{\rm g}$ (Equation \ref{eqn:columndensity}) snapshots after one free-fall time.  Columns represent the three different $\mu_\phi$ values. Rows differ by the direction of integration, either along the $x$-axis (Top) or $z$-axis (Bottom). The arrows in the left column indicate the direction of the initial magnetic field (the $z$-direction). All panels share the same horizontal axis (the $y$-axis from the original simulation output). Formed protostars are labeled as circles. The number of stars and the star formation efficiency (Equation \ref{eqn:sfe}) are shown in the low-left of each panel.}
\end{figure}

In all three runs, stars begin to form around $\sim0.5 t_{\rm ff}$. For MU8 and MU32, formation of the initial protostar is quickly followed by subsequent star formation, whereas in MU2 it takes until $\sim0.8 t_{\rm ff}$ for additional protostars to form. Figure \ref{fig:totalsinks} shows the total protostar count and mass as a function of time. Overall 10--20 protostars form throughout the simulation, and the total protostar mass $\sum_p m_{\rm p}$ ranges from 10--60 $M_\odot$. Equivalently, the total protostellar mass can also be translated to a star formation efficiency
\begin{equation}\label{eqn:sfe}
\epsilon = \frac{\sum_p m_{\rm p}}{M_{\rm gas,0}}\ .
\end{equation}
This efficiency is also shown in Figure \ref{fig:totalsinks}. The star formation efficiency at the end of one free-fall time increases with decreasing magnetic field strength and ranges from a few percent to ten percent.

\begin{figure}\label{fig:totalsinks}
\bcenter
\includegraphics[scale=0.6]{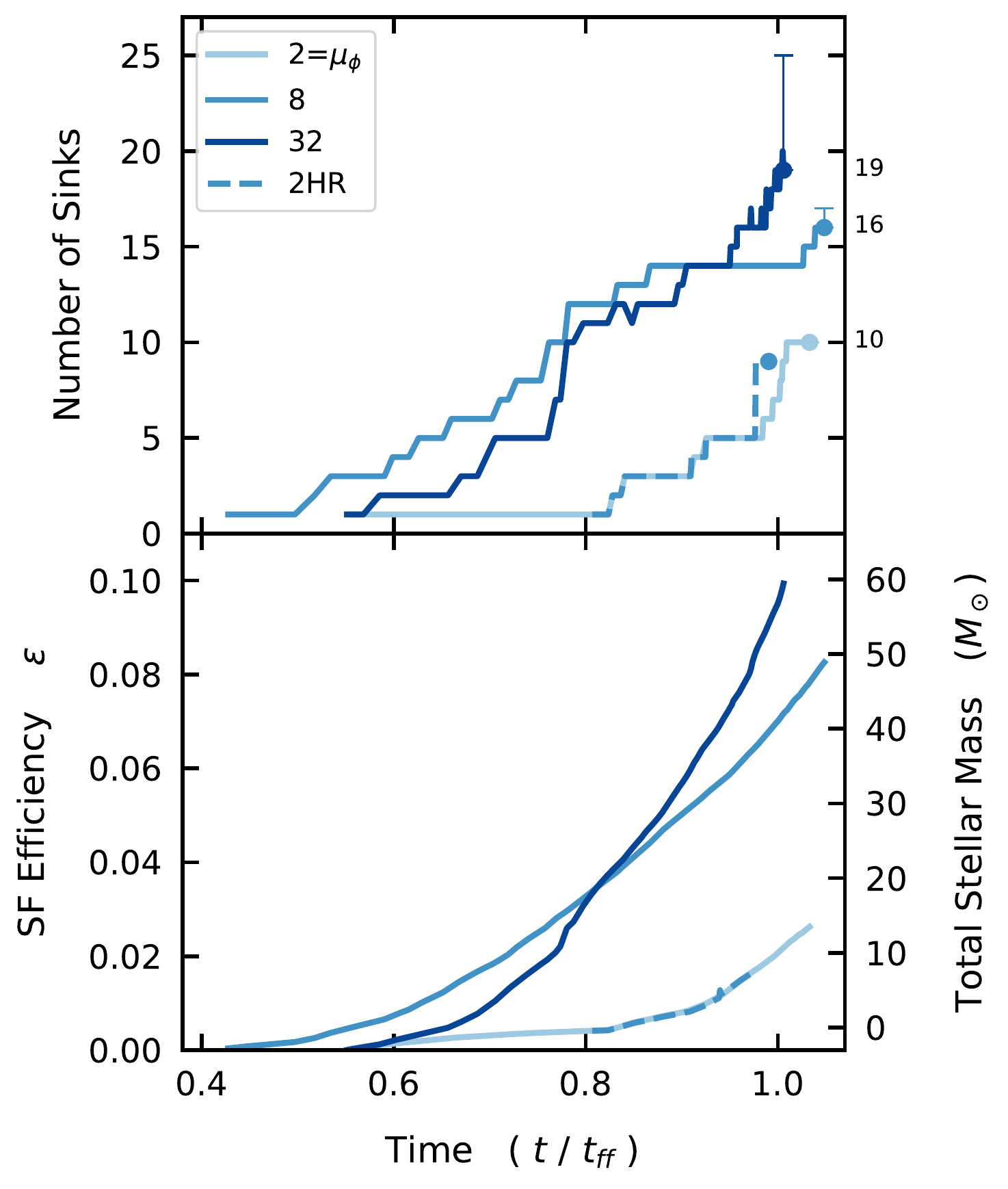}
\ecenter
\caption{{\it Top:} Total number of protostars/sink particles as a function of time. Solid lines show the number of protostars above the minimum required mass for multiplicity considerations ($m_{\rm p}>0.02M_\odot$). Error bars at the end points include smaller sink particles. {\it Bottom:} Star Formation Efficiency, $\epsilon$, calculated using all sink particles and the initial gas mass of $M_{\rm gas,0}=601\, M_\odot$ (Equation \ref{eqn:sfe}). The right $y$-axis converts to the total mass in protostars. \change{In both panels, the MU2HR run is shown as a dashed line, displaying that the normal and HR runs are not diverging from one another.} All times are normalized to the free-fall time.}
\end{figure}

In Figure \ref{fig:totalsinks} and our subsequent multiplicity results, we only consider protostars that are above $0.02\, M_\odot$; however, fragmentation can produce smaller objects. These typically merge with existing protostars but sometimes persist. The whisker bars in Figure \ref{fig:totalsinks} show the numbers when all sink particles are included at the end of the simulations. There are six additional sink particles for MU32 and one additional sink for MU8. In the MU32 panels of Figure \ref{fig:finaloutput}, several of these smaller objects reside in the bowshocks seen in the less-dense gas. For this simulation, the highly-clustered environment collapsed to form almost a dozen sink particles. Typically most of these sink particles merge together within a few time steps to form a low-mass protostar. However, several were tossed out before this occurred. Higher resolution may have prevented these low-mass sinks from escaping before merging. In our case, they remain as artifacts of the collapse but also remain too low-mass to be considered in our multiplicity statistics. 

For our high resolution simulation, we re-run part of the MU2 run starting after the formation of the first protostar. MU2HR is also shown in Figure \ref{fig:totalsinks} as a dashed curve atop the MU2 curves. \change{The two curves agree almost perfectly until around $0.96 t_{\rm ff}$, at which point the high-resolution run begins forming additional protostars. The number of protostars as a function of time in the high resolution run, however, flattens out near $0.99 t_{\rm ff}$, rather than continuing to diverge from the normal-resolution curve. Similar results are anticipated for the other runs as well, though the amount of additional fragmentation would likely increase as $\mu_\phi$ increases. Nonetheless, the agreement between MU2 and MU2HR demonstrates that our results studying turbulent fragmentation should not be sensitive to the resolution of our simulations. Since turbulent fragmentation occurs on scales comparable to the Jeans length ($>10^3$ AU), turbulent fragmentation length scales are well-resolved in all our simulations. Further increasing the resolution would increase the number of protostars formed, but also introduce new physical processes (i.e., disk fragmentation) that are not the focus of this paper.  }

Figure \ref{fig:sfrovertime} shows the instantaneous mass accretion rate onto all the protostars, 
\begin{equation}\label{eqn:sfrinstant}
\text{SFR} = \frac{\sum_p \dot{m}_p }{M_{\rm gas,0}/t_{\rm ff}}\ .
\end{equation}
Outputs subsequent in time are used to estimate the instantaneous mass accretion rate onto all protostars; in this case the accretion rates are actually averaged over $\Delta t_{\rm IO}=0.08$ kyrs. As collapse proceeds and star formation accelerates, this rate reaches  0.2--0.4 after one free-fall time, modulo sporadic bursts of star formation (seen particularly in the MU32 run). An alternative measure of this star formation rate replaces the numerator of Equation (\ref{eqn:sfrinstant}) with $\sum_p m_p /t_{\rm SF}$, where $t_{\rm SF}$ is the total amount of time elapsed since the formation of the first protostar. This is sometimes called the star formation rate per free-fall time \citep{krumholzmckee2005}. With the latter definition, SFR = $(t_{\rm ff}/t_{\rm SF})\epsilon$. Figure \ref{fig:sfrovertime} shows this quantity as dashed curves. This smoother average has the same trend as the original definition of SFR but ranges from 0.05 to 0.22. These values are consistent with results from other works \citep{padoan12,FederrathStarFormLaw}. However, we note the caveat that our simulations do not continue to drive turbulence once gravity is turned on, which increases our star formation efficiencies relative to simulations that continuously drive large-scale motions \citep[e.g.,][]{LiIRDC2018}.

\begin{figure}\label{fig:sfrovertime}
\bcenter
\includegraphics[scale=0.5]{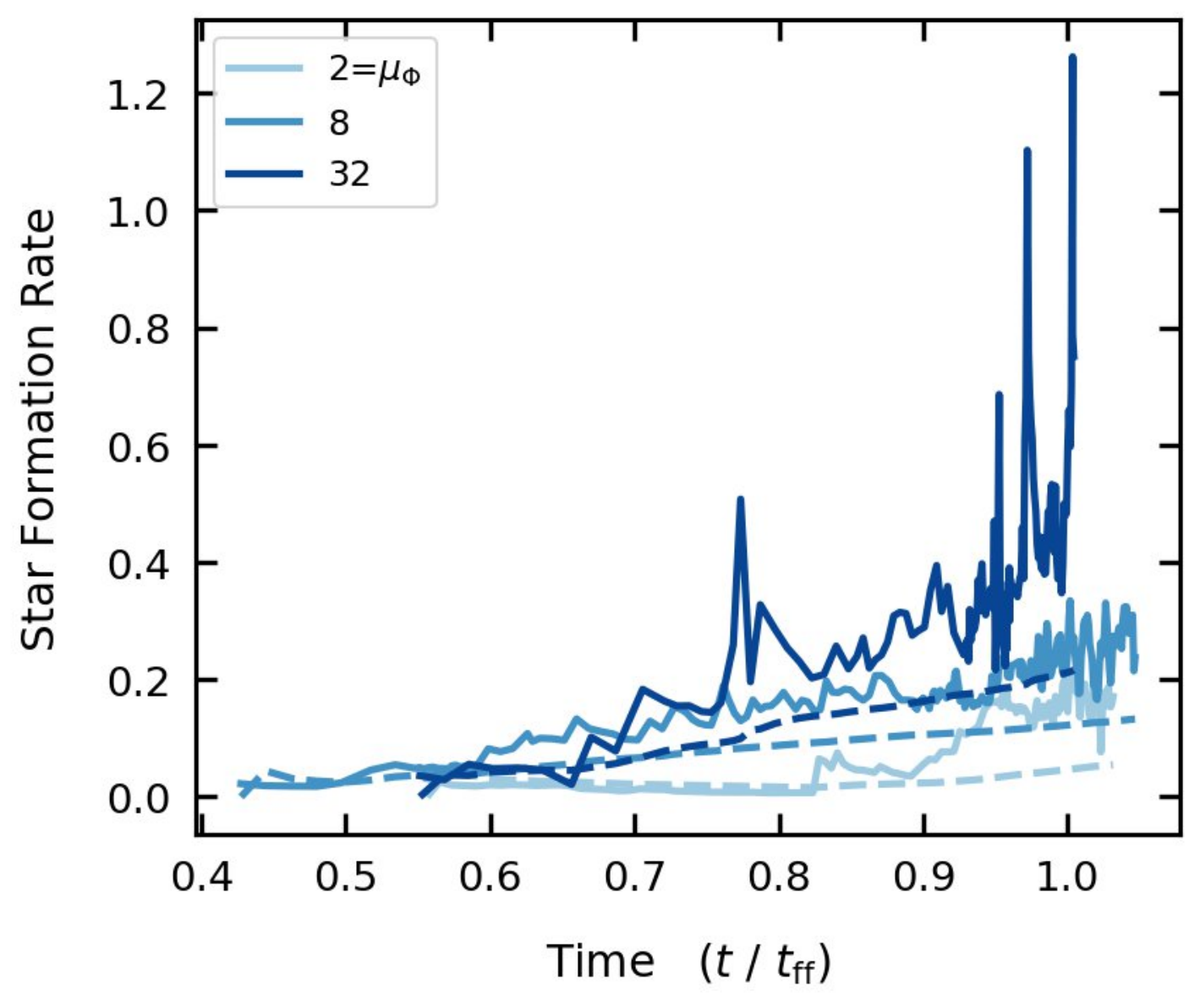}
\ecenter
\caption{The instantaneous total star formation rate, $\sum_p \dot{m}_p$, normalized to $M_{\rm gas, 0}/t_{\rm ff}$. Sporodic rapid accretion in MU32 leads to bursts in the star formation rate, but the overall trend is that the star formation rate increases with $\mu_\phi$ from 10\% to 40\% at the end of one free-fall time. The dashed curves display $\sum_p m_p / t_{\rm SF}$, also normalized to $M_{\rm gas, 0}/t_{\rm ff}$. The same trend is observed.}
\end{figure}

The time evolution of protostar numbers and masses are shown in Figure \ref{fig:sinkmassevo}. After one free-fall time, the most massive sink in MU2 only reaches $6 M_\odot$, whereas MU8 and MU32 have massive protostars above $10 M_\odot$. The accretion rate onto a sink particle inside a collapsing core is 1 to 2 times $10^{-5}\, M_\odot$/yr, on average, for each of the simulations. Protostars that are dynamically ejected from their self-gravitating cores accrete instead at rates that depend on the average strength of the magnetic field, the local gas density, and the relative velocity between the star and the gas \citep[i.e., magnetic Bondi-Hoyle accretion,][]{lee14,Burleighetal2017}. These stars may then compete for the reservoir of gas they roam through \citep{Bonnell2001}. In our simulation these stars are ejected into a low-density environment where their subsequent accreted mass is negligible. 

\begin{figure}\label{fig:sinkmassevo}
\bcenter
\includegraphics[scale=0.5]{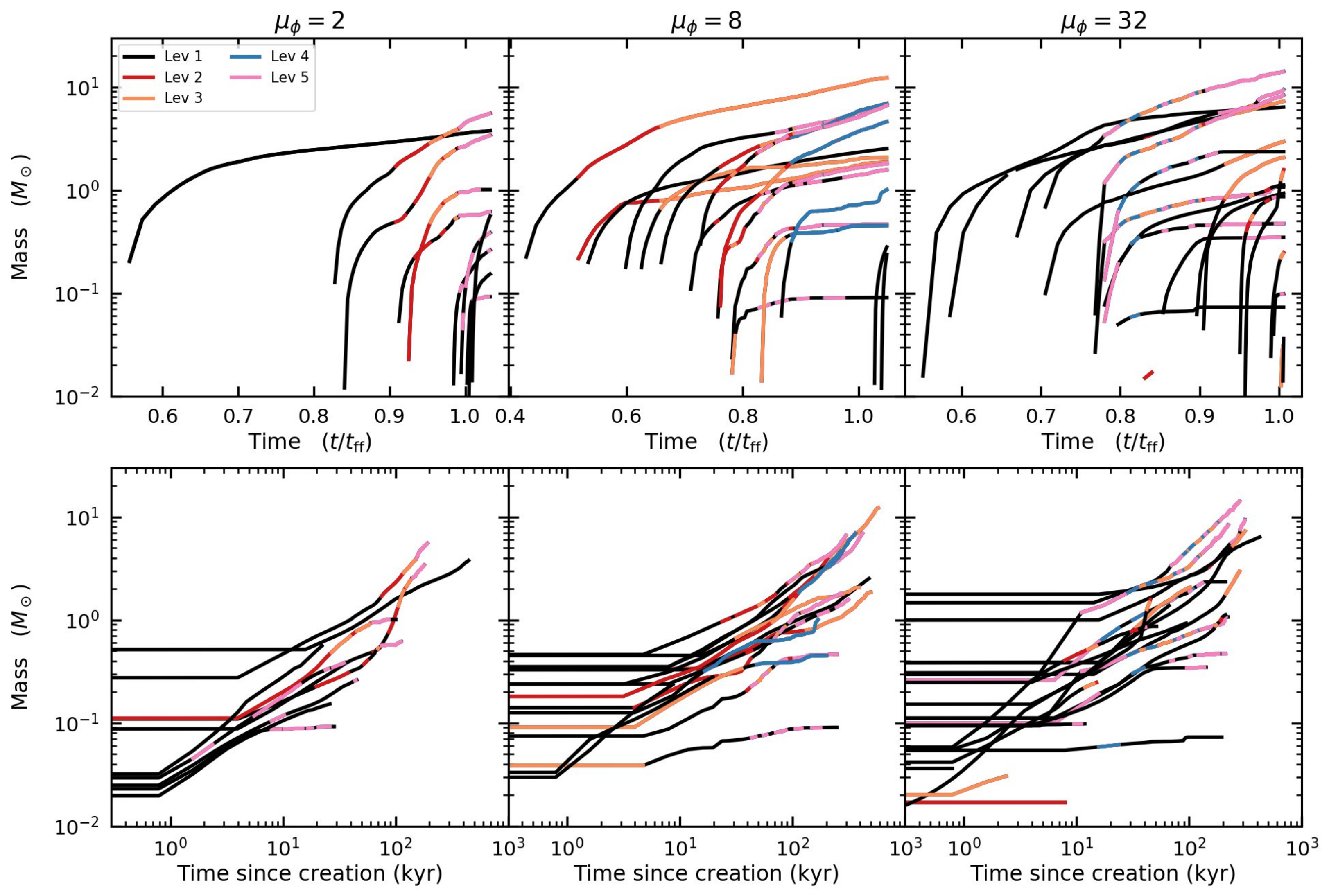}
\ecenter
\caption{Masses of all protostars as a function of time. Columns separate data by simulation. The top row plots mass as a function of simulation time, whereas the bottom row plots mass as a function of protostar age. Color overlays show the highest level multiple the star resides in at that time.} 
\end{figure}

Figure \ref{fig:stellarages} shows the distribution of stellar ages at the end of one free-fall time. Approximately one-third of the stars have formed in the last 10\% of a free-fall time ($0.1t_{\rm ff}\approx100$ kyrs), with the remaining forming between 50\% and 90\% of a free-fall time after the beginning of the collapse phase. Astronomers separate protostars into `classes' based on observed infrared spectral energy distribution \citep{Lada1987}. While there is not a definitive consensus, the timescale between Class 0 objects and Class I objects is $\sim160$ kyr \citep{enoch08b,dunhametal2013,fischeretal2013}. 
Since classes are defined by observations, we instead use a `Stage' designation to separate objects that are younger and older than 160 kyrs (Stage 0/I, respectively). Approximately 60\% of the protostars are Stage I objects, with the remaining 40\% as Stage 0. We will return to Figures \ref{fig:sinkmassevo} and \ref{fig:stellarages} in section \ref{ssec:multipleresults} when we discuss the end-state multiples. 

\begin{figure}\label{fig:stellarages}
\bcenter
\includegraphics[scale=0.5]{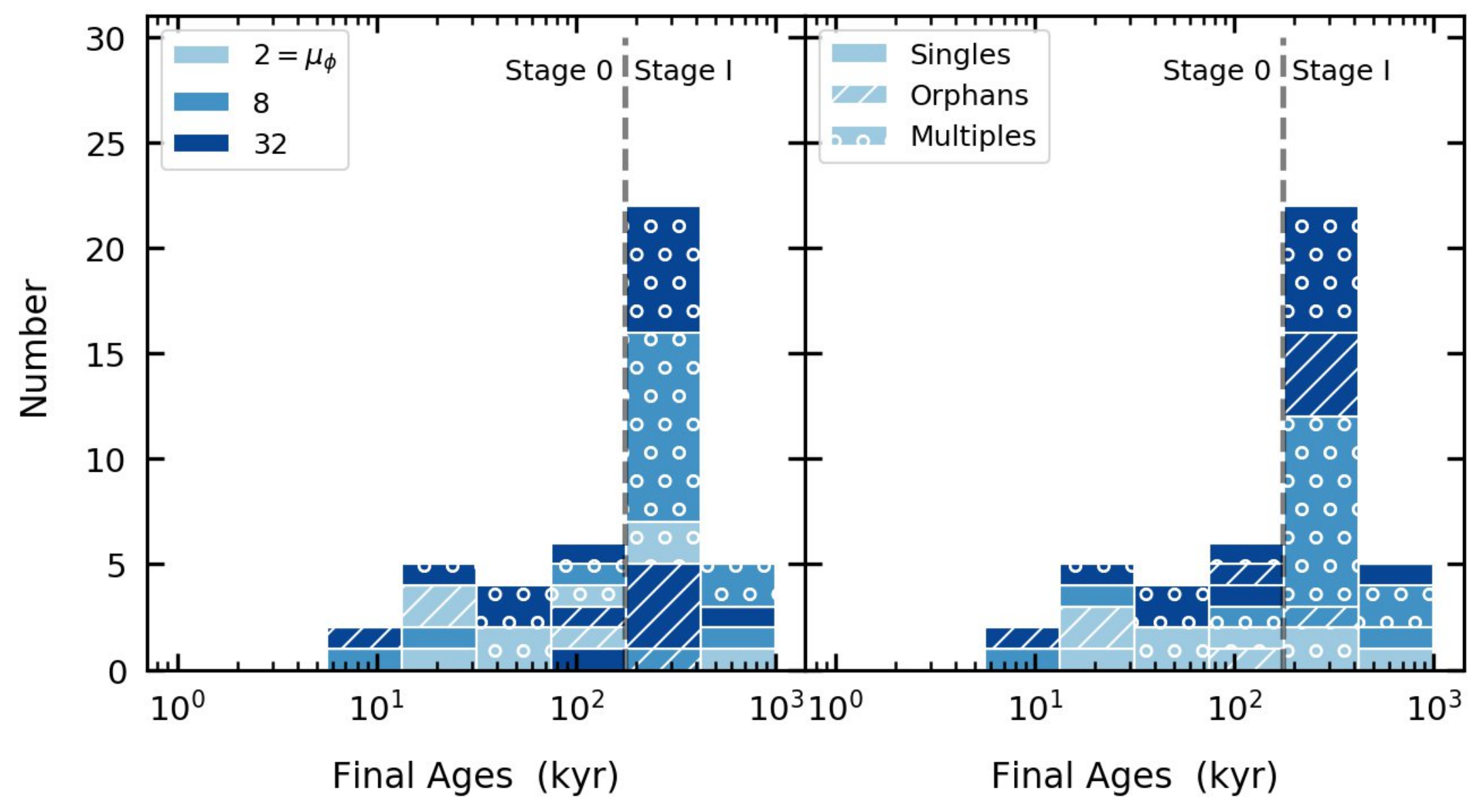}
\ecenter
\caption{Stacked histograms showing the ages of the protostars at the end of the simulations. Colors distinguish between simulations and hatches distinguish between the singles, orphans, and multiples. The same data are shown in both plots. The left plot stacks the data first by type, and the right plot stacks first by $\mu_\phi$. The delineation between Stages marks stars youger and older than 160 kyrs.}
\end{figure}


\subsection{Multiplicity Formation and Evolution}\label{ssec:multipleresults}

\subsubsection{Multiplicity Criterion}\label{sssec:multiples}

To identify multiple star systems, we consider a set of ``objects''--one or more protostars grouped together. For all possible pairs of objects, we identify the most-bound pair as the extremum of the pairs with a negative total orbital energy $E_{\rm tot}$, as measured in the center-of-mass frame. To identify higher-order multiples, we then treat this bound pair as a single  object that has the total mass and center-of-mass position and velocity of the original two objects (which can be protostars or bound pairs themselves). This procedure is recursively applied until no more bound pairs are found. By associating a unique id to each object, we can connect multiples across simulation outputs.

We describe the hierarchy of multiple-star systems by assigning a numerical `level' $l$ to each bound object. Protostars are labeled as level 1 objects. Two bound protostars form a binary system, which we label as a level 2 object. A binary bound to another binary or another single protostar is a level 3 object, and so forth. In general, a level $l$ object contains at least $l$ protostars and at most $2^{l-1}$ protostars. We restrict our classification to objects greater than $0.02\, M_\odot$ and closer than $10^5$ AU. We do not trace hierarchies beyond level 5, as these can encompass the bulk of the cluster and is more indicative of the boundedness of the initial cloud than multiplicity. Additional discussion regarding our multiplicity criterion is given in Appendix \ref{sec:appmultiplicity}.

Figure \ref{fig:cartoonhierarchy} shows an example result of this algorithm from run MU32. Of the 11 sink particles, there are three binary pairs. Two of those binary pairs are bound to a third star, and the resulting two triple systems are gravitationally bound to one another. In highly clustered environments, the hierarchy can change from output to output. The inclusion of the gravitational potential from the local gas can eliminate some output-to-output changes in the hierarchy, but only when the gas mass is comparable to the protostellar masses. We show in Appendix \ref{sec:appmultiplicity} that ignoring the gas mass in determining multiples does not considerably change the overall multiplicity statistics. 

\begin{figure}\label{fig:cartoonhierarchy}
\bcenter
\includegraphics[scale=0.75]{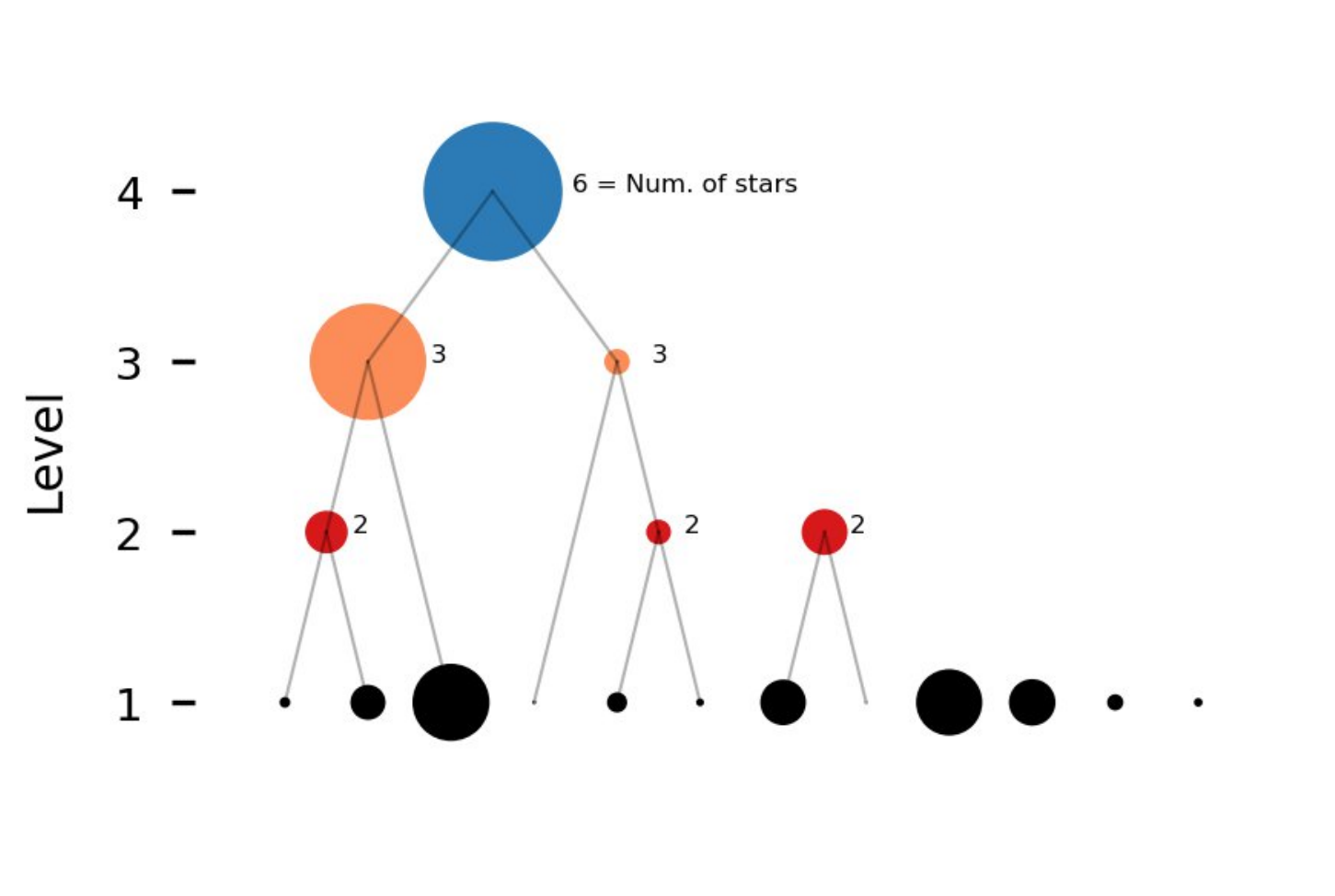}
\ecenter
\caption{A cartoon example of the multiplicity hierarchy. The data is taken from $\mu_\phi=32$ at $t=0.84 t_{\rm ff}$. Individual protostars are at level $l=1$, and binaries are at level $l=2$. There are two level 3 objects; both are binaries bound with a third protostar. Finally, one level 4 object is composed of the two triples. In this case, no level 5 objects exist. Point radius scales with total mass, ranging from 0.02$M_\odot$ to 6$M_\odot$ on level 1.}
\end{figure}

The semi-major axis $a$ of a bound pair's orbit is calculated using the total energy of the pair:
\begin{equation}
    \label{eqn:virialintext}
     a = - \frac{Gm_1m_2}{2 E_{\rm tot}}\ .
\end{equation} 
Figure \ref{fig:semiactualdistcompare} displays the primordial distribution of separations, comparing the instantaneous actual separation between the pair's objects with the calculated semi-major axis. The primordial distribution is not taken from a single simulation output. Instead, we identify all unique pairs that exist throughout the entire simulation. For pairs that exist for at least two subsequent data outputs, we take the distance at the first output as the primordial separation. The average discrepancy between the two measurements increases with increasing actual separation. Initial pairs at these large separations begin with highly eccentric trajectories that substantially evolve before they can complete one full period, after which they settle into closer, more-bound orbits. The primordial semi-major axes of these weakly bound objects in eccentric orbits thus overestimates the actual separations, sometimes by an order of magnitude. 
 
To facilitate comparison with observations, we also show the two-dimensional projected separations in Figure  \ref{fig:semiactualdistcompare} as unfilled symbols. For each pair, three projections are made, one along each Cartesian direction, and the minimum projected distance is plotted. The ratio between actual separations and the projected separations are $\sim$factors of a few, less than the average difference between these separations and the semi-major axis calculations (especially for larger orbits). 

\begin{figure}\label{fig:semiactualdistcompare}
\bcenter
\includegraphics[scale=0.5]{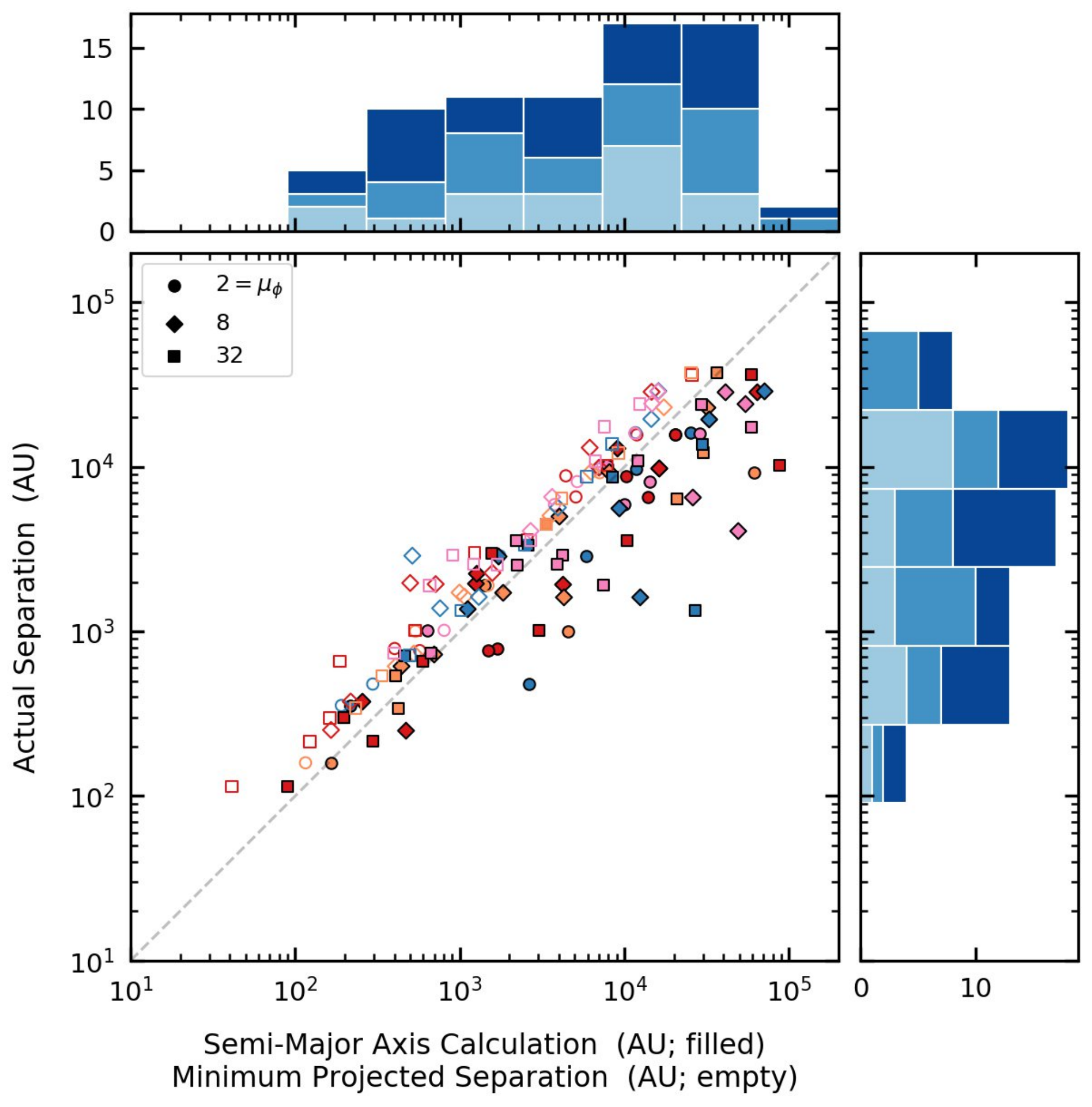}
\ecenter
\caption{ A comparison of the  primordial 3D separations of all multiples that form throughout the simulations to their calculated semi-major axis values (filled symbols) and the minimum projected distance values (empty symbols). The projection distances are calculated along each of the Cartesian directions, and the minimum value is plotted. Symbol shapes denote the simulation and the colors identify the pair's level (level 2: red, 3: orange, 4: blue, 5: pink). Histogram distributions are stacked by $\mu_\phi$ value and show the 3D separations and semi-major axis calculation values.}
\end{figure}

\subsubsection{Multiplicity Statistics}\label{sssec:multiplestats}

We compare our results to observations using two metrics, the multiplicity fraction \begin{equation}
    \label{eqn:MF}
    \text{MF} = \frac{\text{B+T+F+...}}{\text{S+B+T+F+...}}\ ,
\end{equation}
where S, B, T, and F are the number of single, binary systems, triple systems, four-star systems, etc.\footnote{Note that these are not equivalent to our ``levels.''}, and the companion star fraction
\begin{equation}
    \label{eqn:CSF}
    \text{CSF} = \frac{\text{B+2T+3F+...}}{\text{S+B+T+F+...}}\ 
\end{equation}
\citep{Batten1973book,ReipurthZinnecker93}. 
The former calculates the fraction of stellar objects that are members of bound systems, and the latter estimates the average number of stellar companions per system. Figure \ref{fig:mfandcsf} displays the time evolution of MF and CSF for all of our runs. Additionally, this plot shows the observed results from Perseus for Class 0 objects and the full observed sample \citep{tobin16}. The end-state values and uncertainties for MF and CSF are calculated in Table \ref{tab:mf}.\footnote{Since the number of multiple systems is not considerably larger than the total number of systems, binomial statistics are used for the uncertainty in MF. Since the CSF value can be greater than unity, we use Poisson statistics in calculating the CSF uncertainty.} Figure \ref{fig:mfandcsf} shows that there is a clear trend for higher MF and CSF values with stronger magnetic fields, and that the fractions for strong fields better match the multiplicity statistics from \citet{tobin16}. Weaker fields produce multiples with both small and large initial separations (Figure \ref{fig:semiactualdistcompare}), and the stellar environments become crowded almost immediately. The dynamical interactions between protostars ultimately disrupt many of the multiples that form. In contrast, stronger fields form multiples with generally larger initial separations. The separation between multiples is larger as well. These multiples, once formed, have fewer interactions with other protostars, or interact with neighboring protostars only after the stars have existed for some time.

\begin{table}
\bcenter
\caption{End-State Multiplicity Statistics Near One Free-Fall Time compared to \citet{tobin16}} 
\begin{tabular}{|c|c|c|} \hline
Simulation $\mu_\phi$ & MF & CSF  \\ \hline
2 (Strong field) & $0.33 \pm 0.17$ & $1.33 \pm 0.67$ \\
8 & $ 0.5 \pm 0.20$ & $1.33\pm0.47$  \\
32 (Weak field) & $ 0.17 \pm 0.11 $ & $0.5\pm0.20$ \\ \hline
\citet{tobin16} &  &  \\
Class 0 Sample  & $0.57\pm 0.09$     & $1.2\pm 0.2$   \\ 
Full Sample  & $0.4 \pm 0.06$ & $0.71\pm 0.06$  \\ \hline 
\end{tabular} \label{tab:mf}
\ecenter\end{table} 

\begin{figure}\label{fig:mfandcsf}
\bcenter
\includegraphics[scale=0.5]{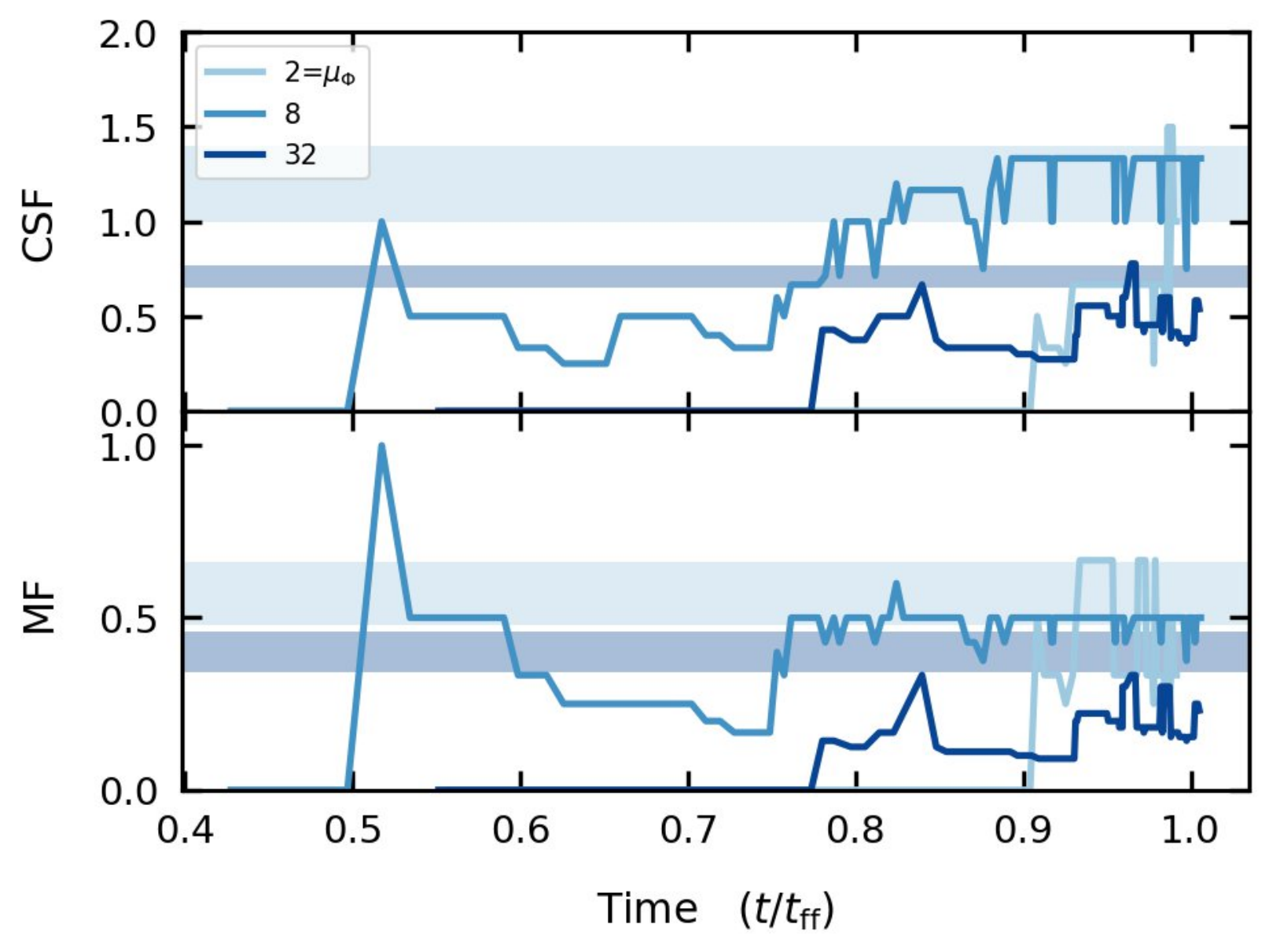}
\ecenter
\caption{Companion star fraction (top; Equation \ref{eqn:CSF}) and multiplicity frequency (bottom; Equation \ref{eqn:MF}). Bars are taken from the observations of Table 6 of \citet{tobin16}. Upper lighter bars are the Class 0 sources, the bottom bars are the full sample.}
\end{figure}

While we stated above that the number of members in each level above $l=2$ can have between $l$ and $2^{l-1}$ protostars, the realized hierarchical structure strongly favors level $l$ having $l$ protostars. Additional stellar companions are added one at a time rather than binding multi-star objects to one another. Comparing all unique pairs from all three simulations, the average number of protostars in objects at levels 3--5 is $3.16$, $4.32$, and $5.05$, respectively. Further weighting by the amount of time a given pairing exists in a simulation reduces these means closer to $l$.

Over time, as gravity causes more gas to undergo collapse, one expects the star formation rate to accelerate and the stellar density to increase \citep{pallastahler2000,leechangmurray2015}. Interestingly, the values of MF remain relatively constant around 0.5 for MU2 and MU8 and 0.25 for MU32, suggesting the multiplicity fraction is a stronger function of the global magnetic field than the stellar density. Stellar density, instead, creates transient variations in MF and CSF when multiples in clustered environments exchange members. We note, however, that the magnitude of these fluctuations appears to decrease with protostar count, and therefore these fluctuations are likely more sensitive to the total number of protostars in the simulation domain. For a given $\mu_\phi$, simulating a larger simulation boxes would produce more protostars and thus reduce the magnitude of these variations. 

The constancy of the MF values does not necessarily imply that multiples are long-lived in all three simulations. While the value of MF is always $\lesssim 0.5$, over 80\% of all the protostars are part of bound pairs at some point during the simulations. For example, while MU32 forms more overall protostars and has a lower MF compared to the MU2 and MU8 runs, only 3 of the 19 protostars were never part of a bound multiple system. Rarely do we see single isolated protostars. The protostar mass distribution at the end of the simulations is shown in Figure \ref{fig:endtimesnapshots}. This figure labels protostars as either `singles,' meaning these stars were never part of a multiple throughout the entire simulation; `orphans,' meaning they are not part of a multiple at the end of the simulation but were in a multiple at some point; and `multiples,' meaning they are part of a multiple at the end of the simulation. Single, orphan, or multiple status is also shown in Figure \ref{fig:stellarages}. The highest level a protostar resides in at a given point in time is shown in Figure \ref{fig:sinkmassevo}. It is clear to see that most protostars enter and leave multiples throughout the simulation.

\change{We briefly note that the resulting distribution of protostellar masses differs from the observed IMF in star forming regions \citep{kroupa01,chabrier2003}. Our distribution lacks the lower mass ($<M_\odot$) protostars needed to match the observed IMF. This is not surprising since these simulations focus on protostars forming through turbulent fragmentation only. At the resolution of these simulations, protostars grow rapidly during the first few thousand years of their lives. As mentioned in Section \ref{ssec:globalevo}, increased resolution would allow for smaller-scale fragmentation and the formation of protostars with even smaller masses. Additionally, protostellar outflows will also reduce star formation efficiency by entraining and expelling dense gas, thereby lowering the median mass of the protostars from $\sim M_\odot$ to the observed value of $\sim0.2 M_\odot$ \citep{offnerarce2014,offnerchaban2017}. }

\begin{figure}\label{fig:endtimesnapshots}
\bcenter
\includegraphics[scale=0.5]{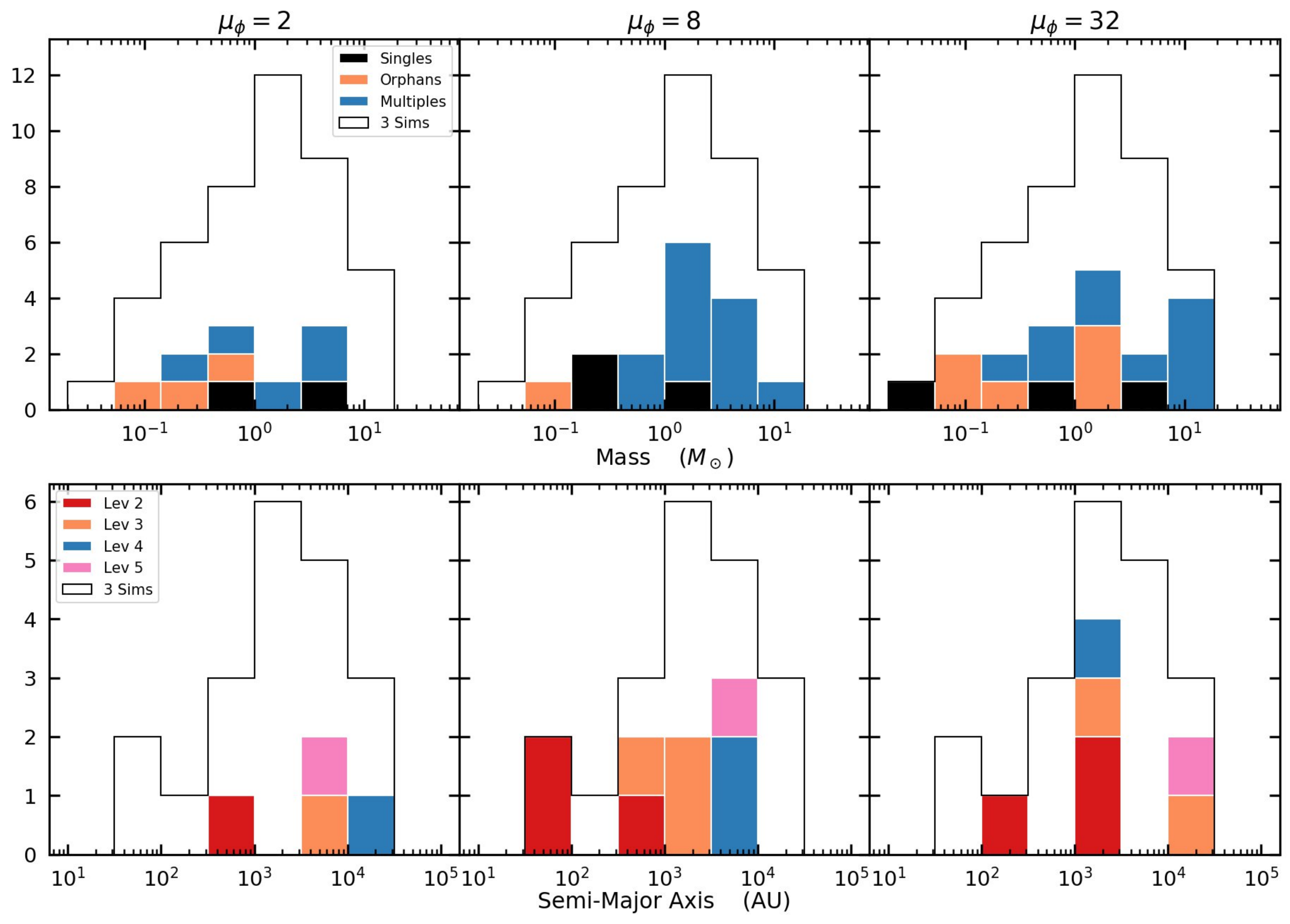}
\ecenter
\caption{ Stacked histograms showing the mass distribution of all protostars and semi-major axis distribution of all multiples at the end of each simulation. The black outline shows the sum of all three simulations. }
\end{figure}

\subsubsection{Spin Mis-alignments of Binaries}\label{sssec:spinmisalignment}

When turbulent fragmentation occurs, single cores fragment into multiple gravitationally bound protostars. For a simple binary star system, the orbit and protostellar spins will retain the angular momentum direction of their local natal core gas. However, dynamical interactions with other protostars, asymmetric mass accretion through magnetic fields, or perturbations from the turbulent gas can reorient the spin directions of the protostars. A number of protobinary
systems with misaligned outflows have been observed \citep{chen2008,leeoutflows2016}. The left and right panels of Figure \ref{fig:misalignment} shows the cumulative distribution function (CDF) of the projected  mis-alignment angles between the two protostars of every binary that forms during the simulations. The left panel shows the primarodial mis-alignment angles and the right panel shows the angles measured at the last available output for each binary (which is either the last output the binary existed or the last output of the simulation). Angles are measured in all three Cartesian planes and we treat each projection as a separate ``observation." Additionally, we create a set of mock observations of preferentially aligned and anti-aligned pairs, for comparison. For the aligned distribution, we generate two randomly-oriented vectors whose true mis-alignment angles (measured in the plane defined by the spin vectors), is between $0^{\circ}$ and $30^{\circ}$. One thousand of these pairs are generated, and their projection CDF is shown as the ``Aligned'' curve in Figure \ref{fig:misalignment}. We repeat this process using a true mis-alignment angle between $150^{\circ}$ and $180^{\circ}$ for the ``Anti-aligned'' curve. A uniform distribution is shown as ``Random.''

\begin{figure}\label{fig:misalignment}
\bcenter
\includegraphics[scale=0.5]{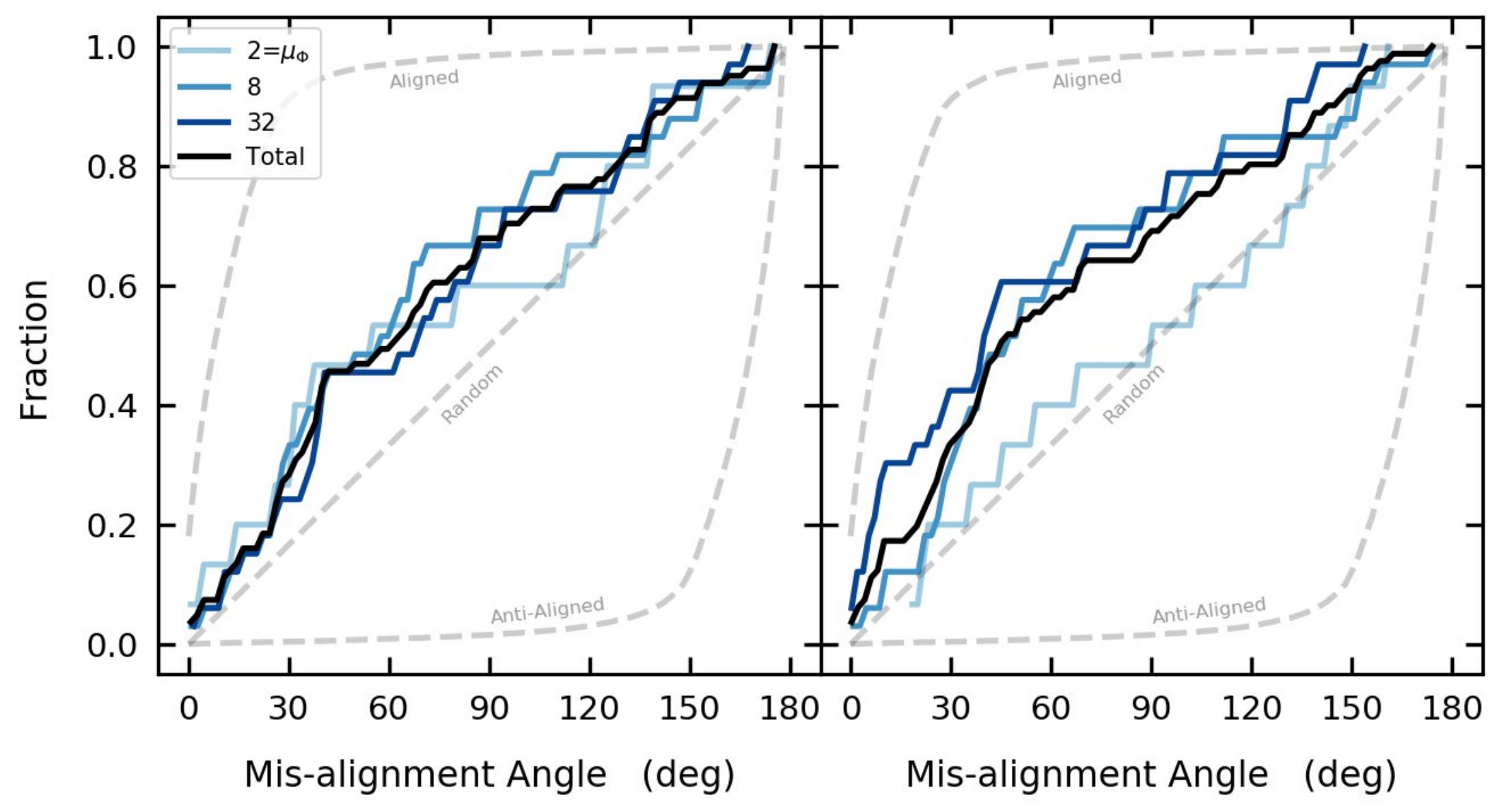}
\ecenter
\caption{The projected mis-alignment angle between the stellar spins of all binaries. The spin vectors of a binary's protostars are projected along each Cartesian direction and the angles are measured in the projected plane. The left plot shows the cumulative distribution function at the time of the binaries' formation. The right plot shows the the distribution at the last available output for each binary (either the end of the simulation or the last output the binary existed). Dashed lines show expected distribution functions for preferentially aligned, random, and anti-aligned pairs.}
\end{figure}

At the time of the binaries' creation, the spin mis-alignment angles range between $0^{\circ}$ and $150^{\circ}$, with a slight tendency toward being aligned. The end-state mis-alignment angles (right panel) show little change compared to the primordial CDF. Slight variation is seen in the MU2 run. However, since star formation occurs later in this simulation, the time elapsed between the formation of the binaries and the last outputs is, on average, shorter compared to the other two runs. Spin alignment evolution can be significant when the protostars are low in mass and accreting from a turbulent gas; even so, the distribution remains consistent with a random distribution of angles. This evidence of spin evolution is similar to the results of  \citet{leehulloffner2017}. Similarly these results agree with \citet{Offner16}, who also found that wider binaries tend to have more mis-aligned spin angles. In contrast, it is expected that binaries forming through disk fragmentation would have more preferentially aligned spin vectors  \citep[e.g.,][]{bate2018}.

\subsubsection{Age Differences in Binaries, Triples, and Quadruples }\label{sssec:agedifferences}

Protostars that form a long-lived binary within the same fragmenting core are expected to be closer to co-eval, whereas binaries formed from dynamical capture or undergo member swapping may have a larger spread in their ages. \citet{tobin16}, for example, finds binaries that are made up of a mix of Class 0, I, and II protostars. Figure \ref{fig:agedifferences} measures the maximum age difference between the stars in binaries, triples, and quadruples (systems with two bound binaries) at the pairs' time of creation. The non-zero starting values give the fraction of systems that are born co-eval. Almost 20\% of binaries are co-eval, but the spread in ages can extend to $\sim200$ kyrs. The average spread in ages increases and the number of co-eval objects decreases when considering triples and quadruples. Figure \ref{fig:stageshistogram} divides the binaries, triples, and quadruples into whether they are composed of all Stage 0, all Stage 1, or a mixture at the end of each multiple. Observations support a picture that turbulent fragmentation typically does not create co-eval protostars \citep{Murillo2018}. Co-evality (or lack thereof) between protostars in bound pairs may be a clue to their formation origins, which we discuss in Section \ref{sec:summary}.

\begin{figure}\label{fig:agedifferences}
\bcenter
\includegraphics[scale=0.5]{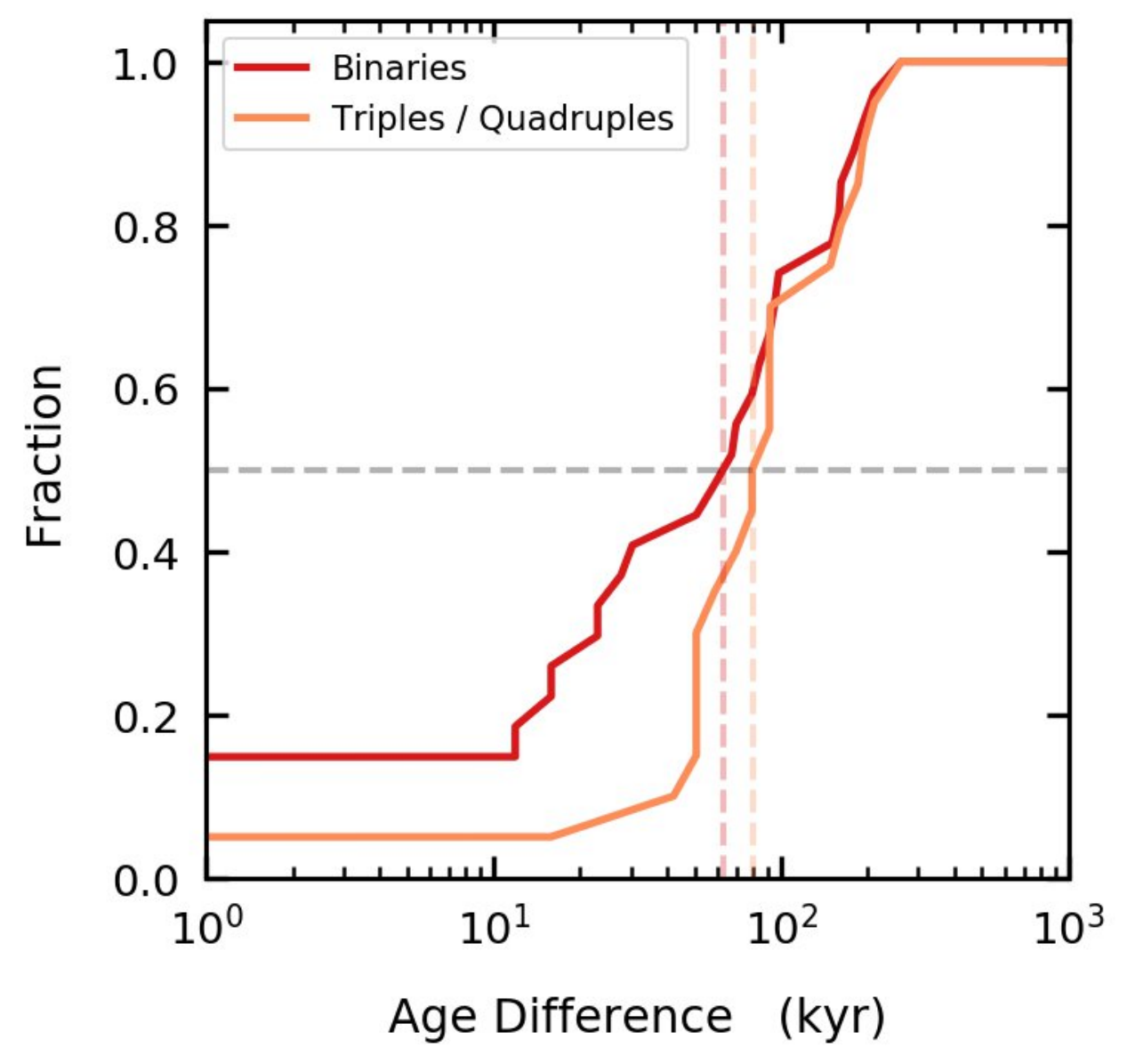}
\ecenter
\caption{Maximum age difference between stars in binaries, triples, and quadruples. The starting non-zero values give the fraction of co-eval stars. The vertical lines identify the 50\% value.}
\end{figure}

\begin{figure}\label{fig:stageshistogram}
\bcenter
\includegraphics[scale=0.5]{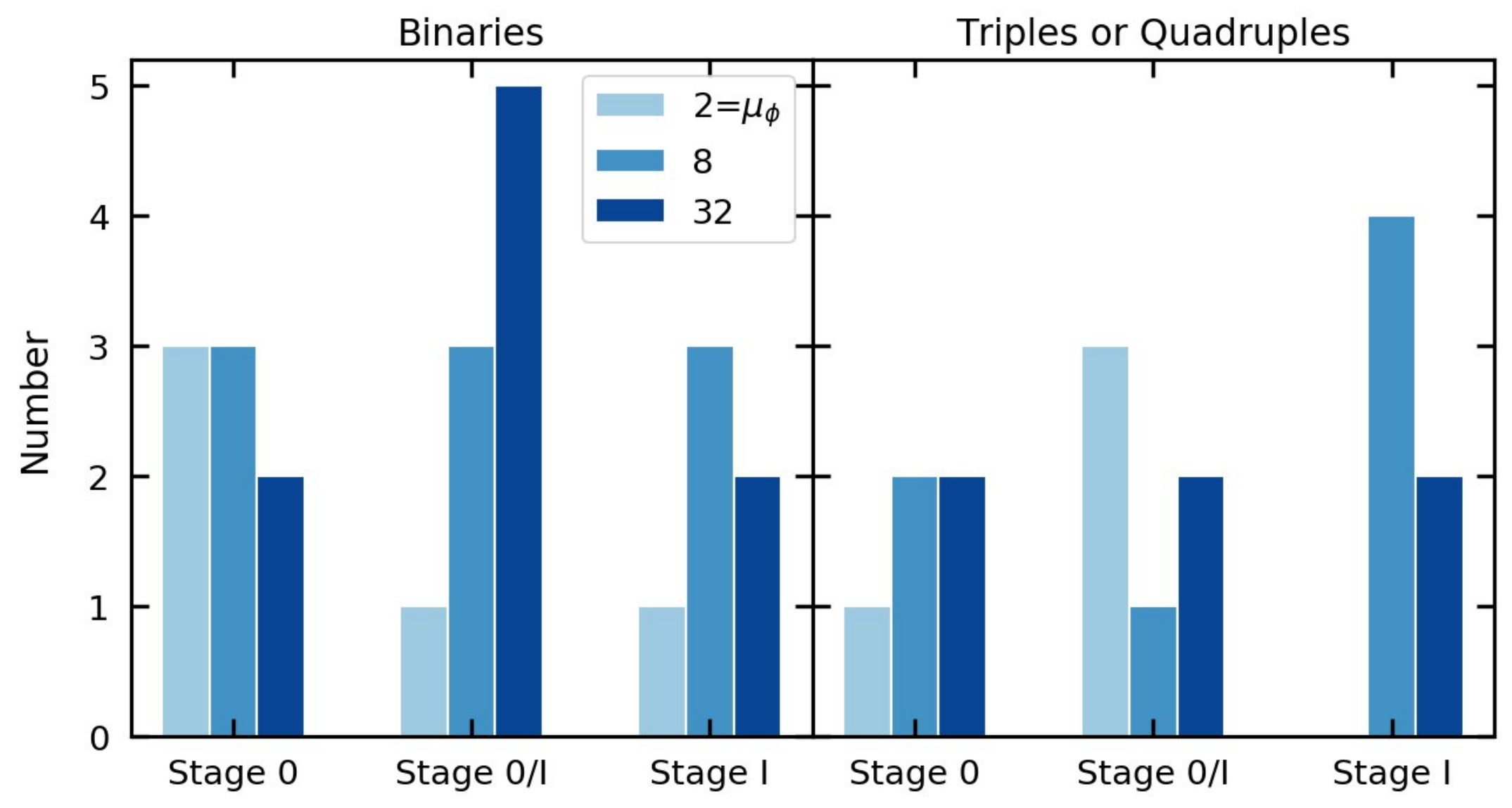}
\ecenter
\caption{Stage of protostars in binaries or triples/quadruples throughout each simulation. Ages of stars are calculated as the simulation time at the last available output for that object minus the times each sink was created. The multiple must exist in at least two outputs to be considered. Stage 0/I means the multiple has protostars in both stages. Stage 0 or Stage I imply all members are of the same stage.}
\end{figure}

\subsubsection{Separation Evolution of Multiples}\label{sssec:semievo}

Figure \ref{fig:semiactualdistcompare} shows the primordial separations of multiples for each simulation. The initial separations for multiples, and binaries in particular, range from a few hundred AU to almost our upper limit of $10^5$ AU. These separations are consistent with the observed wide-separation binaries forming through core fragmentation \citep{chen13,pineda15,tobin16}. Figure \ref{fig:allsemievos} displays the time evolution of separations, measured by the semi-major axis, of all pairs up to level 5. Data points are connected by a line if they represent the same object and the object exists between subsequent outputs. Short-lived transient pairs will appear only as a few dots. In high stellar density environments, membership swapping is common, particularly for high-level objects. For example, in the MU2 run, the various level 5 curves displayed all contain the same set of protostars, but the inner hierarchy changes between the binaries and triples that compose the level 5 object. This results in several broken curves. Figure \ref{fig:mu2multsnapshot} shows the mass evolution of a MU2 quadruple system as well as two column density snapshots. The mass evolution plot associates a unique color with each protostar. Additionally it plots the sum of the masses for protostars identified as binaries; these are shown as multi-colored curves. The column density snapshots show two outputs where identified binaries have changed. Membership swapping occurs frequently, particularly at late times when the average separations between protostars have shrunk to $<10^3$ AU. Figure \ref{fig:mu8multsnapshot} shows snapshots of a binary from MU8 that forms with a separation over $10^4$ AU at $t\approx 0.5 t_{\rm ff}$ and evolves to separations of $10^2$ AU at $t\approx t_{\rm ff}$. This inward spiraling has also spun up a large circumbinary ring around the binary, similar to the observed ring around GG Tauri \citep{ggtau}.

Such evolution has been seen in other simulations \citep{Offner2010,Offner16}. Additionally, we see similar evolution for higher-level multiples in all three simulations. In the MU2 run, the multiple shown in Figure \ref{fig:mu2multsnapshot} behaves more as a cluster of stars, and the separations between any two objects in the system are always comparable. In the MU8 and MU32 runs, the systems are more hierarchical with large differences between separations of different leveled objects. Figure \ref{fig:tobincompare}, shows another example from MU8 of a triple-star system. A binary hosts a circumstellar disk that gravitationally captured a third member with a semi-major axis value $\sim10$ times larger.

\begin{figure}\label{fig:allsemievos}
\bcenter
\includegraphics[scale=0.5]{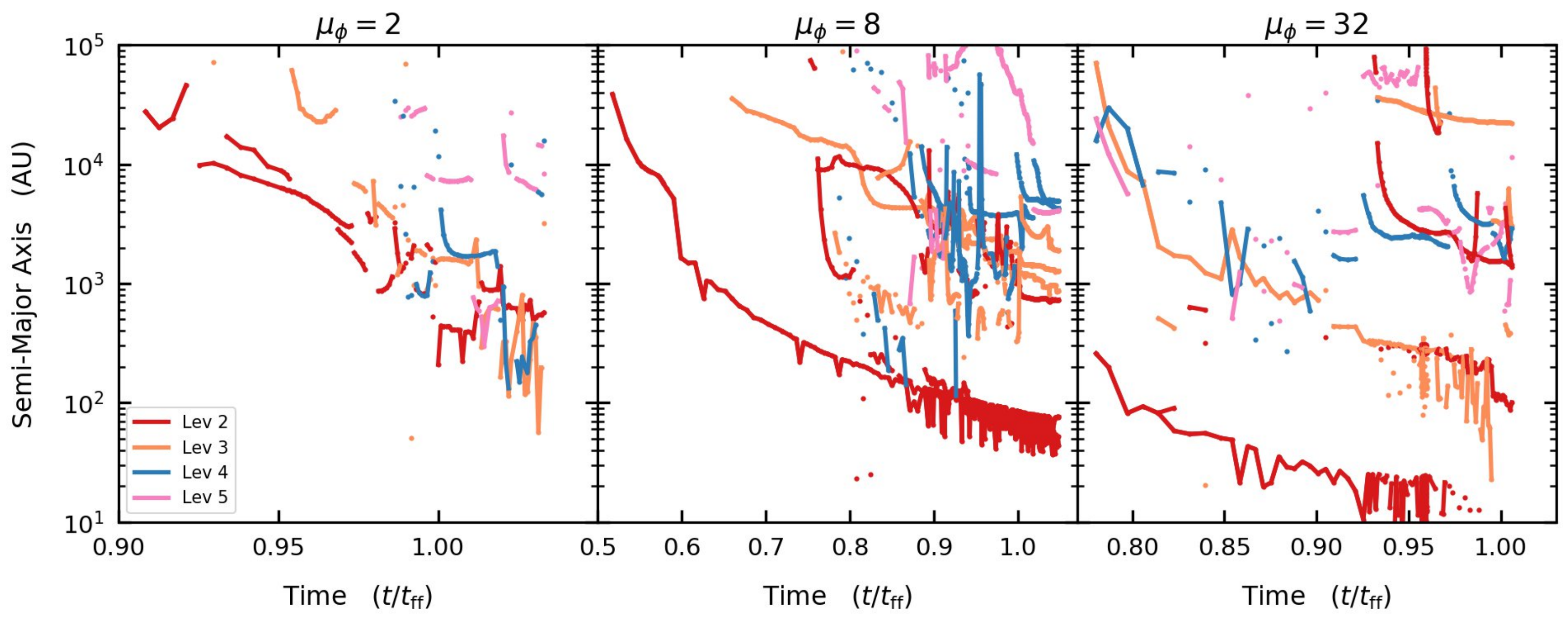}
\ecenter
\caption{ Semi-major axis calculations for all multiples up to level 5 objects. Individual points are connected by a line if they are in subsequent time outputs. Separations tend to shrink over time as the protostars grow in mass and dynamically interact with the local gas reservoir.}
\end{figure}

\begin{figure}\label{fig:mu2multsnapshot}
\bcenter
\includegraphics[scale=0.42]{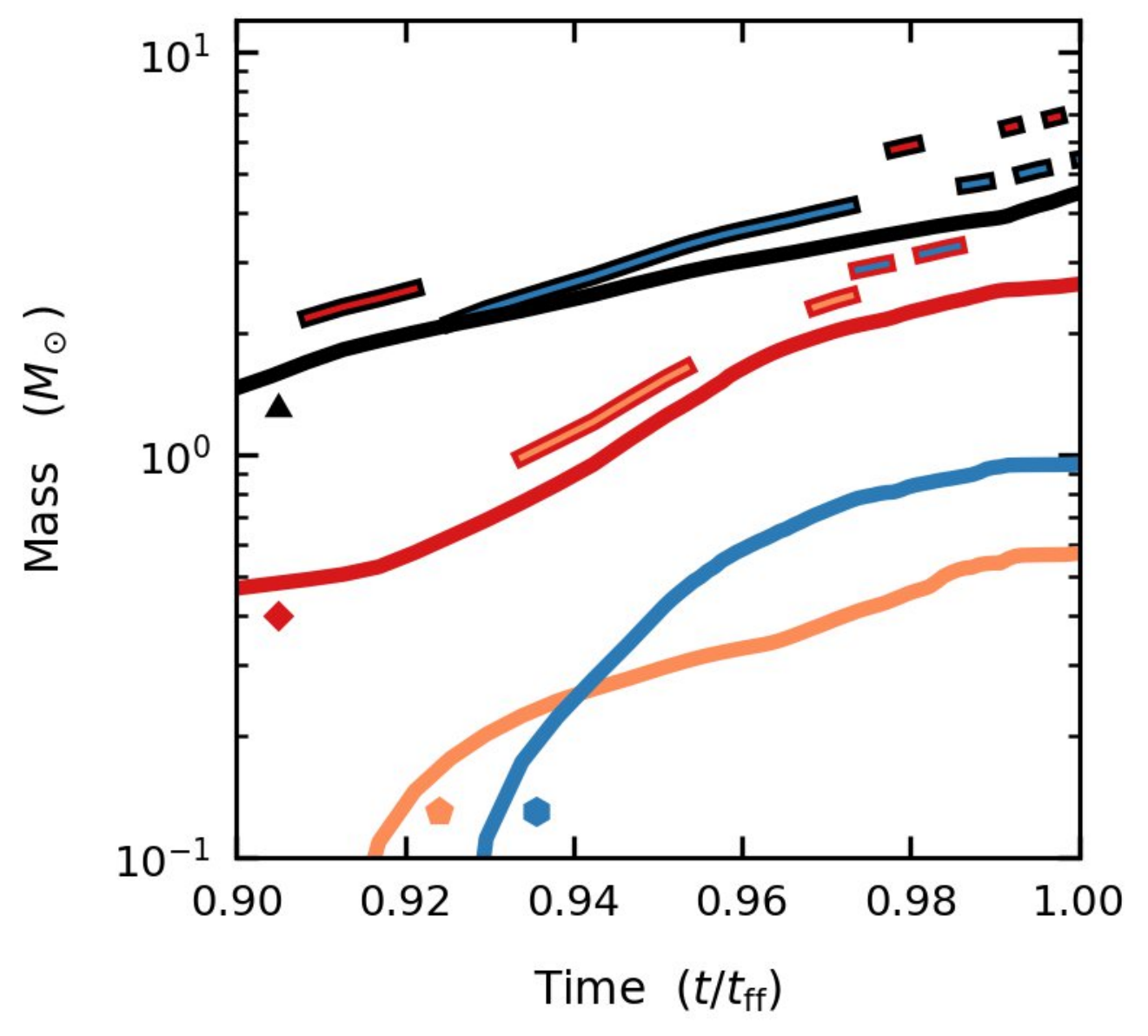}
\includegraphics[scale=0.42]{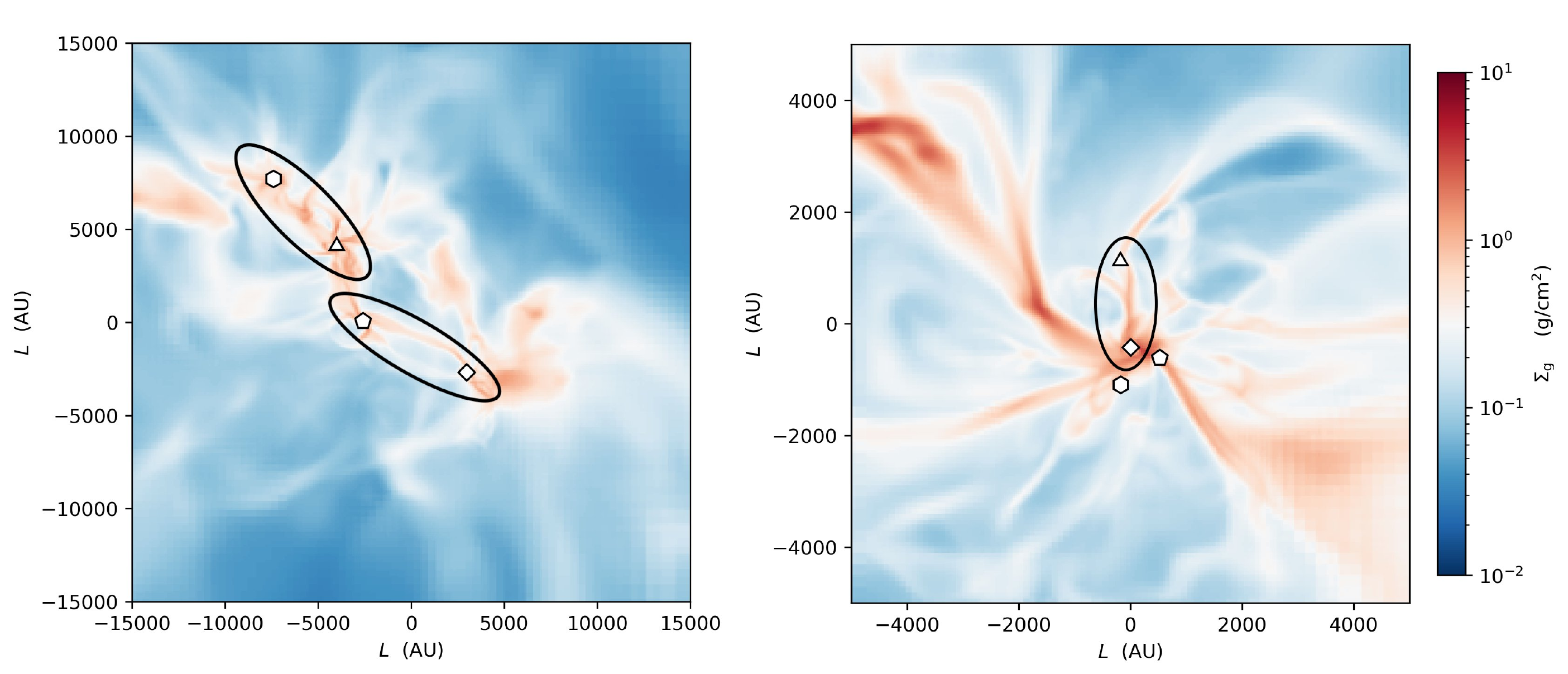}
\ecenter
\caption{Mass evolution and membership swapping for a quadruple star system from MU2. {\it Left:} Mass evolution of the four protostars. Each protostar is given a unique color and symbol. When two protostars are identified as a binary, the sum of their masses is shown as a double-colored line. {\it Right:} Two column density snapshots around 0.94 $t_{\rm ff}$ and 0.975 $t_{\rm ff}$. Symbols match those from the left panel. We circle the binaries identified in each output. In the latter density plot, the two most massive stars are labeled as bound with the remaining two stars paired as a triple and fourth member. }
\end{figure}

\begin{figure}\label{fig:mu8multsnapshot}
\bcenter
\includegraphics[scale=0.5]{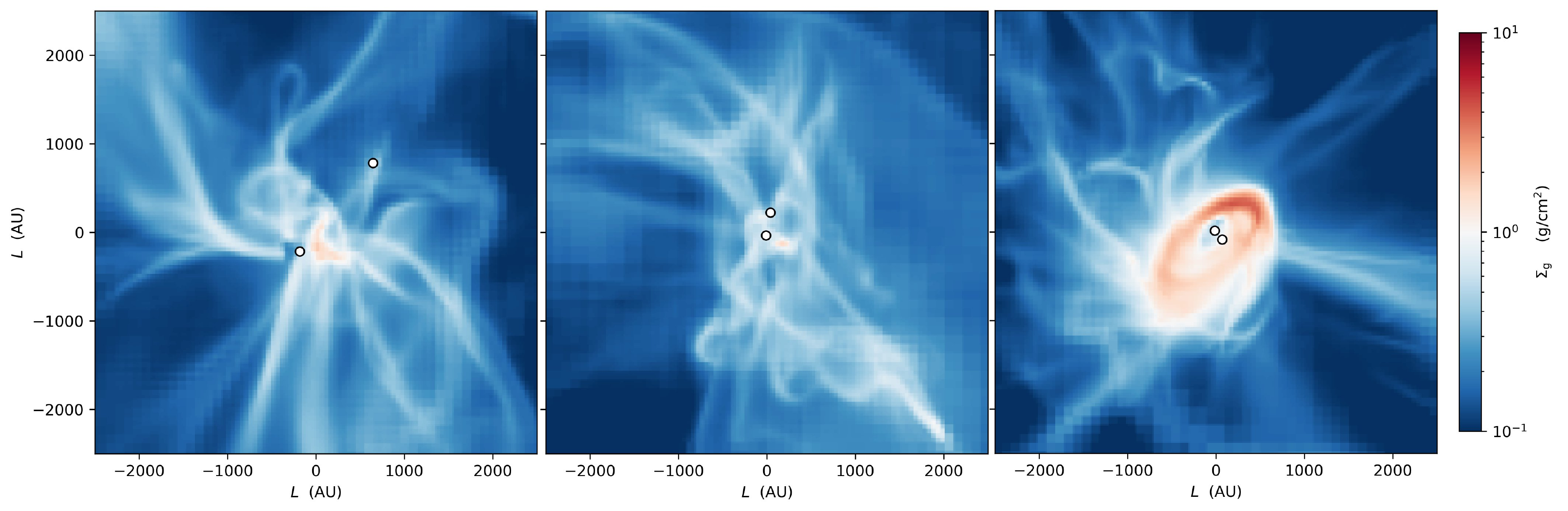}
\ecenter
\caption{Column density snapshots at $t= 0.6$, 0.8, and $1.0 t_{\rm ff}$, from left to right, centered around a binary pair from MU8. Protostars are labeled as circles. The binary's separations shrink by several orders of magnitude over the course of a few hundred kyrs.}
\end{figure}

\begin{figure}\label{fig:tobincompare}
\bcenter
\includegraphics[scale=0.4]{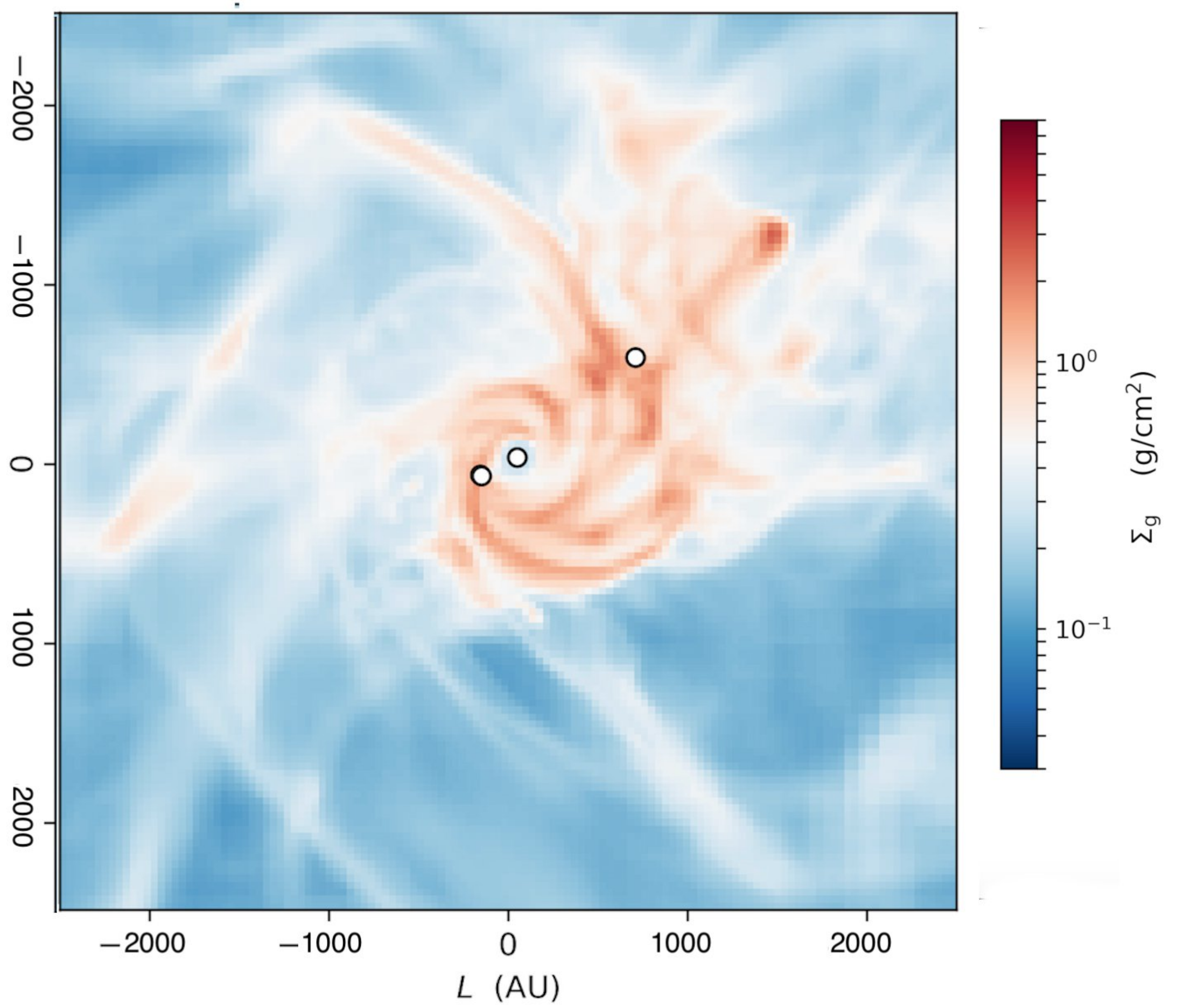}
\ecenter
\caption{Column density around three gravitationally bound protostars (shown as circles). A central binary resides within a developing circumbinary disk. A third member formed beyond $10^4$ AU and was captured and migrated to smaller distances rather than forming within the disk.}
\end{figure}

All multiples tend to evolve to smaller separations unless disrupted. Therefore, in general, closer pairs tend to contain older stars. The left panel of Figure \ref{fig:unstackedages} shows the separations for objects that persist for either 0 kyr (initial), 1 kyr (a few outputs), 10 kyrs, and 100 kyrs. Similar to what is seen in \citet{tobin16}, younger objects span large and small separations, whereas older objects tend to have smaller separations. The right panel shows the primordial separations for the same objects, displaying the original separations for the same objects that last up to 100 kyrs. The objects that have separations $\leq 10^4$ AU after 100 kyrs almost entirely originated with orbits larger than $10^4$ AU.

\begin{figure}\label{fig:unstackedages}
\bcenter
\includegraphics[scale=0.5]{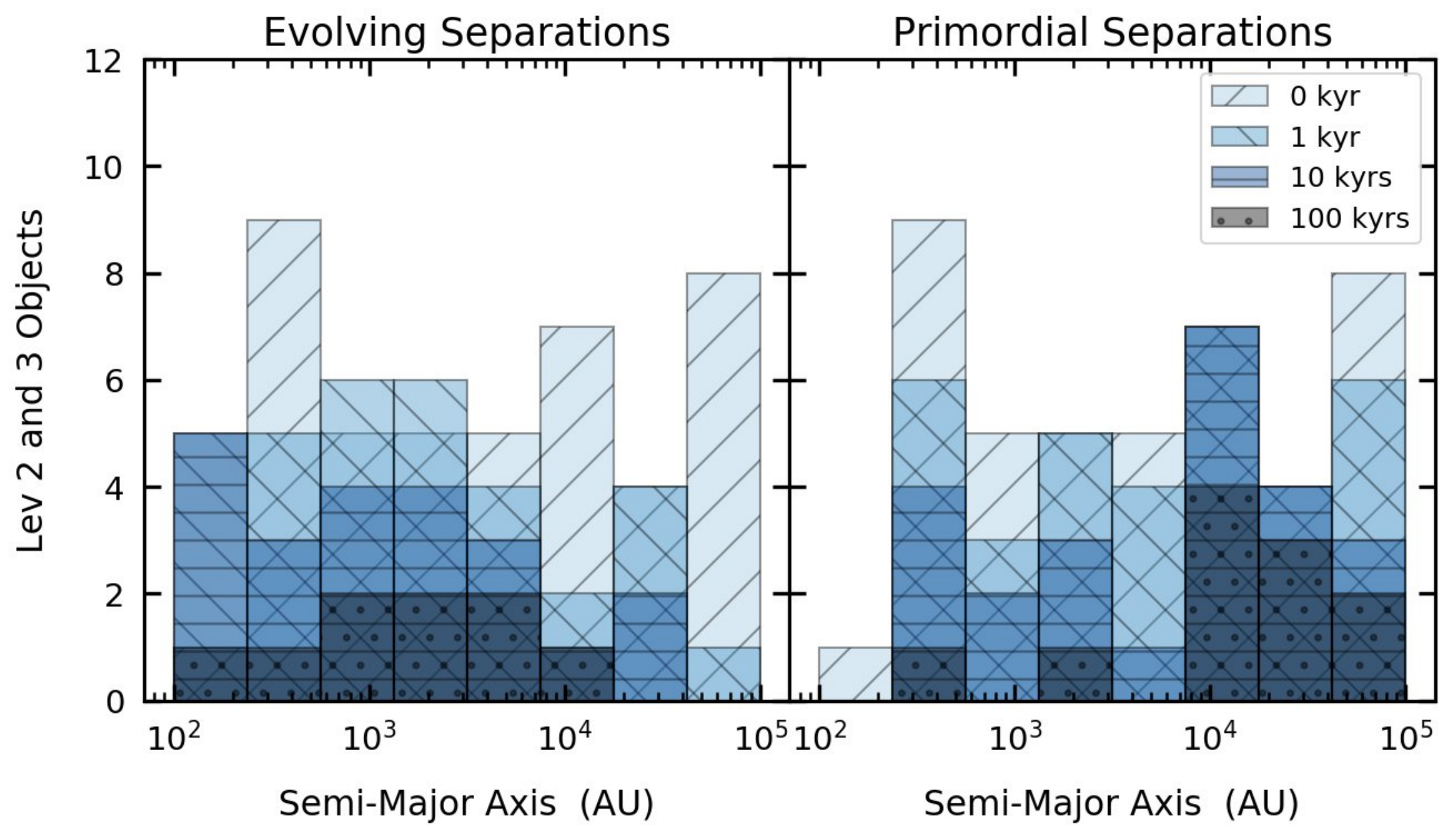}
\ecenter
\caption{Unstacked distributions of multiples at level 2 or 3 from all three simulations. {\it Left:} Semi-major axis separations, measured at the closest possible output after the object has existed for a given amount of time (differentiated by color and hatch pattern). {\it Right:} Separations for the same multiples, but always measured at the time the object was created. In this case, the data is plotted only if the object exists for a given amount of time.}
\end{figure}

\subsubsection{Evolution of binary separations}\label{sssec:evolutionmodel}

As protostars move through the cloud, they gravitationally draw gas toward themselves. Some of the gas gets accreted, but some of this gas also falls into a dense wake behind the object and tugs back on the protostar. In the case of a binary orbit, this wake torques the orbit to smaller separations while transferring angular momentum to the gas. The exact expression of this dynamical friction force, as well as its effects on star clusters and orbits, has been extensively explored \citep[e.g.,][]{Ostriker1999,Bate97b,stahler2010binaries, Leestahler2011,Leestahler2014,Antonietal2019}. Here we give a simple model that can demonstrate how dynamical friction and mass accretion together can explain the evolution of binary orbits. \change{Previous studies have considered the effects of each of these mechanisms individually. For example, \citet{Bate97b} and \citet{Bate2000} considered how  accretion of mass and specific angular momentum changes the separation of binaries, while, e.g., \citet{stahler2010binaries}  considered how dynamical friction alone evolved the separation of binaries. Below we describe our simple model, compare its predictions to our simulations, and then compare our model to the model of \citet{Bate97b}. }

For simplicity, consider two protostars in a circular Keplerian orbit with semi-major axis $a$. The angular momentum of the orbit is $L = m_1 m_2 \sqrt{G M_{\rm tot} a}/M_{\rm tot} $, where  $M_{\rm tot}=m_1+m_2$. The derivative of this expression gives
\begin{equation}
    \label{eqn:angmomderiv}
    \frac{\dot{a}}{a} =   2\frac{\dot{L}}{L} + \frac{\dot{M}_{\rm tot}}{M_{\rm tot}} - 2 \left(\frac{\dot{m}_{\rm 1}}{m_{\rm 1}} + \frac{\dot{m}_{\rm 2}}{m_{\rm 2}} \right)\ .
\end{equation}
Given an initial value for $a$, Equation (\ref{eqn:angmomderiv}) gives a first-order differential equation for the evolution of the semi-major axis. With gas either torquing away angular momentum from the orbit ($\dot{L}<0$) or being accreted ($\dot{m}_i>0$). The right-hand side of this equation is always negative--i.e., the orbit will shrink. \change{ We write the angular momentum derivative as $\dot{L}=\vec{r}_1\times \vec{F}_{\rm 1,DF}+\vec{r}_2 \times \vec{F}_{\rm 2,DF}$, where $\vec{F}_{\rm DF}$ is the expression for the dynamical friction force and $\vec{r}_i$ is the position vector from the pair's center of mass to the individual particle. This expression assumes that torque arises only from gravitational interactions with the gas. We assume the accreted gas carries negligible angular momentum, an assumption we justify below. For the dynamical friction force, we quantify its magnitude using $\vec{F}_{\rm DF} = -\dot{m}_{\rm p}\vec{v}_{\rm rel}$, where $v_{\rm rel}$ is the relative velocity between the gas and the protostar. This is the form used in \citet{Leestahler2011}, who found a connection between the overall mass accretion and the strength of the wake in the case of accreting point-like particles. Other expressions for the friction force \citep[e.g.,][]{Ostriker:1999} have different functional forms for this force but are similar to order-of-magnitude. Given the simplicity of the above expression, we use it for $\vec{F}_{\rm DF}$ in this model. }

To compare the simulation results with this model, we estimate the mass derivatives using a smoothed function $m_i(t)$, derived from the simulation data for each protostar. The relative velocity is calculated by measuring the mass-weighted velocity of the gas in a sphere of radius 500 AU around each star and the star particle's velocity. Figure \ref{fig:dynamicalfriction} shows two examples that compare our model to the actual semi-major axis evolution from the simulations. The left example is the same binary from Figure \ref{fig:mu8multsnapshot}. The biggest discrepancy between the model and the data is at the start of the orbital decay, where the separation drops precipitously. This is to be expected, since at the time of formation the protostars have not settled into an orbit and the circular orbit assumption of Equation (\ref{eqn:angmomderiv}) is poor. At late times the separations are comparable or smaller than the resolution of our grid and therefore may not be accurate. While the sink particles in \ORION\ move independently of the grid, the gas in the sink's accretion zone is altered through accretion. Additionally, in the case of the MU8 example, the binary has a close encounter with another star and is disrupted at times shortly after the data from the plot ends.

\begin{figure}\label{fig:dynamicalfriction}
\bcenter
\includegraphics[scale=0.5]{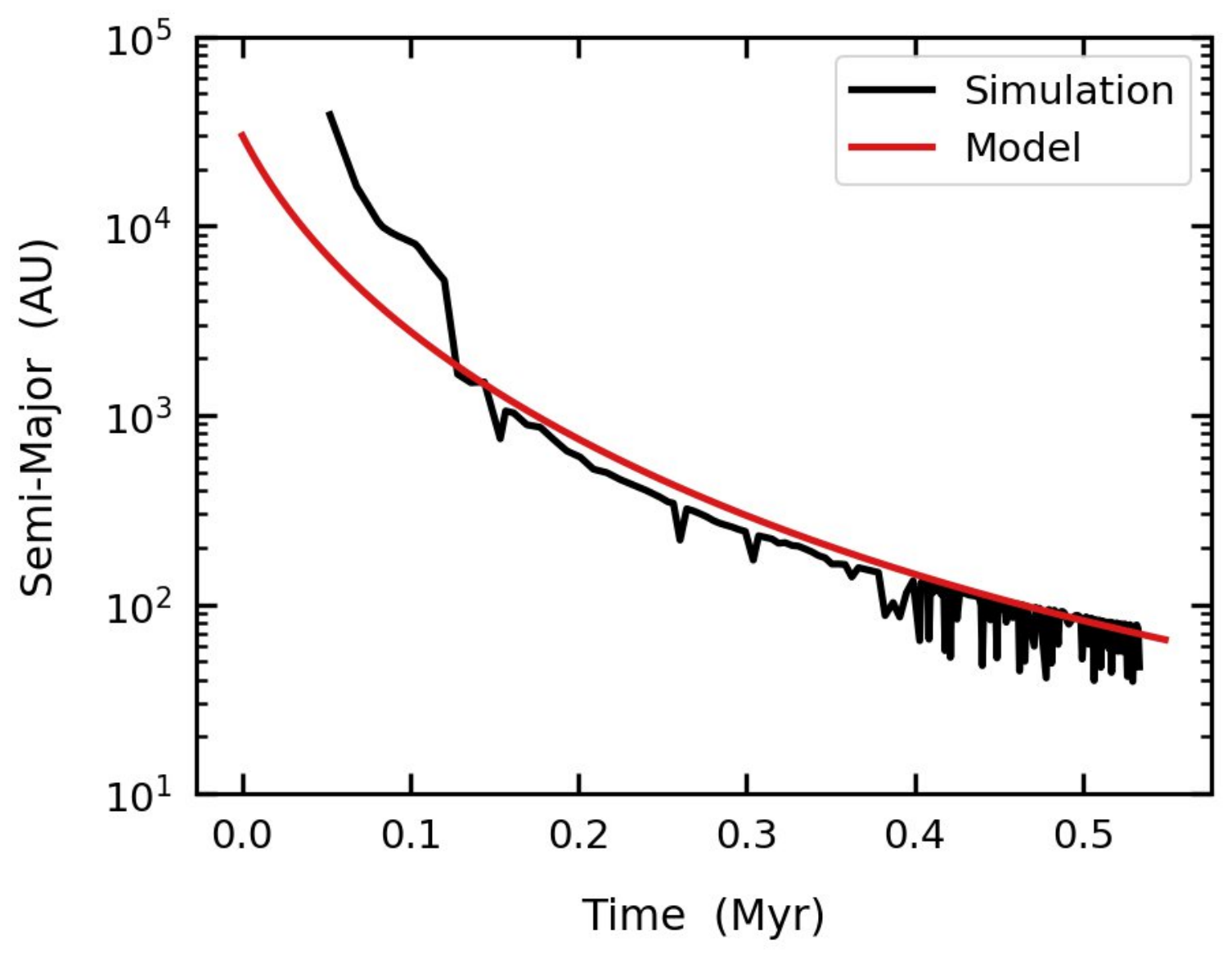}
\includegraphics[scale=0.5]{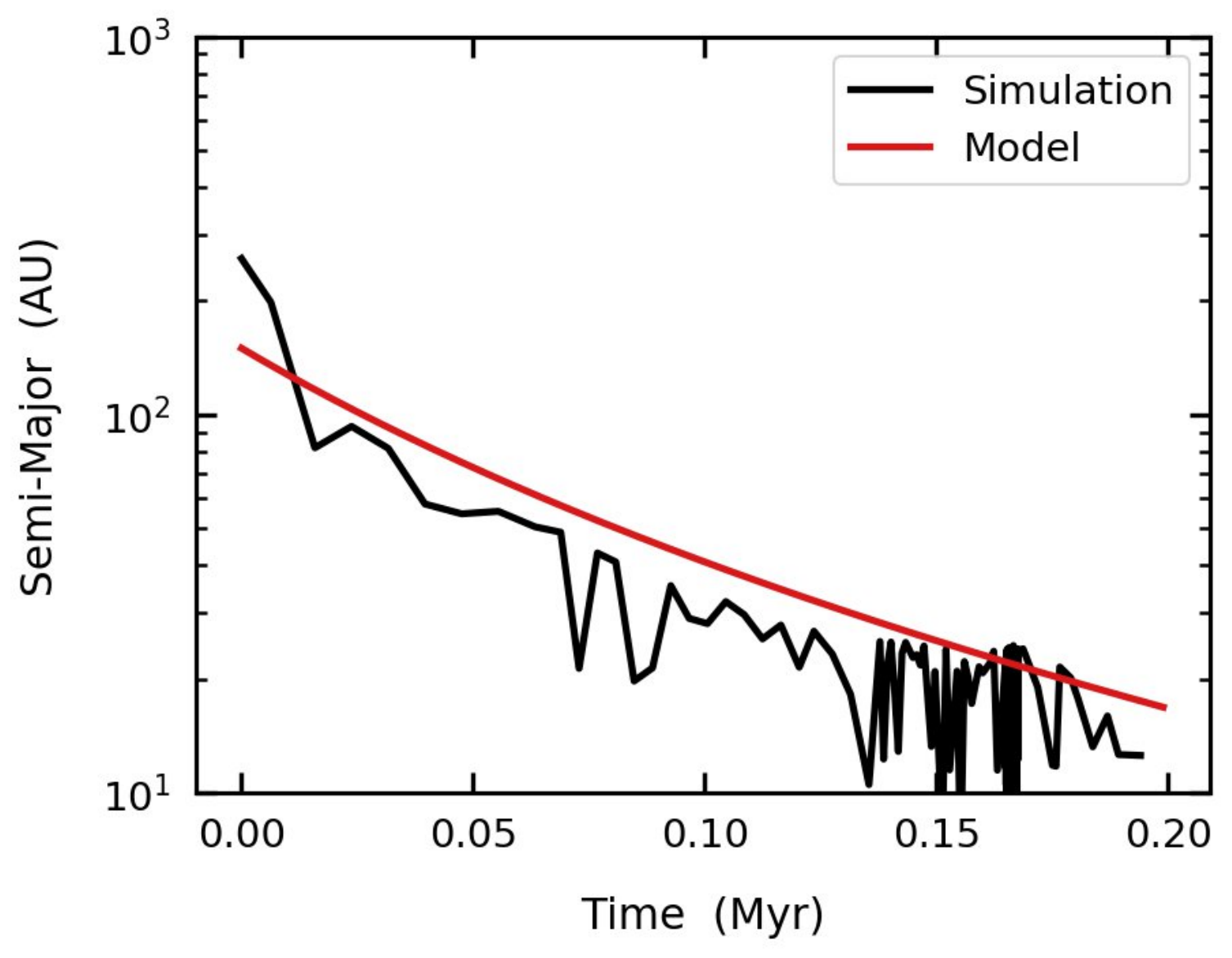}
\ecenter
\caption{Comparing the semi-major axis versus time for two evolving binaries, one from MU8 (left) and from MU32 (right). The data is shown in black. The smooth, red curves show the dynamical friction and mass accretion model from Section \ref{sssec:evolutionmodel}.}
\end{figure}

The relative importance of mass accretion and dynamical friction changes as the binarys' orbit shrink and the protostars grow. The terms on the right-hand side of Equation (\ref{eqn:angmomderiv}) can be broken up into a dynamical friction term and two mass accretion terms. Figure \ref{fig:relativetorques} displays the ratio of these terms for the MU8 binary. The time axis is shared with the left panel of Figure \ref{fig:dynamicalfriction}. At early times, mass accretion dominates the evolution even though there is a larger relative velocity between the gas and the protostars. At this point in the evolution, the protostars are small and the ratios $\dot{m}_i/m_i$ are large. At later times, dynamical friction dominates the orbital evolution despite the fact that the relative velocity between the protostars and the gas has decreased. This decrease arises because the conservation of angular momentum during collapse has spun up both the binary and the gas similarly (e.g., Figure \ref{fig:mu8multsnapshot} shows the formation of a large circumbinary disk after the binary's orbit has shrunk). Nonetheless, the sinks have grown in mass and the mass accretion terms  affect the orbit less than dynamical friction. 

\change{

In this model, we assumed the accreted gas carried a negligible amount of angular momentum. As a result, torque from the gas and accretion both decreased the orbital separation over time. This agrees with \citet{Bate97b}, who developed a similar model to the one above. They find that the accretion of gas with low specific angular momentum decreases the orbit over time. As the specific angular momentum of the gas increases, gas first settles onto a circumstellar disk before being accreted onto the protostar. In this case, the binary separation increased over time. For infalling gas with an even higher specific angular momentum, they found that a circumbinary disk developed and the separations were either constant or decreased with time. 

We see no evidence for orbital separations increasing with time. This is not surprising since we are not resolving the circumstellar disks around protostars (but do capture larger circumbinary disks around shorter-period binaries, e.g., Figure \ref{fig:mu8multsnapshot}). Further studies with higher resolution could assess whether or not circumstellar disks could slow down or reverse the binaries' inward spiral. We note, however, as was also discussed in the followup studies of \citet{Bate2000}, the angular momentum contained in the circumstellar disk is likely to be small compared to the orbital angular momentum of the binary. In the case of circumbinary disks, recent work has shown that torques from the disk increase the binary separations rather than shrink them  \citep{Kratteretal2010,Munoz2019a,Munoz2019b}.
 }

\begin{figure}\label{fig:relativetorques}
\bcenter
\includegraphics[scale=0.5]{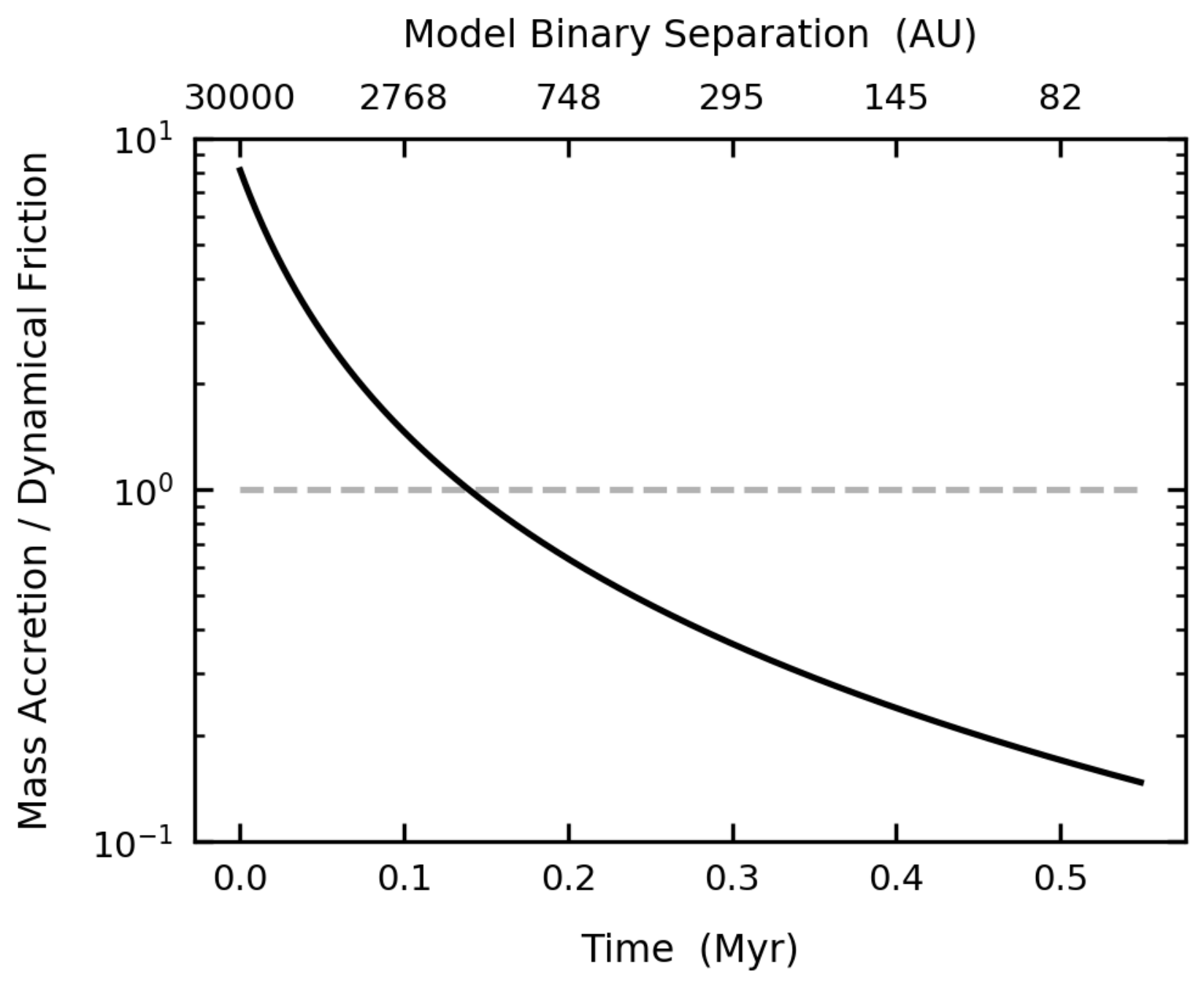}
\ecenter
\caption{The relative magnitude of mass accretion and dynamical friction terms on setting the value of $\dot{a}$ in Equation (\ref{eqn:angmomderiv}). The binary analyzed is the same binary from MU8 shown in the left panel of Figure \ref{fig:dynamicalfriction}. Both that panel and this figure share the same time axis. The upper horizontal axis interpolates the binary separation from our model onto the displayed tick points.}
\end{figure}

\subsubsection{Multiplicity Criteria in Theory and Observations: The Risk of Interlopers}\label{ssec:interlopers}

Our multiplicity criterion groups stars together based on whether or not they are gravitational bound to one another. Doing so requires knowledge of the exact three-dimensional spatial positions and velocities of the protostars, as well as their exact masses. Observers are not privileged to as much information and identify binaries primarily based on their angular separation in the sky, assuming that the objects within a certain distance from one another are gravitationally bound. The potential for interlopers, stars projected as near-neighbors on the sky but are not gravitationally bound, can affect the observed statistics in star-forming regions. 

Figure \ref{fig:semiactualdistcompare} demonstrates that the projected separations generally agree with the actual separations for bound protostellar pairs. Here we consider the projected distances between unbound protostars as well. In Figure \ref{fig:interlopers}, we estimate the potential for interlopers using our simulation data. Since the orbital separations evolve over 100 kyr timescales ($\approx 0.1 t_{\rm ff}$), we calculate the positions over the last tenth of the simulation in the following way. We first create a set of nine logrithmically-spaced bins ranging from $10^2$ AU to $10^5$ AU. For each pair of protostars, we look at every data output available over the last 10\% of the simulation and compute the three projected distances, viewing the pair through each of the Cartesian directions. If the separation is between $10^2$ AU and $10^5$ AU, we count the ``observation" in the appropriate bin. Once this is completed for every data output, we divide the total contributions to each bin by 3, so to average over the three projection points of view. We finally divide each bin contribution by the number of data outputs that were considered in the last 10\% of the simulation. This  also includes data outputs where the particular protostar pair is not present. The sum of the pair's contribution to all bins adds up to a value between 0 and 1, where it equals 0 if its projected distances were never included in any bin and 1 if all three of the projected distances were included in bins for every data output available. We follow this procedure for both pairs that are actually labeled as binaries, triples, or quadruples using our multiplicity criterion (red in the left panel of Figure \ref{fig:interlopers}), and all other pairs, which we call ``interlopers" (yellow). The right panel of Figure \ref{fig:interlopers} shows just the interlopers, separated by simulation. Additionally, we show the full data sample from \citet[][their Figure 5]{tobin16}. Our results match observations between $10^2$ AU and $3\times 10^3$ AU. At lower separations, there are fewer binaries relative to \citet{tobin16}, likely due to our inability to capture disk instabilities.

At larger separations, unsurprisingly, there is a higher chance of mislabeling stars as bound objects, particularly in regions of high stellar density. The ratio of the interlopers to actual objects is less than unity only for the first two bins, reaching a maximum of $\sim40$ around $10^4$ AU. The mean value of this ratio is $\sim 8$. We note that this analysis is solely based on the exact positions of the protostars obtained in the simulations. In the case of MU32, for example, a long filament resides along one of the Cartesian axes, which increases the chances for interlopers in this analysis. A similar result was found in \citet{LiIRDC2018} for star-forming clouds where turbulence was constantly driven throughout the simulation. In their case, even fewer observed pairs were true binaries--and only if the stellar separation was less than $\sim400$ AU.\footnote{However, their criterion for multiplicity is stricter than ours; \change{in addition to being gravitationally bound they also required that the acceleration between two protostars in a binary exceed the tidal acceleration from nearby particles and the gas.}} A further study could mimic the procedures done by observers with synthetic observations of these simulation results to further quantify the potential for mislabeling objects as bound pairs. 

\begin{figure}\label{fig:interlopers}
\bcenter
\includegraphics[scale=0.5]{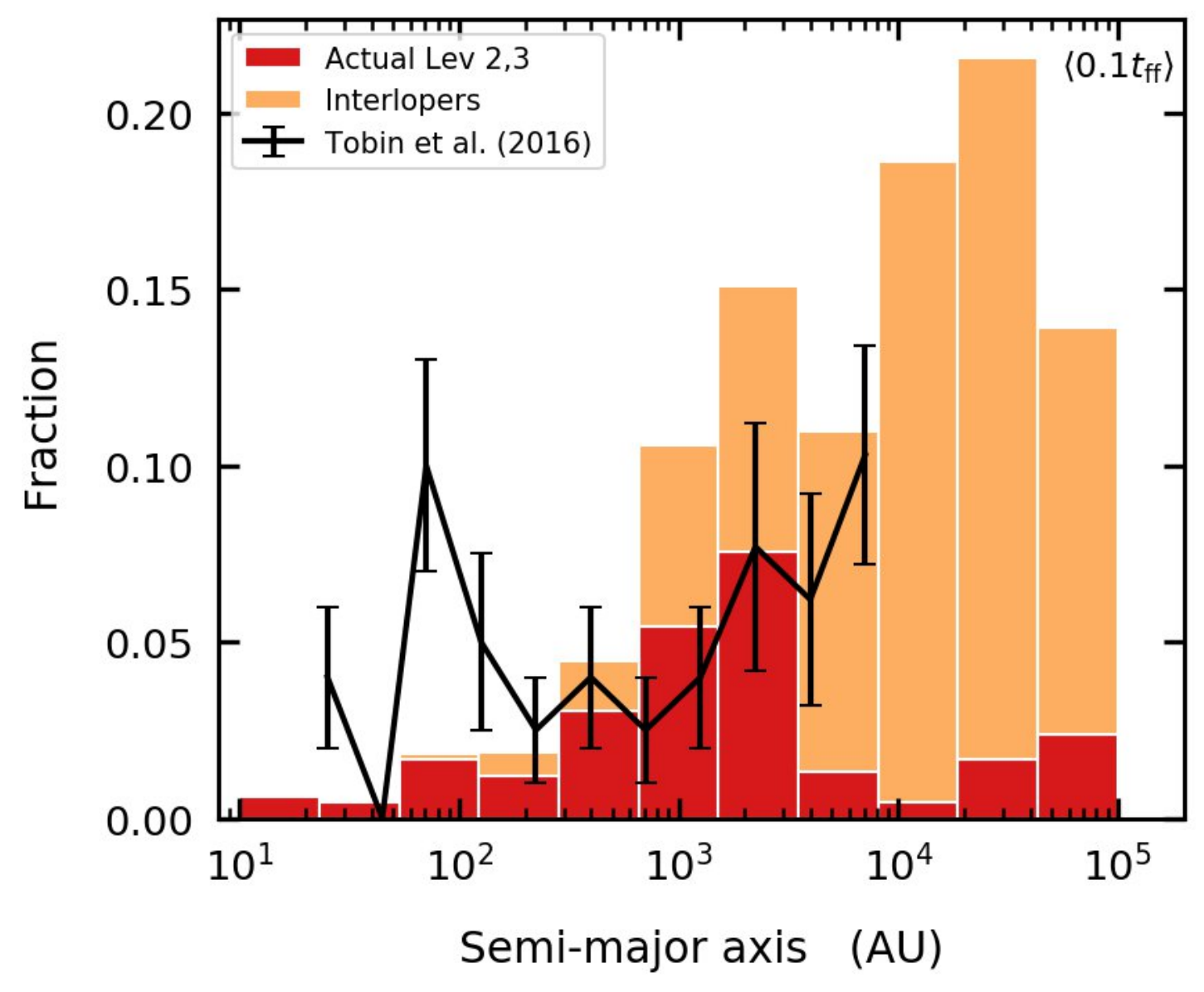}
\includegraphics[scale=0.5]{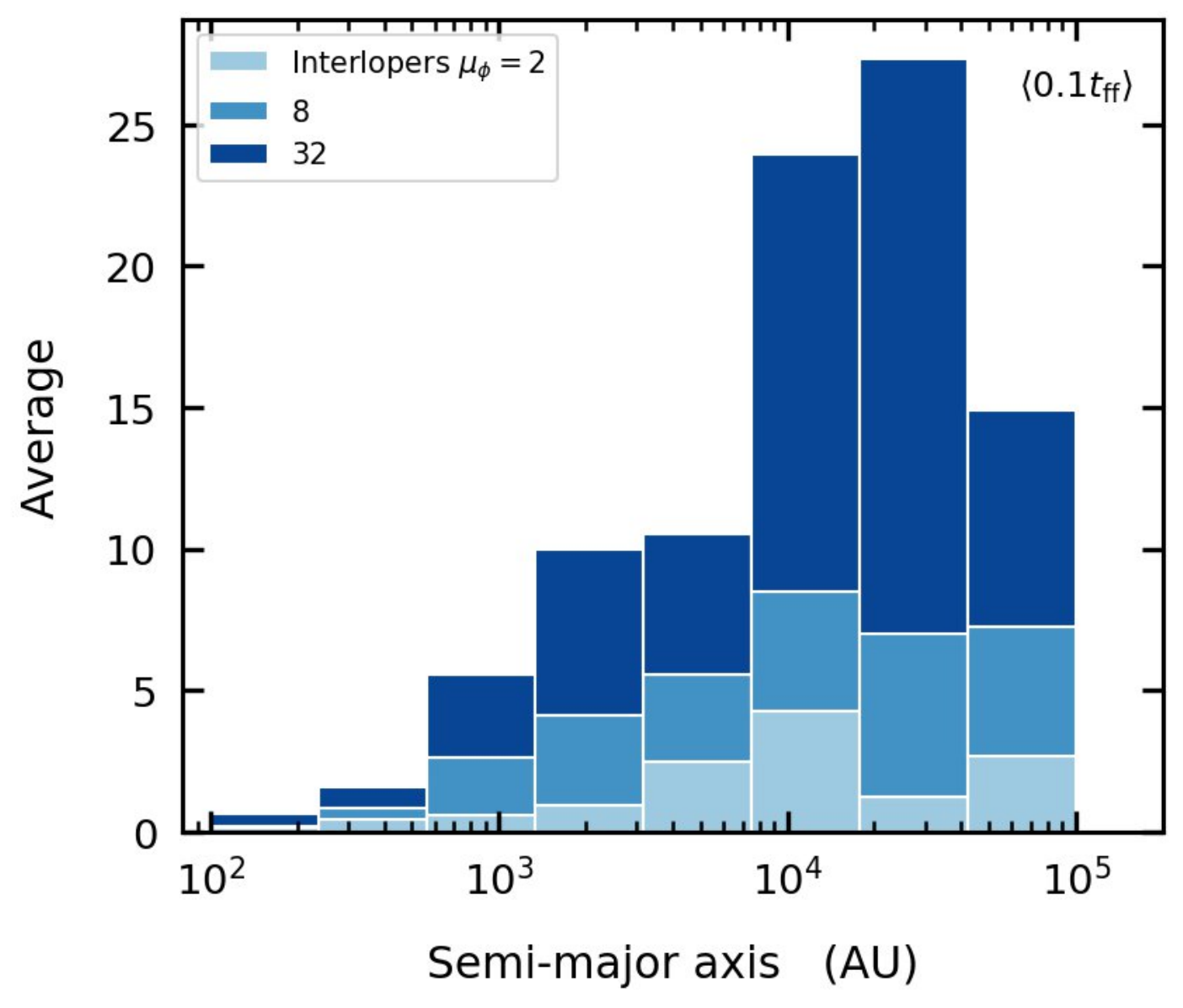}
\ecenter
\caption{ {\it Left:} The average separation of actual binaries, triples, and double binaries (red) and possible interlopers (yellow), averaged over the last 10\% of a free-fall time. The curve and errorbars are taken from the full sample of \citet{tobin16}. The procedure for generating this data is described in the text. To compare to observations, this panel is also normalized. {\it Right:} Only the possible interlopers are shown, differentiated by simulation. Both histograms are stacked. }
\end{figure}

} 

%% file: discussion.tex
%
{
\section{Summary \& Discussion}\label{sec:summary}

We have presented a set of three-dimensional MHD simulations of the collapse of turbulent, magnetized gas that model the formation and evolution of wide-orbit binaries and multiples. The role of the global magnetic field was considered in these three simulations by using three different mass-to-flux ratios $\mu_\phi=2$, 8, and 32. We find that stronger magnetic fields suppress fragmentation in the gas, which results in fewer overall stars compared to runs with weaker magnetic fields (Figure \ref{fig:finaloutput}). However, when looking at the multiplicity fraction as a function of time, we find that stronger magnetic fields ($\mu_\phi = 8$ and 32) produce a larger fraction of multiples and better reproduce the multiplicity statistics observed in nearby star-forming regions (Figure \ref{fig:mfandcsf} and Table \ref{tab:mf}). This result appears robust at all times: while collapse and fragmentation increase the stellar density over time for all our simulations, the value of the multiplicity fraction is relatively constant throughout.

Additionally, we find support for the idea originally put forward by \citet{Larson1972binary} that almost all stars are part of a multiple-star system, at least at some point: over 80\% of the stars formed are part of a multiple at some point during the simulation (Figures \ref{fig:sinkmassevo}, \ref{fig:stellarages}, and \ref{fig:endtimesnapshots}). The fraction of ``orphans," stars that were members of multiples but are single stars at the end of the simulation, is the largest for our weak magnetic field case and tends to decrease with increasing field strength (with the caveat that our multiple statistics are particularly poor for our strongest field run). 

These simulations demonstrate that dynamical evolution is commonplace and important in setting the distribution of separations between protostars. As Figure \ref{fig:allsemievos} displays, pairs that originate at larger separations interact with other protostars and the gas, evolving to smaller separations over 10--100 kyrs (Figure \ref{fig:unstackedages}). In clustered environments, groups of stars interact with each other, and the separations between individual objects can shrink $\sim$1 order of magnitude over these time scales. For more isolated or hierarchical pairings, the evolution can be rapid, with separations changing from $10^4$ AU to less than $10^2$ AU over the same timescales. In Section \ref{sssec:semievo}, we showed that this evolution can be understood as a combination of mass accretion and dynamical friction (Figure \ref{fig:dynamicalfriction}).

These simulations obtained resolutions of 25-50 AU, which is sufficient to resolve turbulent fragmentation of cores but not the vertical structures of accretion disks. By choosing this resolution, we are able to determine the role turbulent fragmentation and dynamical evolution together play in the multiplicity of protostars. Our simulations show that both wide-orbit and short-period binaries can originate through these mechanisms. The observed bi-model distribution of binary separations, with peaks around $\sim100$ AU and $\sim 3000$ AU \citep{tobin16}, has typically been interpreted as displaying two separate formation mechanisms. Instead, we have shown that both peaks include multiples that formed through turbulent fragmentation and evolution. That said, in our simulations turbulent fragmentation does not reproduce this bi-model peak alone; disk fragmentation may still be necessary for shorter-period binaries. Additionally, older star-forming regions show a single-peaked broad distribution of separations, which is also consistent with a blurring of these two mechanisms. Finally, the location of the second peak in \citep{tobin16} appears to be located beyond $10^4$ AU--however, some care must be taken with this interpretation since interloper mis-categorization is also most common at these distances (Figure \ref{fig:interlopers}).

We showed in Figure \ref{fig:tobincompare} a snapshot of a triple star system. Figure 1 of \citet{Tobin2016Nature} displayed a circumbinary disk that appears to have fragmented to form a third companion several 100 AU away from the primarily pair. A direct comparison between these figures reveals a strong similarity in structure except that our physical scale is larger by a factor of $\sim 3$.\footnote{A side-by-side size comparison has subtleties; for example, we are viewing the disk face-on, whereas the disk of \citet{Tobin2016Nature} appears somewhat inclined into the page. Additionally, we are viewing the density directly, versus radiation emitted from the gas and dust.} In our case, an outer third member migrated inward to these separations after forming beyond $10^4$ AU from the central pair. It did not form within the developing disk around the central binary. This figure demonstrates that it can be non-trivial to determine whether short-period multiples formed through disk fragmentation or from turbulent fragmentation and migration. The spin alignments of binaries resulting from disk fragmentation tend to be preferentially more aligned \citep{bate2018}, where we showed that turbulent fragmentation produces more randomly aligned pairs \citep[Figure \ref{fig:misalignment}, also ][]{Offner16}. Observers could distinguish this by observing, for example, the protostellar outflows; however, simulations of outflows have shown that spin orientation can change when there is mass accretion, particularly when the protostars are small in mass \citep{leehulloffner2017}. In these cases, having an estimate of the age of the protostars can be of assistance. Figures \ref{fig:agedifferences} and \ref{fig:stageshistogram} show that binaries and triples are more-often not co-eval when forming through turbulent fragmentation, with age spreads that can exceed 100 kyrs, a result that is supported by recent observations of wide binaries \citep{Murillo2018}. The end-state of our simulations reveal nearly 40\% of the multiples have both Stage 0 and Stage I objects, i.e., at least one object that is less than 160 kyrs and at least one object that is greater than 160 kyrs.
 
In this paper, turbulent driving was shut off at the start of our gravitational collapse phase. As a result, turbulence begins to decay and will dissipate completely after a few free-fall times. Undriven turbulence simulations have higher star formation efficiencies than driven turbulence simulations. They also produce more clustered environments as filaments and cores gravitationally collapse toward one another. Despite these differences, we do not expect that the inclusion of driving during our collapse phase would have produced statistically significant differences in the number of multiples produced, as obtained in \citet{LiIRDC2018}. \citet{Offner2008drive} found that (non-magnetized) driven turbulence simulations produce more cores overall, but these cores have less rotational energy compared to the cores in undriven simulations. There are also fewer core mergers in driven turbulence simulations. The rotational energy particularly affects the rate of disk fragmentation \citep{kratter2006}, which we do not capture here. The core merger rate affects the total number of stars and the stellar density of clustered environments. For us, except perhaps in the MU32 run, core and filament mergers are relatively rare during the first free-fall time. Nonetheless, the consistency of our multiplicity fraction in each simulation suggests that the stellar density plays less of a role in setting multiplicity from turbulent fragmentation compared to other physical properties. 

Our simulations include MHD and gravity and do not consider the role that radiative feedback, protostellar outflows, and chemistry play on fragmentation and multiplicity. Protostellar outflows drive turbulence on small scales and reduce the  accretion rates onto stars and the overall star formation effeciency \citep{offnerarce2014,offnerchaban2017,cunningham18}. A diffuse radiative component can discourage fragmentation on the disk-size scale, but numerical and observational evidence suggest this component does not suppress fragmentation on the scales considered here \citep{Offner2010,Murillo2018}. The coupling between the gas and the magnetic field through the ideal MHD approximation allows angular momentum transport to occur more efficiently than a simulation that employed non-ideal effects. Ideal MHD effects may act to maintain a larger velocity difference between the gas and the embedded binary, which allows orbital evolution to proceed faster. However, non-ideal effects operate most efficiently in dense accretion disks rather than the more-diffuse cores, and we predict this effect will have minimal impact on the results presented here.

} 

%% file: appendix1.tex
%
{
\newcommand{\calO}{$\cal O$}

\section{Multiplicity Hierarchy Method} 
\label{sec:appmultiplicity}
 
\change{In this appendix, we outline our procedure for identifying multiples in our simulations. In the first subsection, we define our procedure. In the second subsection, we assess the sensitivity of our results to assumptions made in this algorithm.} 

\subsection{Labeling Multiples}
Similar to other methods \citep[e.g.,][]{Bate2009} we construct a tree structure by identifying gravitationally bound pairs, replacing these pairs with their center-of-mass equivalent, and recursively iterating until no new bound pairings are found. We associate a ``level'' with each pairing depending on the number of recursive bindings that have been made. For example, two gravitationally bound sink particles both are at level 1, but the ``object'' representing the bound pair is at level 2. If this binary is bound to another binary or another single star, that grouping is at level 3, and so forth. For an object at level $l$, the number of sink particles in the object is at least equal to $l$ and at most equal to $2^{l-1}$. By associating a unique id to each object, we can also create a temporal sequence for multiple systems by comparing the hierarchy from output to output.  

Two objects are gravitationally bound if, once in their center-of-mass frame, their total energy is negative:
\begin{equation}\label{eqn:comenergy}
E_{1,2} = \frac{1}{2}m_{\rm red} v^2 - \frac{Gm_{\rm red} M}{r}\ ,
\end{equation}
where $m_{\rm red}=m_1m_2/M$, $M=m_1+m_2$, $v=|\mathbf{v}_1-\mathbf{v}_2|$, and $r=|\mathbf{r}_1-\mathbf{r}_2|$. This expression ignores the gravitational well of the gas where the sinks are roaming; we address this below. 

We define an ``object'' class, which contains arrays for time, position, velocity, spin angular momentum, the level of the object, a unique id number, and the pair of ids in the case the sink represents a multiple (i.e., when $l$ > 1). This is different from a simulation output $\{S_i\}_k$, $i=1,2,...,N_k$, which outputs the mass, position, velocity, spin, and id for the $N_k$ sink particles at a given output $k$. It is assumed outputs $k$ and $k+1$ are subsequent in time, separated roughly by $\Delta t_{\rm IO}$. Where previous work has focused on the multiplicity statistics at the end of the simulation, our object class here will connect sinks and multiples across multiple data outputs and allow for analysis across simulation time. 

We begin by inputting an ordered list of sink outputs $\{S_i\}_k$ with $k=0,1,...,K$. Here $K$ is the final output of the simulation. Starting with an empty list of objects \calO, we populate \calO\ as follows:\\

\noindent [Time loop] For each $k$:
\begin{enumerate}
    \item If $N_k=0$, continue to the next $k$.
    \item If $\{S_i\}_k$ is the first output containing sink particles (i.e., \calO\ is empty), create a level 1 object for each sink, initializing the arrays with the current times, masses, positions, velocities, and spins for the sink particles. Each object has an id equal to the sink particle id from \ORION. Add them to \calO.
    \item Else if \calO\ is non-empty, for each member of $\{S_i\}_k$ search the level 1 members of \calO\ for an object with the same id number. If it is found, append the current time, mass, position, velocity, and spin to the arrays in that object. If it is not found, create a new level 1 object in \calO.
    \item If $N_k=1$, continue to the next $k$.
    \item Define every item in $\{S_i\}_k$ as level 1 objects. 
    \item Remove objects below $0.02 M_\odot$ from $\{S_i\}_k$.
    \item \ [Multiplicity loop] While $N_k >1$:
        \begin{enumerate}
            \item Using the current set of objects in $\{S_i\}_k$, calculate the minimum value of $E_{1,2}$ of all object pairs using Equation (\ref{eqn:comenergy}). 
            \item If this minimum value is $>0$, break from the [Multiplicity loop]. 
            \item Use the virial theorem
                \begin{equation}
                    \label{eqn:appxvirial}
                    E_{1,2} = -\,\frac{G\mu M}{2a}
                \end{equation}
            to calculate the semi-major axis for the pair. If it is above $10^5$ AU, break from the [Multiplicity loop].
            \item Remove the two objects making up this pair from $\{S_i\}_k$ and replace them with one object with the center-of-mass position, velocity, and spin, the sum of the masses, and a level that is one greater than the maximum level of the two original objects. Label the pair $P$ of this object as the pair of ids of the two original objects. The number of objects in $\{S_i\}_k$ has decreased by one. 
            \item To determine the id of this new object, search objects wtih the same level in \calO\ for one that has the same pair of ids as  $P$. If it exists and existed at time step $k-1$, assign this new object the same id number and append its mass, position, etc. to the object in \calO. If it did not exist at time output $k-1$ (or at all), assign it a unique id and add it to \calO\ as a new object. 
            \item End of [Multiplicity loop].
        \end{enumerate}
    \item End of [Time loop].
\end{enumerate}

This algorithm creates a set \calO\ that includes all the sink and hierarchy information at every output (e.g., Figure \ref{fig:cartoonhierarchy}), and additionally contains information for when multiples remain multiples from output to output. 

A few items to note. By using Equation (\ref{eqn:comenergy}) to compute whether objects are bound, the multiplicity hierarchy is constructed by pairs of objects. However, recall that members of pairs can themselves be multiples, with member numbers not necessarily powers of two. Since higher-level objects are associated with a pair of ids, the multiplicity hierarchy can be easily reconstructed for a given object. In clustered environments, member-swapping can be common, where, for example, three sink particles alternate between which are in a binary that is bound to a third object. These binaries are obviously identified as unique binaries in \calO, and therefore create unique level 3 objects when all three are bound to each other in different combinations. In most circumstances it would not be appropriate to treat these level 3 objects as unique objects because they are composed of the same three level 1 objects. Figure \ref{fig:uniqueornot} compares the non-unique and unique pairings for levels 2--4. Above level 2, membership swapping can create $\sim10$x as many objects in highly-clustered environments, which are more common for larger values of $\mu_\phi$, that is, for weaker global magnetic field strengths. Unless stated, we always use the unique pairings in our statistics throughout the paper.

\subsection{Algorithm Sensitivities}

In several cases, a multiple may be identified as bound for several outputs, become unbound for one or two outputs, and then become bound again. The inclusion of the local gas mass, especially when the sink particles are low-mass, can make these transient unbound events  vanish. Additionally, the gas mass may rearrange members of a hierarchy. To estimate the effects of the gas mass, we take several outputs from our simulations and recompute the hierarchy. For each potential pair, we include the gravitational energy of the gas by computing the total gas mass $M_{\rm gas}$ in a sphere located at the sink pair's center of mass and having a radius equal to half the separation distance $r$ between the two sinks:
\begin{equation}
    \label{eqn:gasenergy}
    E_{\rm gas} \approx -\frac{G M^2_{\rm gas}}{r/2}
\end{equation}
In cases where $M_{\rm gas}/M<1$, the hierarchy calculations remain the same. Only when the ratio is greater than unity do we find that the pairings change or new pairings can arise. Typically, the gas inclusion rearranges objects in a hierarchy but does not change the multiplicity statistics in clustered environments. Figure \ref{fig:gasinclusiontest} shows an example of one such cluster from our $\mu_\phi=8$ run, displaying the hierarchy of seven sink particles with and without including the gas mass. In this case, a small sink particle + the local gas is bound to a nearby sink particle instead of being bound to a six-sink cluster. The multiplicity statistics remain unchanged in this case. 

\change{ In the algorithm above, pairs are identified based on the value of the total energy in the center-of-mass frame, $E_{1,2}$ (Equation \ref{eqn:comenergy}). Prioritization is given to objects that are the most gravitationally bound, regardless of their exact separations. Other algorithms, such as \citet{Bate2009}, prioritize pair protostars that have small separations, as long as the value of $E_{1,2}$ is negative. Other work has required additional constraints when labeling multiples \citep[e.g.,][]{LiIRDC2018}.

It is obvious that the results of a multiplicity study depend on the exact criteria used in defining multiplicity. While it is beyond the scope of this paper to re-analyze our results using several different multiplicity algorithms, we considered whether there was a physical argument for using distance prioritizing algorithms \citep[e.g.,][hereafter called DP algorithms]{Bate2009} versus one that prioritizes orbital energy (e.g., this paper, hereafter called EP algorithms). 

Overall, we argue that both methods generally agree when identifying the multiplicity system in the simulation. We find no clear evidence for valuing one method over another. For one, global averages like the multiplicity fraction should be insensitive to the exact algorithm used. Both methods recursively identify gravitationally bound pairs, and while the exact hierarchy may change between the methods, we do not anticipate that the two methods would grossly disagree on the total number of bound objects. 

A means of comparing the two methods is the following. Consider a binary system (masses $m_1$ and $m_2$) that is bound to a third protostar ($m_3$). Two ratios can be computed: (1) the ratio of the total orbital energy of the triple system to the orbital energy of the inner binary, and (2) the ratio of the exact separation between the center of mass of the binary and the third protostar to the separation between the protostars in the binary. A generalization of this algorithm to higher-order multiples would be to compare two subsequent levels in the hierarchy, calculating the minimum value for each of the separations.

Figure \ref{fig:ratioalgorithm} plots the trajectory of three of our triple-star systems in this parameter space. One data point is drawn for each output. The identification of multiples was done using our algorithm, but the location of the system on this plot allows us to assess the validity of our EP algorithm compared to a DP algorithm. The Figure is divided into four quadrants, depending on whether the distance or energy ratios are greater than or less than unity. The lower-right quadrant is where both methods will always agree on the multiplicity hierarchy. In this case, both the binding energies and separations are sufficiently distinct that either method would draw the same conclusion. A cartoon figure drawn in that quadrant shows an example triple system. The ``Stable Example" data is a triple-star system from our MU8 run. 

Systems that reside in the upper-left quadrant reside in a transient space. In this case, both separations and energies tend to be comparable and the hierarchy is considerably changed from one output to the next. Clusters of stars reside in this quadrant, as well as star groups transitioning into a more-nested hierarchy. The triple star system in our MU2 is drawn as the ``Transition Example.'' These stars are also the protostars identified as the triangle, diamond, and pentagon symbols in Figure \ref{fig:mu2multsnapshot}. With the protostars actually exchanging members with each other from output to output, attempting to define a nested hierarchy is not useful. In this case, membership swapping occurs naturally and is not an artifact of the algorithm. 

The lower-left and upper-right quadrants, however, are where the two algorithms could disagree. In the lower-left quadrant, the drawn cartoon assumes $m_1\approx m_3 \gg m_2$ and that the protostar with mass $m_2$ orbits around $m_1$. A DP algorithm would still identify this hierarchy. However, the mass disparity in the inner-binary and geometry could occasionally pair the two massive protostars with each other first, then identifying the smaller protostar as the third outer member. The ellipses draw the hierarchy identified by our EP algorithm, which does not match reality. We have analyzed all of our triple and quadruple systems and have identified that mis-identification only occurs for one triple-protostar system. This system is shown in the figure as the ``Swapping Example.'' Even then, the multiplicity algorithm oscillates between the correct and incorrect hierarchy between data outputs. 

Similarly, systems in the upper-right quadrant could be mislabeled, but this time by the DP algorithm. The cartoon shows an example of three stars, where the true binary contains the left-most protostars. The binary is assumed to have a fairly eccentric orbit. In this case, the calculated hierarchy by an algorithm favoring distance is drawn with ellipses. Similar to the lower-left quadrant, we anticipate that the algorithm would alternate between the correct and incorrect hierarchy from output to output.

}

\begin{figure}\label{fig:uniqueornot}
\bcenter
\includegraphics[scale=0.5]{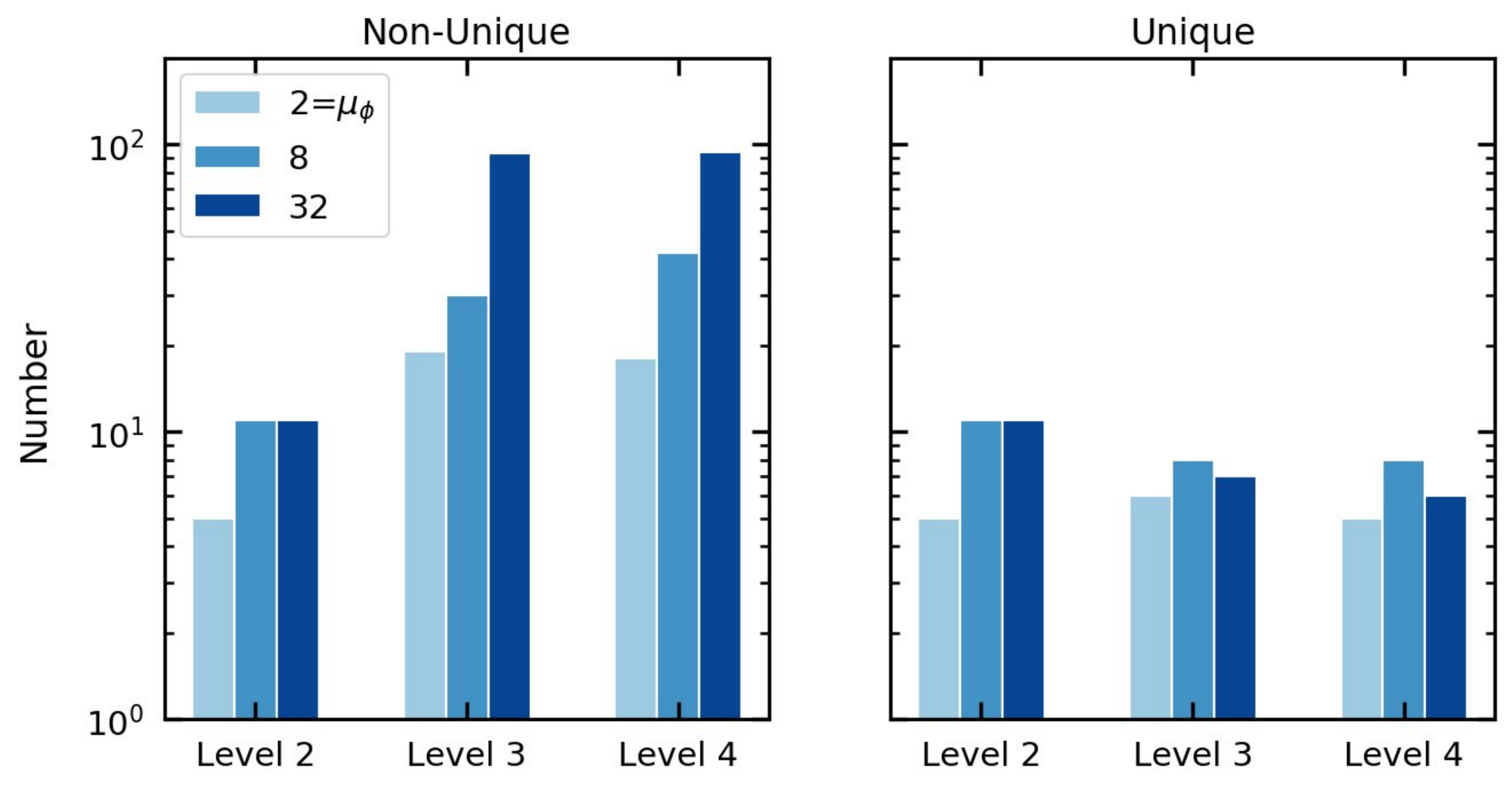}
\ecenter
\caption{A comparison between non-unique and unique pairings for each simulation and for multiples at levels 2, 3, and 4. Clustering and member swapping can produce non-unique pairs above level 2. Note that the $y$-axis is logarithmic.}
\end{figure} 

\begin{figure}\label{fig:gasinclusiontest}
\bcenter
\includegraphics[scale=0.4]{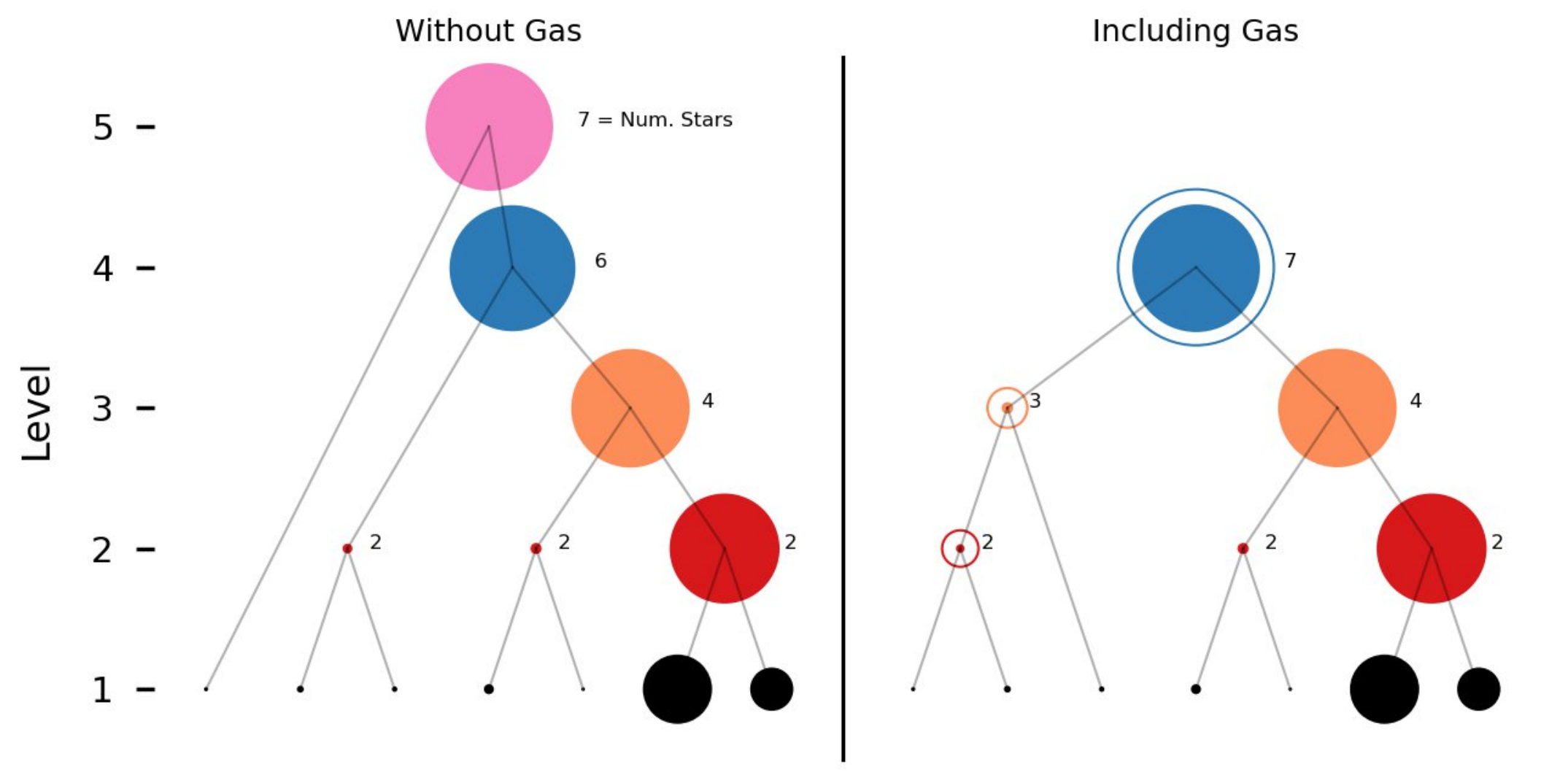}
\includegraphics[scale=0.4]{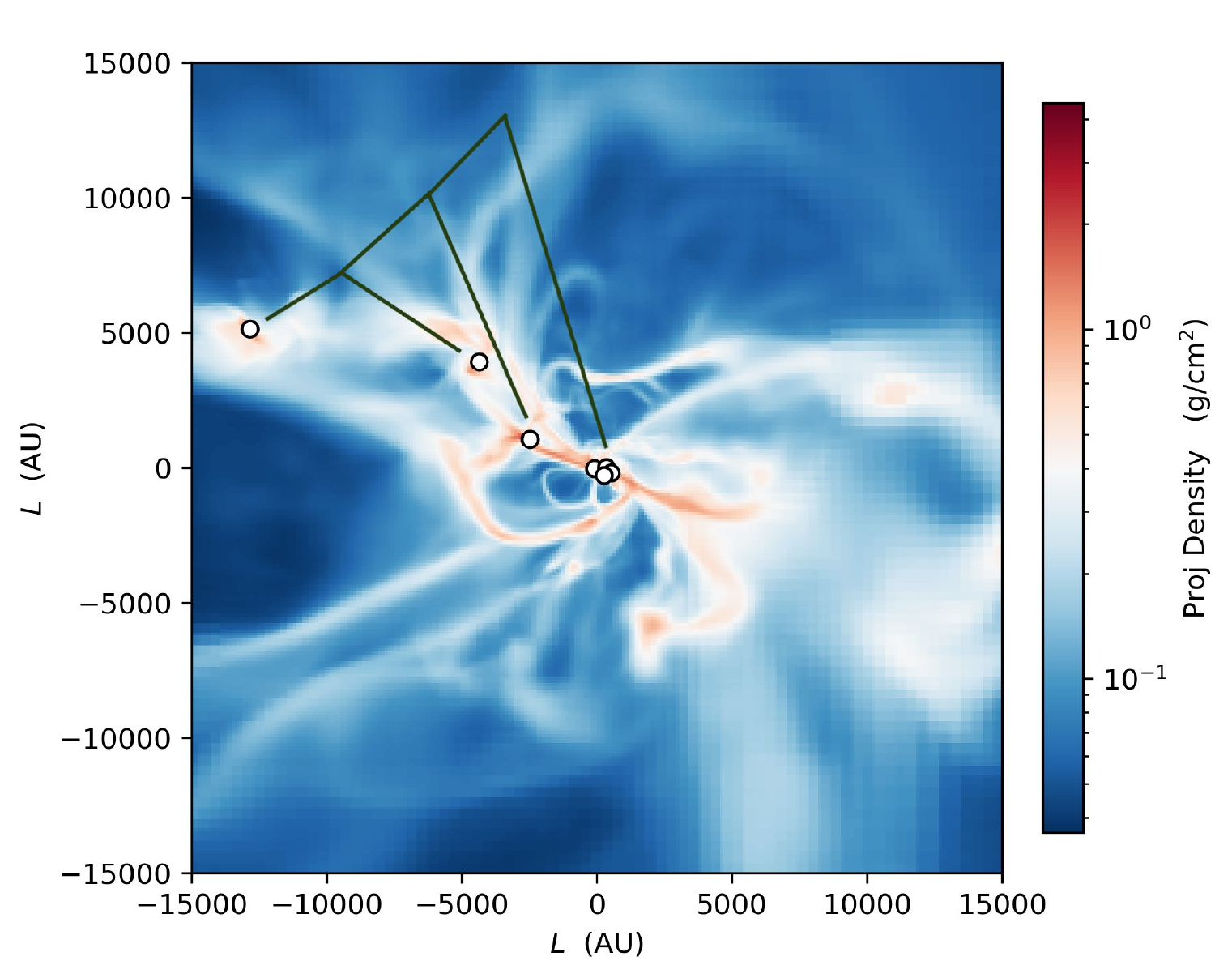}
\ecenter
\caption{Demonstration of multiplicity hierarchy's dependence on the local gas mass. Data is taken from near the end of the $\mu_\phi=8$ simulation. {\it Left:} The multiplicity hierarchy for a cluster of seven sink particles, determined using the method of Appendix \ref{sec:appmultiplicity}. {\it Middle:} The same set of sinks with a hierarchy determined by including an estimate of $M_{\rm gas}$. Point radii scale linearly with total stellar mass. Unfilled circles include the local gas mass, which only is relevant for three of the seven particles in the left-most branch of the hierarchy. {\it Right:} Column density snapshot of the seven sink particles with the left branch of the hierarchy that includes the gas mass overlaid.}
\end{figure}

\begin{figure}\label{fig:ratioalgorithm}
\bcenter
\includegraphics[scale=0.35]{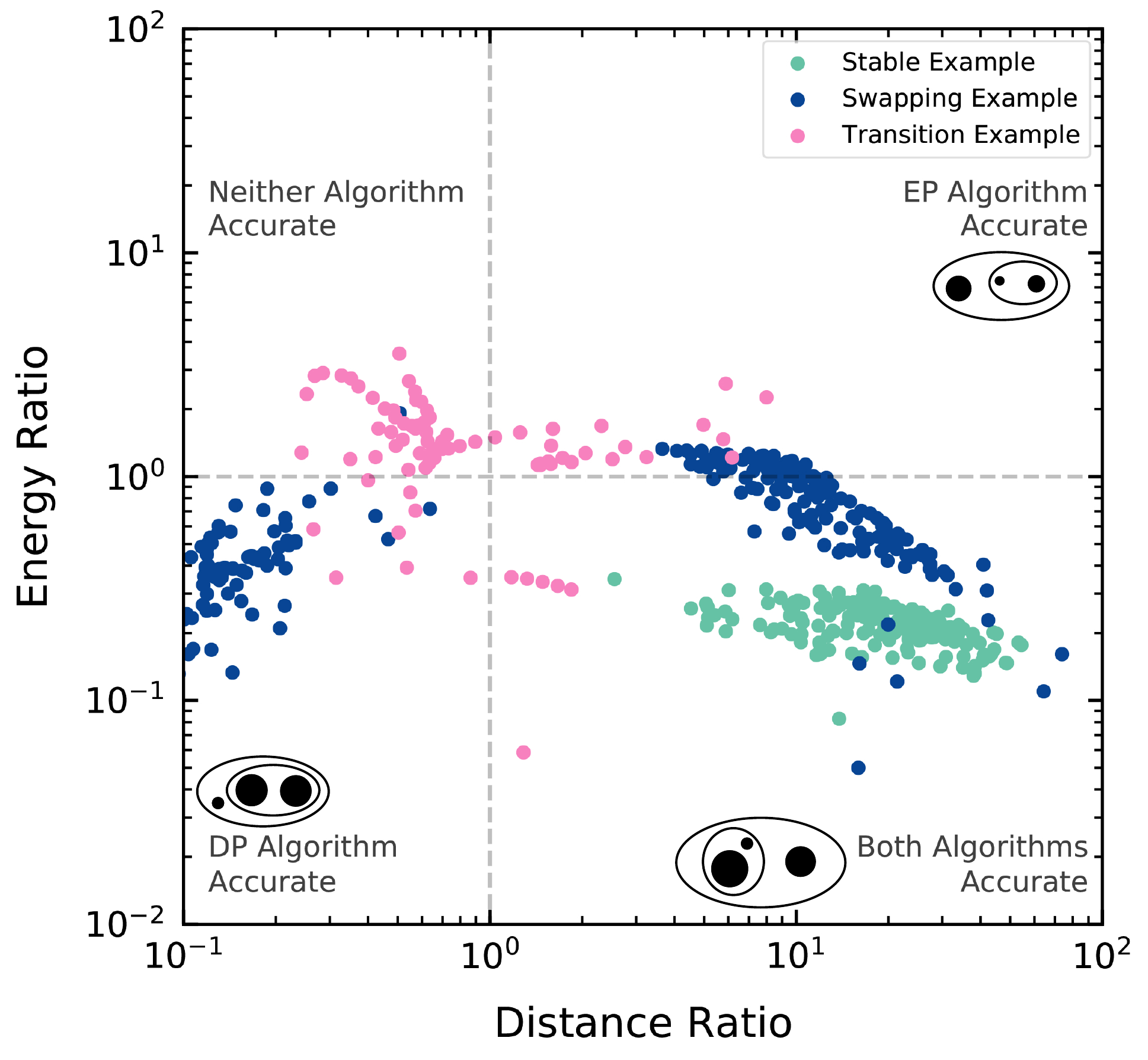}
\ecenter
\caption{Comparison of energy- and distance-prioritizing algorithms (EP and DP, respectively). Three examples of triple-star systems are shown. See the text for discussion.}
\end{figure} 

} 

%% file: main.bbl
\begin{thebibliography}{}
\providecommand\natexlab[1]{#1}
\providecommand\JournalTitle[1]{#1}

\bibitem[{{Adams} {et~al.}(1989){Adams}, {Ruden}, \& {Shu}}]{adamsetal1989}
{Adams}, F.~C., {Ruden}, S.~P., \& {Shu}, F.~H. 1989,
  \href{http://dx.doi.org/10.1086/168187}{\JournalTitle{\apj}, 347, 959}

\bibitem[{{Antoni} {et~al.}(2019){Antoni}, {MacLeod}, \&
  {Ramirez-Ruiz}}]{Antonietal2019}
{Antoni}, A., {MacLeod}, M., \& {Ramirez-Ruiz}, E. 2019,
  \href{http://dx.doi.org/10.3847/1538-4357/ab3466}{\JournalTitle{\apj}, 884,
  22}

\bibitem[{{Arzoumanian} {et~al.}(2011){Arzoumanian}, {Andr{\'e}}, {Didelon},
  {K{\"o}nyves}, {Schneider}, {Men'shchikov}, {Sousbie}, {Zavagno}, {Bontemps},
  {di Francesco}, {Griffin}, {Hennemann}, {Hill}, {Kirk}, {Martin}, {Minier},
  {Molinari}, {Motte}, {Peretto}, {Pezzuto}, {Spinoglio}, {Ward-Thompson},
  {White}, \& {Wilson}}]{Arzoumanian2011}
{Arzoumanian}, D., {Andr{\'e}}, P., {Didelon}, P., {et~al.} 2011,
  \href{http://dx.doi.org/10.1051/0004-6361/201116596}{\JournalTitle{\aap},
  529, L6}

\bibitem[{{Bate}(2000)}]{Bate2000}
{Bate}, M.~R. 2000,
  \href{http://dx.doi.org/10.1046/j.1365-8711.2000.03333.x}{\JournalTitle{\mnras},
  314, 33}

\bibitem[{{Bate}(2009)}]{Bate2009}
---. 2009,
  \href{http://dx.doi.org/10.1111/j.1365-2966.2008.14106.x}{\JournalTitle{\mnras},
  392, 590}

\bibitem[{{Bate}(2018)}]{bate2018}
---. 2018, \href{http://dx.doi.org/10.1093/mnras/sty169}{\JournalTitle{\mnras},
  475, 5618}

\bibitem[{{Bate} \& {Bonnell}(1997)}]{Bate97b}
{Bate}, M.~R., \& {Bonnell}, I.~A. 1997, \JournalTitle{\mnras}, 285, 33

\bibitem[{{Bate} {et~al.}(2003){Bate}, {Bonnell}, \&
  {Bromm}}]{batebonnellbromm}
{Bate}, M.~R., {Bonnell}, I.~A., \& {Bromm}, V. 2003,
  \href{http://dx.doi.org/10.1046/j.1365-8711.2003.06210.x}{\JournalTitle{\mnras},
  339, 577}

\bibitem[{{Batten}(1973)}]{Batten1973book}
{Batten}, A.~H. 1973, {Binary and multiple systems of stars} (Pergamon Press)

\bibitem[{{Bolatto} {et~al.}(2008){Bolatto}, {Leroy}, {Rosolowsky}, {Walter},
  \& {Blitz}}]{bolatto08}
{Bolatto}, A.~D., {Leroy}, A.~K., {Rosolowsky}, E., {Walter}, F., \& {Blitz},
  L. 2008, \href{http://dx.doi.org/10.1086/591513}{\JournalTitle{\apj}, 686,
  948}

\bibitem[{{Bonnell}(1994)}]{Bonnell1994}
{Bonnell}, I.~A. 1994,
  \href{http://dx.doi.org/10.1093/mnras/269.3.837}{\JournalTitle{\mnras}, 269,
  837}

\bibitem[{{Bonnell} {et~al.}(2001){Bonnell}, {Bate}, {Clarke}, \&
  {Pringle}}]{Bonnell2001}
{Bonnell}, I.~A., {Bate}, M.~R., {Clarke}, C.~J., \& {Pringle}, J.~E. 2001,
  \href{http://dx.doi.org/10.1046/j.1365-8711.2001.04270.x}{\JournalTitle{\mnras},
  323, 785}

\bibitem[{{Boss}(1986)}]{Boss1986v}
{Boss}, A.~P. 1986,
  \href{http://dx.doi.org/10.1086/191150}{\JournalTitle{\apjs}, 62, 519}

\bibitem[{{Boss}(1988)}]{Boss1988}
---. 1988, \JournalTitle{Comments on Astrophysics}, 12, 169

\bibitem[{{Burkert} \& {Bodenheimer}(2000)}]{Burkert:2000}
{Burkert}, A., \& {Bodenheimer}, P. 2000, \JournalTitle{\apj}, 543, 822

\bibitem[{{Burleigh} {et~al.}(2017){Burleigh}, {McKee}, {Cunningham}, {Lee}, \&
  {Klein}}]{Burleighetal2017}
{Burleigh}, K.~J., {McKee}, C.~F., {Cunningham}, A.~J., {Lee}, A.~T., \&
  {Klein}, R.~I. 2017,
  \href{http://dx.doi.org/10.1093/mnras/stx439}{\JournalTitle{\mnras}, 468,
  717}

\bibitem[{{Chabrier}(2003)}]{chabrier2003}
{Chabrier}, G. 2003,
  \href{http://dx.doi.org/10.1086/376392}{\JournalTitle{\pasp}, 115, 763}

\bibitem[{{Chen} {et~al.}(2008){Chen}, {Bourke}, {Launhardt}, \&
  {Henning}}]{chen2008}
{Chen}, X., {Bourke}, T.~L., {Launhardt}, R., \& {Henning}, T. 2008,
  \href{http://dx.doi.org/10.1086/593033}{\JournalTitle{\apjl}, 686, L107}

\bibitem[{{Chen} {et~al.}(2013){Chen}, {Arce}, {Zhang}, {Bourke}, {Launhardt},
  {J{\o}rgensen}, {Lee}, {Foster}, {Dunham}, {Pineda}, \& {Henning}}]{chen13}
{Chen}, X., {Arce}, H.~G., {Zhang}, Q., {et~al.} 2013,
  \href{http://dx.doi.org/10.1088/0004-637X/768/2/110}{\JournalTitle{\apj},
  768, 110}

\bibitem[{{Clarke} \& {Pringle}(1991)}]{ClarkePringle1991}
{Clarke}, C.~J., \& {Pringle}, J.~E. 1991,
  \href{http://dx.doi.org/10.1093/mnras/249.4.584}{\JournalTitle{\mnras}, 249,
  584}

\bibitem[{{Conroy} \& {Kratter}(2012)}]{ConroyKratter2012}
{Conroy}, C., \& {Kratter}, K.~M. 2012,
  \href{http://dx.doi.org/10.1088/0004-637X/755/2/123}{\JournalTitle{\apj},
  755, 123}

\bibitem[{{Correia} {et~al.}(2006){Correia}, {Zinnecker}, {Ratzka}, \&
  {Sterzik}}]{Correiaetal2006}
{Correia}, S., {Zinnecker}, H., {Ratzka}, T., \& {Sterzik}, M.~F. 2006,
  \href{http://dx.doi.org/10.1051/0004-6361:20065545}{\JournalTitle{\aap}, 459,
  909}

\bibitem[{{Cunningham} {et~al.}(2011){Cunningham}, {Klein}, {Krumholz}, \&
  {McKee}}]{cunningham11}
{Cunningham}, A.~J., {Klein}, R.~I., {Krumholz}, M.~R., \& {McKee}, C.~F. 2011,
  \href{http://dx.doi.org/10.1088/0004-637X/740/2/107}{\JournalTitle{\apj},
  740, 107}

\bibitem[{{Cunningham} {et~al.}(2018){Cunningham}, {Krumholz}, {McKee}, \&
  {Klein}}]{cunningham18}
{Cunningham}, A.~J., {Krumholz}, M.~R., {McKee}, C.~F., \& {Klein}, R.~I. 2018,
  \href{http://dx.doi.org/10.1093/mnras/sty154}{\JournalTitle{\mnras}, 476,
  771}

\bibitem[{{Dubinski} {et~al.}(1995){Dubinski}, {Narayan}, \&
  {Phillips}}]{Dubinski:1995}
{Dubinski}, J., {Narayan}, R., \& {Phillips}, T.~G. 1995,
  \href{http://dx.doi.org/10.1086/175954}{\JournalTitle{\apj}, 448, 226}

\bibitem[{{Duch{\^e}ne} \& {Kraus}(2013)}]{duchene13}
{Duch{\^e}ne}, G., \& {Kraus}, A. 2013,
  \href{http://dx.doi.org/10.1146/annurev-astro-081710-102602}{\JournalTitle{\araa},
  51, 269}

\bibitem[{{Dunham} {et~al.}(2013){Dunham}, {Arce}, {Allen}, {Evans},
  {Broekhoven-Fiene}, {Chapman}, {Cieza}, {Gutermuth}, {Harvey}, \&
  {Hatchell}}]{dunhametal2013}
{Dunham}, M.~M., {Arce}, H.~G., {Allen}, L.~E., {et~al.} 2013,
  \href{http://dx.doi.org/10.1088/0004-6256/145/4/94}{\JournalTitle{\aj}, 145,
  94}

\bibitem[{{Duquennoy} \& {Mayor}(1991)}]{Duquennoy1991}
{Duquennoy}, A., \& {Mayor}, M. 1991, \JournalTitle{\aap}, 500, 337

\bibitem[{{Dutrey} {et~al.}(1994){Dutrey}, {Guilloteau}, \& {Simon}}]{ggtau}
{Dutrey}, A., {Guilloteau}, S., \& {Simon}, M. 1994, \JournalTitle{\aap}, 286,
  149

\bibitem[{{Eldridge} \& {Stanway}(2009)}]{eldridge2009}
{Eldridge}, J.~J., \& {Stanway}, E.~R. 2009,
  \href{http://dx.doi.org/10.1111/j.1365-2966.2009.15514.x}{\JournalTitle{\mnras},
  400, 1019}

\bibitem[{{Enoch} {et~al.}(2008){Enoch}, {Evans}, {Sargent}, {Glenn},
  {Rosolowsky}, \& {Myers}}]{enoch08b}
{Enoch}, M.~L., {Evans}, II, N.~J., {Sargent}, A.~I., {et~al.} 2008,
  \href{http://dx.doi.org/10.1086/589963}{\JournalTitle{\apj}, 684, 1240}

\bibitem[{{Federrath} {et~al.}(2011{\natexlab{a}}){Federrath}, {Chabrier},
  {Schober}, {Banerjee}, {Klessen}, \& {Schleicher}}]{federrathetal11a}
{Federrath}, C., {Chabrier}, G., {Schober}, J., {et~al.} 2011{\natexlab{a}},
  \href{http://dx.doi.org/10.1103/PhysRevLett.107.114504}{\JournalTitle{PhRvL},
  107, 114504}

\bibitem[{{Federrath} \& {Klessen}(2012)}]{FederrathStarFormLaw}
{Federrath}, C., \& {Klessen}, R.~S. 2012,
  \href{http://dx.doi.org/10.1088/0004-637X/761/2/156}{\JournalTitle{\apj},
  761, 156}

\bibitem[{{Federrath} {et~al.}(2011{\natexlab{b}}){Federrath}, {Sur},
  {Schleicher}, {Banerjee}, \& {Klessen}}]{federrathetal11b}
{Federrath}, C., {Sur}, S., {Schleicher}, D.~R.~G., {Banerjee}, R., \&
  {Klessen}, R.~S. 2011{\natexlab{b}},
  \href{http://dx.doi.org/10.1088/0004-637X/731/1/62}{\JournalTitle{\apj}, 731,
  62}

\bibitem[{{Fischer} {et~al.}(2013){Fischer}, {Megeath}, {Stutz}, {Tobin},
  {Ali}, {Stanke}, {Osorio}, {Furlan}, {HOPS Team}, \& {Orion Protostar
  Survey}}]{fischeretal2013}
{Fischer}, W.~J., {Megeath}, S.~T., {Stutz}, A.~M., {et~al.} 2013,
  \href{http://dx.doi.org/10.1002/asna.201211761}{\JournalTitle{Astronomische
  Nachrichten}, 334, 53}

\bibitem[{{Fisher}(2004)}]{Fisher04}
{Fisher}, R.~T. 2004, \JournalTitle{\apj}, 600, 769

\bibitem[{{Ghez} {et~al.}(1993){Ghez}, {Neugebauer}, \& {Matthews}}]{Ghez1993}
{Ghez}, A.~M., {Neugebauer}, G., \& {Matthews}, K. 1993,
  \href{http://dx.doi.org/10.1086/116782}{\JournalTitle{\aj}, 106, 2005}

\bibitem[{{Goodwin} \& {Kroupa}(2005)}]{goodwinkroupa2005}
{Goodwin}, S.~P., \& {Kroupa}, P. 2005,
  \href{http://dx.doi.org/10.1051/0004-6361:20052654}{\JournalTitle{\aap}, 439,
  565}

\bibitem[{{Goodwin} {et~al.}(2004){Goodwin}, {Whitworth}, \&
  {Ward-Thompson}}]{Goodwin2004}
{Goodwin}, S.~P., {Whitworth}, A.~P., \& {Ward-Thompson}, D. 2004,
  \href{http://dx.doi.org/10.1051/0004-6361:20031594}{\JournalTitle{\aap}, 414,
  633}

\bibitem[{{Heyer} {et~al.}(2009){Heyer}, {Krawczyk}, {Duval}, \&
  {Jackson}}]{heyerlarson08}
{Heyer}, M., {Krawczyk}, C., {Duval}, J., \& {Jackson}, J.~M. 2009,
  \href{http://dx.doi.org/10.1088/0004-637X/699/2/1092}{\JournalTitle{\apj},
  699, 1092}

\bibitem[{{Indulekha}(2013)}]{Indulekha2013}
{Indulekha}, K. 2013,
  \href{http://dx.doi.org/10.1007/s12036-013-9175-7}{\JournalTitle{JApA}, 34,
  207}

\bibitem[{{King} {et~al.}(2012{\natexlab{a}}){King}, {Goodwin}, {Parker}, \&
  {Patience}}]{king2012paper2}
{King}, R.~R., {Goodwin}, S.~P., {Parker}, R.~J., \& {Patience}, J.
  2012{\natexlab{a}},
  \href{http://dx.doi.org/10.1111/j.1365-2966.2012.22108.x}{\JournalTitle{\mnras},
  427, 2636}

\bibitem[{{King} {et~al.}(2012{\natexlab{b}}){King}, {Parker}, {Patience}, \&
  {Goodwin}}]{king2012paper1}
{King}, R.~R., {Parker}, R.~J., {Patience}, J., \& {Goodwin}, S.~P.
  2012{\natexlab{b}},
  \href{http://dx.doi.org/10.1111/j.1365-2966.2012.20437.x}{\JournalTitle{\mnras},
  421, 2025}

\bibitem[{{Kounkel} {et~al.}(2019){Kounkel}, {Covey}, {Moe}, {Kratter},
  {Su{\'a}rez}, {Stassun}, {Rom{\'a}n-Z{\'u}{\~n}iga}, {Hernand ez}, {Kim}, \&
  {Pe{\~n}a Ram{\'\i}rez}}]{Kounkel2019}
{Kounkel}, M., {Covey}, K., {Moe}, M., {et~al.} 2019,
  \href{http://dx.doi.org/10.3847/1538-3881/ab13b1}{\JournalTitle{\aj}, 157,
  196}

\bibitem[{{Kratter}(2011)}]{Kratter2011ASPC}
{Kratter}, K.~M. 2011, in 2011ASPC447, ed. L.~{Schmidtobreick}, M.~R.
  {Schreiber}, \& C.~{Tappert}

\bibitem[{{Kratter} \& {Matzner}(2006)}]{kratter2006}
{Kratter}, K.~M., \& {Matzner}, C.~D. 2006,
  \href{http://dx.doi.org/10.1111/j.1365-2966.2006.11103.x}{\JournalTitle{\mnras},
  373, 1563}

\bibitem[{{Kratter} {et~al.}(2010){Kratter}, {Matzner}, {Krumholz}, \&
  {Klein}}]{Kratteretal2010}
{Kratter}, K.~M., {Matzner}, C.~D., {Krumholz}, M.~R., \& {Klein}, R.~I. 2010,
  \href{http://dx.doi.org/10.1088/0004-637X/708/2/1585}{\JournalTitle{\apj},
  708, 1585}

\bibitem[{{Kraus} \& {Hillenbrand}(2007)}]{kraushillenbrand2007}
{Kraus}, A.~L., \& {Hillenbrand}, L.~A. 2007,
  \href{http://dx.doi.org/10.1086/516835}{\JournalTitle{\apj}, 662, 413}

\bibitem[{{Kraus} {et~al.}(2011){Kraus}, {Ireland}, {Martinache}, \&
  {Hillenbrand}}]{krausetal2011}
{Kraus}, A.~L., {Ireland}, M.~J., {Martinache}, F., \& {Hillenbrand}, L.~A.
  2011, \href{http://dx.doi.org/10.1088/0004-637X/731/1/8}{\JournalTitle{\apj},
  731, 8}

\bibitem[{{Kroupa}(2001)}]{kroupa01}
{Kroupa}, P. 2001,
  \href{http://dx.doi.org/10.1046/j.1365-8711.2001.04022.x}{\JournalTitle{\mnras},
  322, 231}

\bibitem[{{Krumholz} {et~al.}(2007){Krumholz}, {Klein}, \&
  {McKee}}]{krumholz07}
{Krumholz}, M.~R., {Klein}, R.~I., \& {McKee}, C.~F. 2007,
  \href{http://dx.doi.org/10.1086/510664}{\JournalTitle{\apj}, 656, 959}

\bibitem[{{Krumholz} {et~al.}(2012){Krumholz}, {Klein}, \&
  {McKee}}]{krumholz12}
---. 2012,
  \href{http://dx.doi.org/10.1088/0004-637X/754/1/71}{\JournalTitle{\apj}, 754,
  71}

\bibitem[{{Krumholz} \& {McKee}(2005)}]{krumholzmckee2005}
{Krumholz}, M.~R., \& {McKee}, C.~F. 2005,
  \href{http://dx.doi.org/10.1086/431734}{\JournalTitle{\apj}, 630, 250}

\bibitem[{{Krumholz} {et~al.}(2004){Krumholz}, {McKee}, \&
  {Klein}}]{Krumholz04}
{Krumholz}, M.~R., {McKee}, C.~F., \& {Klein}, R.~I. 2004,
  \href{http://dx.doi.org/10.1086/421935}{\JournalTitle{\apj}, 611, 399}

\bibitem[{{Lada}(1987)}]{Lada1987}
{Lada}, C.~J. 1987, in IAU Symposium, Vol. 115, Star Forming Regions, ed.
  M.~{Peimbert} \& J.~{Jugaku}, 1

\bibitem[{{Larson}(1972)}]{Larson1972binary}
{Larson}, R.~B. 1972,
  \href{http://dx.doi.org/10.1093/mnras/156.4.437}{\JournalTitle{\mnras}, 156,
  437}

\bibitem[{{Larson}(1981)}]{larson81}
---. 1981, \JournalTitle{\mnras}, 194, 809

\bibitem[{{Lee} {et~al.}(2014){Lee}, {Cunningham}, {McKee}, \& {Klein}}]{lee14}
{Lee}, A.~T., {Cunningham}, A.~J., {McKee}, C.~F., \& {Klein}, R.~I. 2014,
  \href{http://dx.doi.org/10.1088/0004-637X/783/1/50}{\JournalTitle{\apj}, 783,
  50}

\bibitem[{{Lee} \& {Stahler}(2011)}]{Leestahler2011}
{Lee}, A.~T., \& {Stahler}, S.~W. 2011,
  \href{http://dx.doi.org/10.1111/j.1365-2966.2011.19273.x}{\JournalTitle{\mnras},
  416, 3177}

\bibitem[{{Lee} \& {Stahler}(2014)}]{Leestahler2014}
---. 2014,
  \href{http://dx.doi.org/10.1051/0004-6361/201322829}{\JournalTitle{\aap},
  561, A84}

\bibitem[{{Lee} {et~al.}(2015){Lee}, {Chang}, \& {Murray}}]{leechangmurray2015}
{Lee}, E.~J., {Chang}, P., \& {Murray}, N. 2015,
  \href{http://dx.doi.org/10.1088/0004-637X/800/1/49}{\JournalTitle{\apj}, 800,
  49}

\bibitem[{{Lee} {et~al.}(2017){Lee}, {Hull}, \& {Offner}}]{leehulloffner2017}
{Lee}, J. W.~Y., {Hull}, C. L.~H., \& {Offner}, S. S.~R. 2017,
  \href{http://dx.doi.org/10.3847/1538-4357/834/2/201}{\JournalTitle{\apj},
  834, 201}

\bibitem[{{Lee} {et~al.}(2016){Lee}, {Dunham}, {Myers}, {Arce}, {Bourke},
  {Goodman}, {J{\o}rgensen}, {Kristensen}, {Offner}, {Pineda}, {Tobin}, \&
  {Vorobyov}}]{leeoutflows2016}
{Lee}, K.~I., {Dunham}, M.~M., {Myers}, P.~C., {et~al.} 2016,
  \href{http://dx.doi.org/10.3847/2041-8205/820/1/L2}{\JournalTitle{\apjl},
  820, L2}

\bibitem[{{Leinert} {et~al.}(1993){Leinert}, {Zinnecker}, {Weitzel},
  {Christou}, {Ridgway}, {Jameson}, {Haas}, \& {Lenzen}}]{Leinert1993}
{Leinert}, C., {Zinnecker}, H., {Weitzel}, N., {et~al.} 1993,
  \JournalTitle{\aap}, 278, 129

\bibitem[{{Li} {et~al.}(2018){Li}, {Klein}, \& {McKee}}]{LiIRDC2018}
{Li}, P.~S., {Klein}, R.~I., \& {McKee}, C.~F. 2018,
  \href{http://dx.doi.org/10.1093/mnras/stx2611}{\JournalTitle{\mnras}, 473,
  4220}

\bibitem[{{Li} {et~al.}(2012){Li}, {Martin}, {Klein}, \& {McKee}}]{li12}
{Li}, P.~S., {Martin}, D.~F., {Klein}, R.~I., \& {McKee}, C.~F. 2012,
  \href{http://dx.doi.org/10.1088/0004-637X/745/2/139}{\JournalTitle{\apj},
  745, 139}

\bibitem[{{Ma} {et~al.}(2016){Ma}, {Hopkins}, {Kasen}, {Quataert},
  {Faucher-Gigu{\`e}re}, {Kere{\v{s}}}, {Murray}, \& {Strom}}]{Maetal2016}
{Ma}, X., {Hopkins}, P.~F., {Kasen}, D., {et~al.} 2016,
  \href{http://dx.doi.org/10.1093/mnras/stw941}{\JournalTitle{\mnras}, 459,
  3614}

\bibitem[{{Mac Low}(1999)}]{maclow99}
{Mac Low}, M.-M. 1999,
  \href{http://dx.doi.org/10.1086/307784}{\JournalTitle{\apj}, 524, 169}

\bibitem[{{Marzari} \& {Gallina}(2016)}]{marzarigallina2016}
{Marzari}, F., \& {Gallina}, G. 2016,
  \href{http://dx.doi.org/10.1051/0004-6361/201628342}{\JournalTitle{\aap},
  594, A89}

\bibitem[{{Masunaga} \& {Inutsuka}(2000)}]{Masunaga:2000}
{Masunaga}, H., \& {Inutsuka}, S.-i. 2000,
  \href{http://dx.doi.org/10.1086/308901}{\JournalTitle{\apj}, 536, 406}

\bibitem[{{McKee} {et~al.}(2010){McKee}, {Li}, \& {Klein}}]{mckeeliklein10}
{McKee}, C.~F., {Li}, P.~S., \& {Klein}, R.~I. 2010,
  \href{http://dx.doi.org/10.1088/0004-637X/720/2/1612}{\JournalTitle{\apj},
  720, 1612}

\bibitem[{{McKee} \& {Ostriker}(2007)}]{mckeeostriker07}
{McKee}, C.~F., \& {Ostriker}, E.~C. 2007,
  \href{http://dx.doi.org/10.1146/annurev.astro.45.051806.110602}{\JournalTitle{\araa},
  45, 565}

\bibitem[{{Mignone} {et~al.}(2012){Mignone}, {Zanni}, {Tzeferacos}, {van
  Straalen}, {Colella}, \& {Bodo}}]{mignone12}
{Mignone}, A., {Zanni}, C., {Tzeferacos}, P., {et~al.} 2012,
  \href{http://dx.doi.org/10.1088/0067-0049/198/1/7}{\JournalTitle{\apjs}, 198,
  7}

\bibitem[{{Moe} \& {Di Stefano}(2013)}]{MoeStefano2013}
{Moe}, M., \& {Di Stefano}, R. 2013,
  \href{http://dx.doi.org/10.1088/0004-637X/778/2/95}{\JournalTitle{\apj}, 778,
  95}

\bibitem[{{Moe} \& {Di Stefano}(2017)}]{moedistefano17}
---. 2017,
  \href{http://dx.doi.org/10.3847/1538-4365/aa6fb6}{\JournalTitle{\apjs}, 230,
  15}

\bibitem[{{Moe} \& {Kratter}(2018)}]{MoeKratter2018}
{Moe}, M., \& {Kratter}, K.~M. 2018,
  \href{http://dx.doi.org/10.3847/1538-4357/aaa6d2}{\JournalTitle{\apj}, 854,
  44}

\bibitem[{{Moeckel} \& {Bally}(2007)}]{MoeckelBally2007}
{Moeckel}, N., \& {Bally}, J. 2007,
  \href{http://dx.doi.org/10.1086/518738}{\JournalTitle{\apjl}, 661, L183}

\bibitem[{{Mu{\~n}oz} {et~al.}(2019{\natexlab{a}}){Mu{\~n}oz}, {Lai},
  {Kratter}, \& {Mirand a}}]{Munoz2019b}
{Mu{\~n}oz}, D., {Lai}, D., {Kratter}, K., \& {Mirand a}, R.
  2019{\natexlab{a}}, \JournalTitle{arXiv e-prints}, arXiv:1910.04763,
  submitted to ApJ

\bibitem[{{Mu{\~n}oz} {et~al.}(2019{\natexlab{b}}){Mu{\~n}oz}, {Miranda}, \&
  {Lai}}]{Munoz2019a}
{Mu{\~n}oz}, D.~J., {Miranda}, R., \& {Lai}, D. 2019{\natexlab{b}},
  \href{http://dx.doi.org/10.3847/1538-4357/aaf867}{\JournalTitle{\apj}, 871,
  84}

\bibitem[{{Murillo} {et~al.}(2018){Murillo}, {van Dishoeck}, {Tobin},
  {Mottram}, \& {Karska}}]{Murillo2018}
{Murillo}, N.~M., {van Dishoeck}, E.~F., {Tobin}, J.~J., {Mottram}, J.~C., \&
  {Karska}, A. 2018,
  \href{http://dx.doi.org/10.1051/0004-6361/201832954}{\JournalTitle{\aap},
  620, A30}

\bibitem[{{Myers} {et~al.}(2013){Myers}, {McKee}, {Cunningham}, {Klein}, \&
  {Krumholz}}]{myers13}
{Myers}, A.~T., {McKee}, C.~F., {Cunningham}, A.~J., {Klein}, R.~I., \&
  {Krumholz}, M.~R. 2013,
  \href{http://dx.doi.org/10.1088/0004-637X/766/2/97}{\JournalTitle{\apj}, 766,
  97}

\bibitem[{{Offner} \& {Arce}(2014)}]{offnerarce2014}
{Offner}, S. S.~R., \& {Arce}, H.~G. 2014,
  \href{http://dx.doi.org/10.1088/0004-637X/784/1/61}{\JournalTitle{\apj}, 784,
  61}

\bibitem[{{Offner} \& {Chaban}(2017)}]{offnerchaban2017}
{Offner}, S. S.~R., \& {Chaban}, J. 2017,
  \href{http://dx.doi.org/10.3847/1538-4357/aa8996}{\JournalTitle{\apj}, 847,
  104}

\bibitem[{{Offner} {et~al.}(2016){Offner}, {Dunham}, {Lee}, {Arce}, \&
  {Fielding}}]{Offner16}
{Offner}, S.~S.~R., {Dunham}, M.~M., {Lee}, K.~I., {Arce}, H.~G., \&
  {Fielding}, D.~B. 2016,
  \href{http://dx.doi.org/10.3847/2041-8205/827/1/L11}{\JournalTitle{\apjl},
  827, L11}

\bibitem[{{Offner} {et~al.}(2008){Offner}, {Klein}, \&
  {McKee}}]{Offner2008drive}
{Offner}, S. S.~R., {Klein}, R.~I., \& {McKee}, C.~F. 2008,
  \href{http://dx.doi.org/10.1086/590238}{\JournalTitle{\apj}, 686, 1174}

\bibitem[{{Offner} {et~al.}(2010){Offner}, {Kratter}, {Matzner}, {Krumholz}, \&
  {Klein}}]{Offner2010}
{Offner}, S. S.~R., {Kratter}, K.~M., {Matzner}, C.~D., {Krumholz}, M.~R., \&
  {Klein}, R.~I. 2010,
  \href{http://dx.doi.org/10.1088/0004-637X/725/2/1485}{\JournalTitle{\apj},
  725, 1485}

\bibitem[{{Ostriker}(1999)}]{Ostriker1999}
{Ostriker}, E.~C. 1999,
  \href{http://dx.doi.org/10.1086/306858}{\JournalTitle{\apj}, 513, 252}

\bibitem[{{Ostriker} {et~al.}(1999){Ostriker}, {Gammie}, \&
  {Stone}}]{Ostriker:1999}
{Ostriker}, E.~C., {Gammie}, C.~F., \& {Stone}, J.~M. 1999,
  \href{http://adsabs.harvard.edu/cgi-bin/nph-bib_query?bibcode=1999ApJ...513..259O&db_key=AST}{\JournalTitle{\apj},
  513, 259}

\bibitem[{{Padoan} {et~al.}(2012){Padoan}, {Haugb{\o}lle}, \&
  {Nordlund}}]{padoan12}
{Padoan}, P., {Haugb{\o}lle}, T., \& {Nordlund}, {\AA}. 2012,
  \href{http://dx.doi.org/10.1088/2041-8205/759/2/L27}{\JournalTitle{\apjl},
  759, L27}

\bibitem[{{Padoan} \& {Nordlund}(2002)}]{Padoan:2002}
{Padoan}, P., \& {Nordlund}, {\AA}. 2002, \JournalTitle{\apj}, 576, 870

\bibitem[{{Palla} \& {Stahler}(2000)}]{pallastahler2000}
{Palla}, F., \& {Stahler}, S.~W. 2000,
  \href{http://dx.doi.org/10.1086/309312}{\JournalTitle{\apj}, 540, 255}

\bibitem[{{Pineda} {et~al.}(2015){Pineda}, {Offner}, {Parker}, {Arce},
  {Goodman}, {Caselli}, {Fuller}, {Bourke}, \& {Corder}}]{pineda15}
{Pineda}, J.~E., {Offner}, S.~S.~R., {Parker}, R.~J., {et~al.} 2015,
  \href{http://dx.doi.org/10.1038/nature14166}{\JournalTitle{\nat}, 518, 213}

\bibitem[{{Price} \& {Bate}(2008)}]{PriceBate2008MNRAS}
{Price}, D.~J., \& {Bate}, M.~R. 2008,
  \href{http://dx.doi.org/10.1111/j.1365-2966.2008.12976.x}{\JournalTitle{\mnras},
  385, 1820}

\bibitem[{{Price} \& {Bate}(2009)}]{PriceBate2009MNRAS}
---. 2009,
  \href{http://dx.doi.org/10.1111/j.1365-2966.2009.14969.x}{\JournalTitle{\mnras},
  398, 33}

\bibitem[{{Pringle}(1989)}]{Pringle1989}
{Pringle}, J.~E. 1989,
  \href{http://dx.doi.org/10.1093/mnras/239.2.361}{\JournalTitle{\mnras}, 239,
  361}

\bibitem[{{Raghavan} {et~al.}(2010){Raghavan}, {McAlister}, {Henry}, {Latham},
  {Marcy}, {Mason}, {Gies}, {White}, \& {ten Brummelaar}}]{Raghavan2010}
{Raghavan}, D., {McAlister}, H.~A., {Henry}, T.~J., {et~al.} 2010,
  \href{http://dx.doi.org/10.1088/0067-0049/190/1/1}{\JournalTitle{\apjs}, 190,
  1}

\bibitem[{{Reipurth} \& {Mikkola}(2012)}]{Reipurth2012}
{Reipurth}, B., \& {Mikkola}, S. 2012,
  \href{http://dx.doi.org/10.1038/nature11662}{\JournalTitle{\nat}, 492, 221}

\bibitem[{{Reipurth} \& {Zinnecker}(1993)}]{ReipurthZinnecker93}
{Reipurth}, B., \& {Zinnecker}, H. 1993, \JournalTitle{\aap}, 278, 81

\bibitem[{{Rosdahl} {et~al.}(2018){Rosdahl}, {Katz}, {Blaizot}, {Kimm},
  {Michel-Dansac}, {Garel}, {Haehnelt}, {Ocvirk}, \& {Teyssier}}]{Rosdahl2018}
{Rosdahl}, J., {Katz}, H., {Blaizot}, J., {et~al.} 2018,
  \href{http://dx.doi.org/10.1093/mnras/sty1655}{\JournalTitle{\mnras}, 479,
  994}

\bibitem[{{Sana} \& {Evans}(2011)}]{SanaMassive2011}
{Sana}, H., \& {Evans}, C.~J. 2011, in 2011IAUS272, ed. C.~{Neiner}, G.~{Wade},
  G.~{Meynet}, \& G.~{Peters}

\bibitem[{{Solomon} {et~al.}(1987){Solomon}, {Rivolo}, {Barrett}, \&
  {Yahil}}]{solomon87}
{Solomon}, P.~M., {Rivolo}, A.~R., {Barrett}, J., \& {Yahil}, A. 1987,
  \href{http://dx.doi.org/10.1086/165493}{\JournalTitle{\apj}, 319, 730}

\bibitem[{{Stahler}(2010)}]{stahler2010binaries}
{Stahler}, S.~W. 2010,
  \href{http://dx.doi.org/10.1111/j.1365-2966.2009.15994.x}{\JournalTitle{\mnras},
  402, 1758}

\bibitem[{{Stamatellos} \& {Whitworth}(2009)}]{Stamatellos2009}
{Stamatellos}, D., \& {Whitworth}, A.~P. 2009,
  \href{http://dx.doi.org/10.1111/j.1365-2966.2008.14069.x}{\JournalTitle{\mnras},
  392, 413}

\bibitem[{Takeda {et~al.}(2008)Takeda, Kita, \& Rasio}]{Takeda_2008}
Takeda, G., Kita, R., \& Rasio, F.~A. 2008,
  \href{http://dx.doi.org/10.1086/589852}{\JournalTitle{\apj}, 683, 1063}

\bibitem[{{Tobin} {et~al.}(2016{\natexlab{a}}){Tobin}, {Kratter}, {Persson},
  {Looney}, {Dunham}, {Segura-Cox}, {Li}, {Chandler}, {Sadavoy}, \&
  {Harris}}]{Tobin2016Nature}
{Tobin}, J.~J., {Kratter}, K.~M., {Persson}, M.~V., {et~al.}
  2016{\natexlab{a}},
  \href{http://dx.doi.org/10.1038/nature20094}{\JournalTitle{\nat}, 538, 483}

\bibitem[{{Tobin} {et~al.}(2016{\natexlab{b}}){Tobin}, {Looney}, {Li},
  {Chandler}, {Dunham}, {Segura-Cox}, {Sadavoy}, {Melis}, {Harris}, {Kratter},
  \& {Perez}}]{tobin16}
{Tobin}, J.~J., {Looney}, L.~W., {Li}, Z.-Y., {et~al.} 2016{\natexlab{b}},
  \href{http://dx.doi.org/10.3847/0004-637X/818/1/73}{\JournalTitle{\apj}, 818,
  73}

\bibitem[{{Tohline}(2002)}]{Tohline2002}
{Tohline}, J.~E. 2002,
  \href{http://dx.doi.org/10.1146/annurev.astro.40.060401.093810}{\JournalTitle{\araa},
  40, 349}

\bibitem[{{Tokovinin}(2017)}]{Tokovinin2017}
{Tokovinin}, A. 2017,
  \href{http://dx.doi.org/10.1093/mnras/stx707}{\JournalTitle{\mnras}, 468,
  3461}

\bibitem[{{Truelove} {et~al.}(1998){Truelove}, {Klein}, {McKee}, {Holliman},
  {Howell}, {Greenough}, \& {Woods}}]{truelove98}
{Truelove}, J.~K., {Klein}, R.~I., {McKee}, C.~F., {et~al.} 1998,
  \href{http://dx.doi.org/10.1086/305329}{\JournalTitle{\apj}, 495, 821}

\bibitem[{{Turk} {et~al.}(2011){Turk}, {Smith}, {Oishi}, {Skory}, {Skillman},
  {Abel}, \& {Norman}}]{turkyt}
{Turk}, M.~J., {Smith}, B.~D., {Oishi}, J.~S., {et~al.} 2011,
  \href{http://dx.doi.org/10.1088/0067-0049/192/1/9}{\JournalTitle{\apjs}, 192,
  9}

\bibitem[{{Whitworth} {et~al.}(1995){Whitworth}, {Chapman}, {Bhattal},
  {Disney}, {Pongracic}, \& {Turner}}]{Whitworth1995}
{Whitworth}, A.~P., {Chapman}, S.~J., {Bhattal}, A.~S., {et~al.} 1995,
  \href{http://dx.doi.org/10.1093/mnras/277.2.727}{\JournalTitle{\mnras}, 277,
  727}

\bibitem[{{Wong} {et~al.}(2011){Wong}, {Hughes}, {Ott}, {Muller}, {Pineda},
  {Bernard}, {Chu}, {Fukui}, {Gruendl}, {Henkel}, {Kawamura}, {Klein},
  {Looney}, {Maddison}, {Mizuno}, {Paradis}, {Seale}, \& {Welty}}]{wongetal11}
{Wong}, T., {Hughes}, A., {Ott}, J., {et~al.} 2011,
  \href{http://dx.doi.org/10.1088/0067-0049/197/2/16}{\JournalTitle{\apjs},
  197, 16}

\bibitem[{{Xie} {et~al.}(2011){Xie}, {Payne}, {Th{\'e}bault}, {Zhou}, \&
  {Ge}}]{Xie2011paper1}
{Xie}, J.-W., {Payne}, M.~J., {Th{\'e}bault}, P., {Zhou}, J.-L., \& {Ge}, J.
  2011,
  \href{http://dx.doi.org/10.1088/0004-637X/735/1/10}{\JournalTitle{\apj}, 735,
  10}

\end{thebibliography}
